\documentclass[twocolumn]{aastex631}

\usepackage{amsmath}
\usepackage{enumitem}

\newcommand{\lSect}[1]{{\label{sec:#1}}}

\newcommand{\lEq}[1]{{\label{eq:#1}}}
\newcommand{\lTab}[1]{{\label{tab:#1}}}

\newcommand{\Tabff}[1]{{\ref{tab:#1}}}
\newcommand{\Tab}[1]{{Table~\Tabff{#1}}}

\newcommand{\pan}[1]{{\textit{#1}}}

\newcommand{\FIGFF}[2]{{\ref{fig:#2}\pan{#1}}}

\newcommand{\Figure}[1]{{Figure~\FIGFF{}{#1}}}

\newcommand{\Sectff}[1]{{\ref{sec:#1}}}
\newcommand{\Sect}[1]{{\S\Sectff{#1}}}

\newcommand{\isofont}[1]{{\mathrm{#1}}}
\newcommand{\isomass}[1]{{\ensuremath{\isofont{^{#1}}}}}
\newcommand{\isocharge}[1]{{\ensuremath{\isofont{_{#1}}}}}
\newcommand{\isotope}[3]{{\ensuremath{\isocharge{#1}\isomass{#2}\isofont{#3}}}}

\newcommand{\I}[2]{{\isotope{}{#1}{#2}}}

\newcommand{\Ep}[1]{{\ensuremath{10^{#1}}}}
\newcommand{\E}[1]{{\ensuremath{\powersep\Ep{#1}}}}

\newcommand{\powersep}{\times}

\newcommand{\tqs}{\textquotesingle}

\newcommand{\unitspace}{\ensuremath{\,}}
\newcommand{\usp}{\unitspace}

\newcommand{\unitstyle}[1]{\ensuremath{\mathrm{#1}}}
\newcommand{\power}[2]{\ensuremath{{#1}^{#2}}}

\newcommand{\centi}{\unitstyle{c}}

\newcommand{\Mega}{\unitstyle{M}}

\newcommand{\meter}{\unitstyle{m}}

\newcommand{\second}{\unitstyle{s}}

\newcommand{\Kelvin}{\unitstyle{K}}
\newcommand{\K}{\Kelvin}  

\newcommand{\cm}{\centi\meter}
\newcommand{\gram}{\unitstyle{g}}

\newcommand{\grampercc}{\gram\usp\power{\cm}{-3}} 

\newcommand{\cmpersecondSc}{\cm\usp\power{\second}{-2}} 

\newcommand{\eV}{\unitstyle{eV}}        
\newcommand{\MeV}{\Mega\eV} 

\newcommand{\Msun}{\ensuremath{\unitstyle{M}_\odot}}
\newcommand{\Lsun}{\ensuremath{\unitstyle{L}_{\odot}}}

\newcommand{\Msunyr}{\Msun\,\power{\yr}{-1}}

\newcommand{\yr}{\unitstyle{yr}}        

\newcommand{\code}[1]{\texttt{#1}}
\newcommand{\mesa}{\code{MESA}}
\newcommand{\MESA}{\mesa}

\newcommand{\ADIPLS}{\code{ADIPLS}}
\newcommand{\GYRE}{\code{GYRE}}

\newcommand{\rsp}{\code{RSP}}

\newcommand{\REACLIB}{\code{REACLIB}}

\newcommand{\mesastar}{\mesa\code{star}}

\newcommand{\MESAstar}{\mesastar}
\newcommand{\num}{\code{num}}

\newcommand{\kap}{\code{kap}}
\newcommand{\eos}{\code{eos}}

\newcommand{\atm}{\code{atm}}

\newcommand{\astero}{\code{astero}}

\newcommand{\kB}{\ensuremath{k_\mathrm{B}}} 
\newcommand{\NA}{\ensuremath{N_\mathrm{\!A}}} 

\input{vectors.tex}

\newcommand{\dif}{\ensuremath{\mathrm{d}}}
\newcommand{\Dif}{\ensuremath{\mathrm{D}}}

\newcommand{\D}{{\mathrm d}}

\newcommand{\dd}[2]{\frac{\partial #1}{\partial #2}} 
\newcommand{\ddp}[2]{\left(\frac{\partial #1}{\partial #2}\right)} 
\newcommand{\DDt}[1]{\frac{\Dif #1}{\Dif t}} 
\newcommand{\ddm}[1]{\frac{\partial #1}{\partial m}} 

\newcommand{\nuclei}[2]{\ensuremath{\mathrm{^{#1}#2}}}

\newcommand{\lithium}[1][7]{\nuclei{#1}{Li}}
\newcommand{\beryllium}[1][9]{\nuclei{#1}{Be}}

\newcommand{\carbon}[1][12]{\nuclei{#1}{C}}

\newcommand{\nickel}[1][58]{\nuclei{#1}{Ni}}
\newcommand{\copper}[1][63]{\nuclei{#1}{Cu}}

\newcommand{\dm}[1]{\ensuremath{dm_{#1}}} 

\newcommand{\timestep}{\ensuremath{\delta t}} 

\newcommand{\dt}{\ensuremath{\delta t}}
\newcommand{\area}{\ensuremath{\mathcal{A}}}

\newcommand{\bvv}{Brunt-V\"ais\"al\"a}

\newcommand{\start}[1]{\ensuremath{#1_{\rm start}}} 
\newcommand{\epsnuc}{\ensuremath{\epsilon_{\mathrm{nuc}}}} 
\newcommand{\epsgrav}{\ensuremath{\epsilon_{\mathrm{grav}}}} 
\newcommand{\epsgravX}{\ensuremath{\epsilon_{{\rm grav},X_i}}}
\newcommand{\epsnu}{\ensuremath{\epsilon_{\mathrm{\nu}}}} 

\newcommand{\grada}{\ensuremath{\nabla_{{\rm ad}}}}

\newcommand{\gradL}{\ensuremath{\nabla_{{\rm L}}}}
\newcommand{\gradr}{\ensuremath{\nabla_{{\rm rad}}}}
\newcommand{\gradT}{\ensuremath{\nabla}}
\newcommand{\Teff}{\ensuremath{T_{\rm eff}}}	

\newcommand{\BV}{Brunt-V\"{a}is\"{a}l\"{a}}

\newcommand{\rhoc}{\ensuremath{\rho_{\mathrm{c}}}} 
\newcommand{\gcc}{\ensuremath{\mathrm{g\,cm^{-3}}}}
\newcommand{\logrhoc}{\ensuremath{\log(\rhoc/\gcc)}}

\newcommand{\Tsurf}{T_{\rm s}}
\newcommand{\Psurf}{P_{\rm s}}
\newcommand{\tausurf}{\tau_{\rm s}}

\newcommand{\gammaone}{\ensuremath{\Gamma_1}}

\newcommand{\Tbcz}{T_{\rm BCZ}} 
\newcommand{\Pbcz}{P_{\rm BCZ}}
\newcommand{\logTbcz}{\ensuremath{\log(T_{\rm BCZ}/\mathrm{K})}}

\newcommand{\theat}{t_{\rm heat}} 
\newcommand{\tdyn}{t_{\rm dyn}} 
\newcommand{\taccel}{t_{\rm accel}} 
\newcommand{\dlnTdt}{\left| \dif \ln T / \dif t \right|^{-1}} 

\newcommand{\mesaone}{MESA~I}  
\newcommand{\mesatwo}{MESA~II} 
\newcommand{\mesathree}{MESA~III} 
\newcommand{\mesafour}{MESA~IV} 
\newcommand{\mesafive}{MESA~V} 

\newcommand{\revision}[1]{{#1}} 

\newlength{\apjcolwidth}
\newlength{\figwidth}
\newlength{\doublewide}
\setlist[itemize]{noitemsep, topsep=0pt, leftmargin=*}

\begin{document}
\title{Modules for Experiments in Stellar Astrophysics (MESA):\\ Time-Dependent Convection, Energy Conservation,\\ Automatic Differentiation, and Infrastructure}

\author[0000-0001-5048-9973]{Adam S. Jermyn}
\affiliation{Center for Computational Astrophysics, Flatiron Institute, New York, NY 10010, USA}

\author[0000-0002-4791-6724]{Evan B. Bauer}
\affiliation{Center for Astrophysics $\vert$ Harvard \& Smithsonian, 60 Garden St, Cambridge, MA 02138, USA}

\author[0000-0002-4870-8855]{Josiah Schwab}
\affiliation{Department of Astronomy and Astrophysics, University of California, Santa Cruz, CA 95064, USA}

\author[0000-0003-3441-7624]{R. Farmer}
\affiliation{Max-Planck-Institut f{\"u}r Astrophysik, Karl-Schwarzschild-Stra{\ss}e 1, 85741 Garching, Germany}

\author[0000-0002-4773-1017]{Warrick H. Ball}
\affiliation{School of Physics and Astronomy, University of Birmingham, Edgbaston, Birmingham B15 2TT, UK}

\author[0000-0003-4456-4863]{Earl P. Bellinger}
\affiliation{Max-Planck-Institut f{\"u}r Astrophysik, Karl-Schwarzschild-Stra{\ss}e 1, 85741 Garching, Germany}
\affiliation{Stellar Astrophysics Centre, Aarhus University, 8000 Aarhus C, Denmark}

\author[0000-0002-4442-5700]{Aaron Dotter}
\affiliation{Department of Physics and Astronomy, Dartmouth College, Hanover, NH 03755 USA}

\author[0000-0002-8717-127X]{Meridith Joyce}
\affiliation{Space Telescope Science Institute, 3700 San Martin Drive, Baltimore, MD 21218, USA}

\author[0000-0002-0338-8181]{Pablo Marchant}
\affiliation{Institute of Astrophysics, KU Leuven, Celestijnenlaan 200D, 3001 Leuven, Belgium}

\author[0000-0002-9901-3113]{Joey S. G. Mombarg}
\affiliation{Institute of Astronomy, KU Leuven, Celestijnenlaan 200D, Leuven, Belgium}

\author[0000-0002-6828-0630]{William M.~Wolf}
\affiliation{Department of Physics and Astronomy, University of Wisconsin-Eau Claire, Eau Claire, WI 54701, USA}

\author[0000-0001-9195-7390]{Tin Long Sunny Wong}
\affiliation{Department of Physics, University of California, Santa Barbara, CA 93106, USA}

\author[0000-0001-7902-8134]{Giulia C. Cinquegrana}
\affiliation{School of Physics \& Astronomy, Monash University, Clayton VIC 3800, Australia}
\affiliation{ARC Centre of Excellence for All Sky Astrophysics in 3 Dimensions (ASTRO 3D)}

\author[0000-0001-5631-5878]{Eoin Farrell}
\affiliation{School of Physics, Trinity College Dublin, The University of Dublin, Dublin 2, Ireland}

\author[0000-0001-7217-4884]{R.~Smolec}
\affiliation{Nicolaus Copernicus Astronomical Center of the Polish Academy of Sciences, Bartycka 18, PL-00-716 Warszawa, Poland}  

\author[0000-0002-8107-118X]{Anne Thoul}
\affiliation{Space sciences, Technologies and Astrophysics Research (STAR) Institute, Universit\'e de Li\`ege, All\'ee du 6 Ao$\hat{u}$t 19C, Bat. B5C, 4000 Li\`ege, Belgium}

\author[0000-0002-8171-8596]{Matteo Cantiello}
\affiliation{Center for Computational Astrophysics, Flatiron Institute, 162 5th Avenue, New York, NY 10010, USA}
\affiliation{Department of Astrophysical Sciences, Princeton University, Princeton, NJ 08544, USA}

\author[0000-0001-8087-9278]{Falk Herwig}
\affiliation{Dept. of Physics and Astronomy, University of Victoria, Victoria, BC V8P5C2, Canada}

\author[0000-0002-2398-719X]{Odette Toloza}
\affiliation{Millennium Nucleus for Planet Formation, NPF, Valpara\'iso, Chile}
\affiliation{Departamento de F\'isica, Universidad T\'ecnica Federico Santa Mar\'ia, Avenida Espa\~na 1680, Valpara\'iso, Chile}

\author[0000-0001-8038-6836]{Lars Bildsten}
\affiliation{Kavli Institute for Theoretical Physics, University of California, Santa Barbara, CA 93106, USA}
\affiliation{Department of Physics, University of California, Santa Barbara, CA 93106, USA}

\author[0000-0002-2522-8605]{Richard H. D. Townsend}
\affiliation{Department of Astronomy, University of Wisconsin-Madison, Madison, WI 53706, USA}

\author[0000-0002-0474-159X]{F.X.~Timmes}
\affiliation{School of Earth and Space Exploration, Arizona State University, Tempe, AZ 85287, USA}


\begin{abstract}
\noindent
We update the capabilities of the open-knowledge software instrument Modules for Experiments in Stellar Astrophysics (\MESA).
The new \code{auto\_diff} module implements automatic differentiation in \MESA, an enabling capability that alleviates
the need for hard-coded analytic expressions or finite difference approximations.
We significantly enhance the treatment of the growth and decay of convection in \MESA\ with a new model for time-dependent convection,
which is particularly important during late-stage nuclear burning in massive stars and electron degenerate ignition events.
We strengthen \MESA's implementation of the equation of state, and we quantify continued improvements to energy accounting
and solver accuracy through a discussion of different energy equation features and enhancements.
To improve the modeling of stars in \MESA\, we describe key updates to the 
treatment of stellar atmospheres, molecular opacities, Compton opacities, conductive opacities, element diffusion coefficients, and nuclear reaction rates.
We introduce treatments of starspots, an important consideration for low-mass stars, and 
modifications for superadiabatic convection in radiation-dominated regions.
We describe new approaches for increasing the efficiency of calculating monochromatic opacities and radiative levitation,
and for increasing the efficiency of evolving the late stages of massive stars with a new operator split nuclear burning mode.
We close by discussing major updates to \MESA's software infrastructure that enhance source code development and community engagement.

\end{abstract}


\tableofcontents

\section{Introduction}
\label{sec:introduction}

A resurgence of stellar astrophysics research is being fueled by the
transformative capabilities in space- and ground-based hardware
instruments providing an unprecedented volume of high-quality
measurements of stars, significantly strengthening and extending the
observational data upon which stellar astrophysics ultimately rests \citep{decadal_2021_aa}.
Examples include:

Several individual stars at redshifts of $\simeq$ 1 have been discovered
by temporary magnification factors of $\simeq$ 1000 from microlensing
\citep{kelly_2018_aa,rodney_2018_aa,chen_2019_aa}.
A more persistent and highly magnified star at a redshift of 6.2 has also been discovered with the
{\it Hubble Space Telescope} \citep{welch_2022_aa}
by a fortuitous alignment with a foreground galaxy cluster lens caustic \citep{windhorst_2018_aa}.
The infrared instruments aboard the {\it James Webb Space Telescope}
\citep{gardner_2006_aa,beichman_2012_aa,artigau_2014_aa,rieke_2015_aa,labiano_2021_aa}
will search for confirmation and spectral classification of this distant star 
\citep{welch_2022_ab} to define its place on the Hertzsprung–Russell Diagram (HRD),
assess how galaxies evolve from their formation \citep{zackrisson_2011_aa,robertson_2021_aa},
observe the formation of stars \citep{senarath_2018_aa,boquien_2021_aa},
and measure the properties of stellar-planetary systems including the Solar System \citep{sarkar_2021_aa,patapis_2022_aa}.

In the late 2020s, kilometer-scale gravitational wave detectors such as
{\it Advanced Laser Interferometer Gravitational Observatory} \citep{ligo-scientific-collaboration_2015_aa},
{\it Advanced Virgo} \citep{acernese_2015_aa} and
{\it Kamioka Gravitational Wave Detector} \citep{akutsu_2021_aa}
will routinely detect tens of binary neutron-star mergers with kilonovae annually \citep{abbott_2018_aa},
probe how kilonova r-process nucleosynthetic yields vary with environment \citep[e.g.,][]{barnes_2021_aa},
and assess the populations that contribute to the stellar black hole mass distribution,
including the presence of any gaps in the distribution \citep{perna_2019_aa,zevin_2021_aa,renzo_2020_aa,renzo_2021_aa,mandel_2022_aa}.

The next core-collapse supernova in the Milky Way or its satellites will be
a unique opportunity to observe the explosion of a star. The {\it SuperNova Early Warning System}
is a global network of neutrino experiments sensitive to supernova neutrinos
\citep{al-kharusi_2021_aa} that includes multi-kiloton detectors such as
{\it KamLAND} \citep{Araki:2004mb},
{\it Borexino} \citep{borexino-collaboration_2018_aa,borexino-collaboration_2020_aa},
{\it SNO+} \citep{Andringa:2015tza},
{\it Daya Bay} \citep{Guo:2007ug},
{\it SuperKamiokande} \citep{simpson_2019_aa}, and the upcoming
{\it HyperKamiokande} \citep{abe_2016_aa},
{\it DUNE} \citep{Acciarri:2016ooe} and
{\it JUNO} \citep{juno-collaboration_2022_aa}.
Searching for pre-supernova neutrinos is ongoing, and of interest as they allow
tests of stellar and neutrino physics \citep[e.g.,][]{kosmas_2022_aa} and enable an early alert of
an impending core-collapse supernova to the electromagnetic and gravitational wave communities
\citep{beacom_1999_aa, vogel_1999_aa, mukhopadhyay_2020_aa,al-kharusi_2021_aa}.

Sky surveys that probe ever-larger areas of the dynamic sky and ever-fainter transient sources
include
the {\it Imaging X-ray Polarimetry Explorer} \citep{soffitta_2021_aa},
the {\it Compton Spectrometer and Imager} \citep{tomsick_2022_aa},
{\it eROSITA} \citep{predehl_2021_aa},
{\it Gaia} \citep{gaia-collaboration_2016_aa,gaia-collaboration_2018_aa,gaia-collaboration_2021_aa},
the {\it Sloan Digital Sky Survey} \citep{york_2000_aa,abdurrouf_2022_aa},
the {\it All-Sky Automated Survey for Supernovae} \citep{chen_2022_aa},
{\it Pan-STARRS1} \citep{flewelling_2020_aa},
the {\it Zwicky Transient Factory} \citep{bellm_2019_aa,dhawan_2022_aa},
{\it Gattini-IR} \citep{moore_2016_aa}, and
the {\it Nancy Grace Roman Space Telescope} \citep{akeson_2019_aa}.
{\it Roman} will measure proper motions of stars several magnitudes
fainter than the {\it Gaia} mission \citep{brandt_2021_aa,dorn-wallenstein_2021_aa}, which is
sufficient to probe the main sequence turnoff to distances of $\simeq$\,10\,kpc
and red giants throughout the Galactic halo \citep{spergel_2015_aa}.

Wide-field spectroscopic surveys in the coming decade will resolve
stellar populations and the Milky Way's structure \citep{bolton_2019_aa}
at facilities such as
{\it Gaia} DR3 \citep{gaia-collaboration_2021_aa},
{\it SDSS-V} \citep{kollmeier_2017_aa},
{\it FOBOS} \citep{bundy_2019_aa},
{\it Maunakea Spectroscopic Explorer} \citep{marshall_2019_aa}, and
{\it SpecTel} \citep{ellis_2019_aa}.
For example, {\it FOBOS}
is a next-generation spectroscopic facility at the {\it W.M. Keck Observatory} that
will provide multi-epoch, high-multiplex, and deep spectroscopic
follow-up of panoramic deep-imaging surveys.

The {\it Vera C. Rubin Observatory}
will conduct a multicolor optical survey of the Southern Hemisphere sky, the Legacy Survey of Space and Time
\citep{lsst-science-collaboration_2017_aa, ivezic_2019_aa}, to
probe dark energy and dark matter \citep{sanchez_2021_aa, zhang_2022_aa},
explore the transient optical sky
\citep{bianco_2022_aa, li_2022_aa, raiteri_2022_aa, hernitschek_2022_aa, andreoni_2022_aa, bellm_2022_aa},
and build a catalog of solar system objects with an order of magnitude more objects
\citep{lsst-solar-system-science-collaboration_2020_aa, schwamb_2021_aa}.

The {\it TESS} mission \citep{ricker_2016_aa} is providing
systematic measurements of the radii, masses, and ages of 200,000 individual stars
sampled at a 2 minute cadence to open a new era of stellar variability
exploration \citep[e.g.,][]{huang_2018_aa,ball_2018_aa,dragomir_2019_aa,wang_2019_aa}.
Within the next decade, the {\it Planetary Transits and Oscillations of Stars} mission
\citep[PLATO;][]{rauer_2014_exa}
will search for planetary transits across up to one million stars,
characterize rocky extrasolar planets around yellow dwarf stars, subgiant stars, and red dwarf stars \citep{montalto_2021_aa}, and
investigate the seismology of stars \citep{miglio_2017_aa,cunha_2021_aa,nascimbeni_2022_aa}.

In partnership with this ongoing explosion of activity in stellar
astrophysics, revolutionary advances in software infrastructure, computer processing power,
data storage capability, and open-knowledge software instruments are
transforming how stellar theory, modeling, and simulations interact with experiments and observations.
Examples include
\code{Astropy} \citep{astropy-collaboration_2018_aa,Astropy2022},
\code{Athena++} \citep{stone_2020_aa,jiang_2021_aa},
\code{Castro} \citep{almgren_2020_aa},
\code{Dedalus} \citep{burns_2020_aa},
\code{emcee} \citep{foreman-mackey_2013_aa},
\code{Flash-X} \citep{dubey_2022_aa},
\GYRE\ \citep{Townsend:2013,Townsend:2018},
\code{MAESTROeX} \citep{fan_2019_aa},
\code{MESA2Hydro} \citep{Joyce2019},
\code{MSG} \citep{townsend_2022_aa},
\code{Phantom} \citep{price_2018_aa},
\code{PHOEBE} \citep{conroy_2020_aa},
\code{Starlib} \citep{sallaska_2013_aa},
\code{TARDIS} \citep{vogl_2019_aa},
\code{TULIPS} \citep{laplace_2022_aa},
and \code{yt} \citep{turk_2011_aa}.

The previous Modules for Experiments in Stellar Astrophysics software instrument papers
(\citealt[][\mesaone]{paxton_2011_aa};
\citealt[][\mesatwo]{paxton_2013_aa};
\citealt[][\mesathree]{paxton_2015_aa};
\citealt[][\mesafour]{paxton_2018_aa};
\citealt[][\mesafive]{paxton_2019_aa}),
as well as this one, describe new capabilities and limitations of \MESA\ while
also comparing to other available numerical or analytic results.
We do not fully explore the science implications
in this software instrument paper.  The scientific potential of these
new capabilities will be unlocked by future efforts of
the \MESA\ research community.

This MESA VI software instrument paper is organized as follows.
Section~\ref{sec:auto_diff} describes the implementation of automatic differentiation and
\S\ref{sec:tdc} introduces a new model for time dependent convection.
Section~\ref{sec:eos} describes improvements to \MESA's implementation of the equation of state (EOS) and
\S\ref{sec:energy_eq} discusses treatments of the energy equation.
Section~\ref{sec:atmosphere} describes treatments of the stellar atmosphere and
\S\ref{sec:mltconv} introduces new models of starspots and a superadiabatic convection.
Section~\ref{sec:opacity} reports improvements to the opacities,
\S\ref{sec:diffusion} to the element diffusion coefficients,
\S\ref{sec:rates} to the nuclear physics,  and
\S\ref{sec:const} to the physical constants.
Section~\ref{sec:infrastructure} discusses \MESA's infrastructure. Finally,
\S\ref{sec:summary} summarizes MESA VI.

Important symbols are defined in Table \ref{table:list-of-symbols}.
Acronyms are defined in Table \ref{tab:acronym}.
Components of \MESA, such as modules and routines, are in typewriter font e.g., \texttt{tdc}.

\startlongtable

\begin{deluxetable}{clc}
  \tablecolumns{3}
  \tablewidth{1.0\apjcolwidth}
  \tablecaption{Important symbols.
   \label{table:list-of-symbols}}
  \tablehead{\colhead{Name} & \colhead{Description} & \colhead{Appears}}
  \startdata
  $a$            & Radiation constant                  & \ref{sec:superad}        \\
\area          & $4 \pi r^2$ \ \ Area of face        & \ref{sec:tdc}        \\
$c$            & Speed of light in a vacuum          & \ref{sec:opacity}     \\
$D$            & Element diffusion coefficient       & \ref{sec:diffusion}  \\
$e$            & Specific internal energy            & \ref{sec:eos}        \\ 
$E$            & Energy                              & \ref{sec:energy_eq}  \\
$f_\text{spot}$ & Filling factor                     & \ref{sec:starspots}  \\
$g$            & Gravitational acceleration          & \ref{sec:atmosphere} \\
$G$            & Gravitational constant              & \ref{sec:atmosphere} \\
$h$            & Pressure scale height               & \ref{sec:tdc}        \\
$\kappa$       & Opacity                             & \ref{sec:atmosphere} \\
$\lambda$      & Reaction rate                       & \ref{sec:rates}      \\
\kB            & Boltzmann constant                  & \ref{sec:diffusion}  \\
$L$            & Luminosity                          & \ref{sec:tdc}        \\
$m$            & Mass coordinate                     & \ref{sec:tdc}        \\ 
$M$            & Stellar mass                        & \ref{sec:tdc}        \\ 
$n$            & Number density                      & \ref{sec:diffusion}  \\ 
\NA            & Avogadro number                     & \ref{sec:opacity}    \\
$P$            & Pressure                            & \ref{sec:tdc}        \\ 
$Q$            & Thermal expansion                   & \ref{sec:tdc}        \\ 
$r$            & Radial coordinate                   & \ref{sec:atmosphere} \\ 
$R$            & Stellar radius                      & \ref{sec:atmosphere} \\ 
$\rho$         & Mass density                        & \ref{sec:tdc}        \\ 
$s$            & Specific entropy                    & \ref{sec:eos}        \\
$\sigma$       & Stefan-Boltzmann constant           & \ref{sec:tdc}        \\
$t$            & Time                                & \ref{sec:tdc}        \\ 
$T$            & Temperature                         & \ref{sec:tdc}        \\
$u$            & Velocity                            & \ref{sec:energy_eq}  \\ 
$w$            & Turbulent velocity                  & \ref{sec:tdc}        \\ 
$x_\text{spot}$ & Temperature contrast               & \ref{sec:starspots}   \\
X              & Hydrogen mass fraction              & \ref{sec:tdc}        \\ 
Y              & Helium mass fraction                & \ref{sec:tdc}        \\ 
$\mathcal{Y}$  & Superadiabaticity  \gradT$-$\grada  & \ref{sec:tdc}        \\ 
$Z$            & Charge                              & \ref{sec:diffusion}  \\
Z              & Metal mass fraction                 & \ref{sec:diffusion}  \\


$a_e$              & Electron spacing (4$\pi n_e$/3)$^{-1/3}$                               & \ref{sec:diffusion}  \\ 
$\alpha$           & Convective flux parameter                      & \ref{sec:tdc}        \\ 
$\alpha_D$         & Convective flux parameter                      & \ref{sec:tdc}        \\ 
$\alpha_r$         & Convective flux parameter                      & \ref{sec:tdc}        \\ 
$\alpha_{P_t}$     & Convective flux parameter                      & \ref{sec:tdc}        \\ 
$c_P$              & Specific heat at constant pressure             & \ref{sec:tdc}        \\ 
$\alpha_t$         & \revision{Convective flux parameter}                      & \ref{sec:tdc}        \\
$c_V$              & Specific heat at constant volume  $\partial e/\partial T|_\rho$ & \ref{sec:energy_eq} \\ 
\timestep          & Numerical time step                            & \ref{sec:energy_eq}  \\ 
\dm{}              & Mass of cell                                   & \ref{sec:energy_eq}  \\ 
$\epsilon$         & Energy generation rate                         & \ref{sec:energy_eq}  \\ 
\epsgrav           & Gravitational heating rate                     & \ref{sec:eos}        \\
\epsnuc            & Nuclear energy generation rate                 & \ref{sec:rates}        \\
$\epsilon_q$       & Viscous heating rate                            & \ref{sec:tdc}        \\ 
$e_t$              & Specific kinetic energy of turbulence          & \ref{sec:tdc}        \\ 
$\Gamma$           & Efficiency of convection                       & \ref{sec:tdc}        \\ 
$\Gamma_{1}$       & First adiabatic exponent $(\partial \ln P/\partial \ln \rho)_{\rm ad}$ & \ref{sec:tdc} \\ 
$\Gamma_{3}$       & Third adiabatic exponent                       & \ref{sec:tdc}        \\ 
$\Gamma_{\rm MCP}$ & Multi-component plasma coupling parameter      & \ref{sec:diffusion}  \\ 
\grada             & Adiabatic temperature gradient                 & \ref{sec:tdc}        \\
$\nabla_e$         & Temperature gradient of convective eddy        & \ref{sec:tdc}        \\
\gradL             & Ledoux temperature gradient                    & \ref{sec:tdc}        \\ 
\gradr             & Radiative temperature gradient                 & \ref{sec:tdc}        \\ 
\gradT             & Temperature gradient                           & \ref{sec:tdc}  \\
$K_{ij}$           & Resistance coefficients                        & \ref{sec:diffusion}  \\
$\lambda_e$        & Electron screening length                      & \ref{sec:diffusion}  \\
$L_t$              & Luminosity of turbulent kinetic energy         & \ref{sec:tdc}        \\ 
$L_{\rm edd}$      & Eddington luminosity                           & \ref{sec:superad}        \\ 
$L_{\rm rad}$      & Radiative luminosity                           & \ref{sec:superad}        \\ 
$\mu$              & Molecular weight                               & \ref{sec:opacity}    \\ 
$\dot{M}$          & Mass transfer rate                             & \ref{sec:tdc}    \\ 
$P_t$              & Turbulent pressure                             & \ref{sec:tdc}        \\ 
$q_e$              & Electric charge                                & \ref{sec:diffusion} \\
$\tau$             & Optical depth                                  & \ref{sec:atmosphere} \\
\Teff              & Effective temperature                          & \ref{sec:atmosphere} \\
$u$                & Cell velocity                                  & \ref{sec:energy_eq}        \\ 
$v_c$              & Convection velocity                            & \ref{sec:tdc}        \\ 
$\chi_{\rho}$      & Adiabatic index ($\partial$log$P$/$\partial$log$\rho)|_{T,X}$  & \ref{sec:tdc}        \\ 
$\chi_T$           & Adiabatic index ($\partial$log$P$/$\partial$log$T)|_{\rho,X}$  & \ref{sec:tdc}        \\ 
$X_i$              & Mass fraction                                                  & \ref{sec:tdc}        \\

  \enddata
\tablenotetext{}
{\hsize \apjcolwidth {\bf Note:} Single character symbols are listed first,
symbols with modifiers are listed second.
Some symbols may be further subscripted, for example,
by $\mathrm{c}$ (indicating a central quantity),
by $\mathrm{s}$ (indicating a surface quantity),
by a cell index $k$, or by
species index $i$.
}
\end{deluxetable}

\startlongtable
\begin{deluxetable}{clc}
  \tablecolumns{3}
  \tablewidth{0.9\apjcolwidth}
  \tablecaption{Acronyms used in this article.\label{tab:acronym}}
  \tablehead{ \colhead{Acronym} & \colhead{Description} & \colhead{Appears} }
  \startdata
  AGB    & Asymptotic Giant Branch        & \ref{sec:opacity} \\
BCZ    & Base of the Convection Zone    & \ref{sec:tdc} \\  
DA     & White dwarf spectral type      & \ref{sec:opacity} \\  
DB     & White dwarf spectral type      & \ref{sec:atmosphere} \\  
EOS    & Equation of State              & \ref{sec:introduction} \\  
HRD    & Hertzsprung Russell Diagram    & \ref{sec:introduction} \\      
MLT    & Mixing Length Theory           & \ref{sec:tdc} \\
RSG    & Red Super Giant                & \ref{sec:superad} \\  
TAMS   & Terminal Age Main Sequence     & \ref{sec:superad} \\  
TDC    & Time-Dependent Convection      & \ref{sec:tdc} \\  
TP     & Thermal Pulse                  & \ref{sec:opacity} \\
WD     & White Dwarf                    & \ref{sec:tdc} \\
ZAMS   & Zero Age Main Sequence         & \ref{sec:opacity}\\

  \enddata
\end{deluxetable}

\section{Automatic Differentiation}
\label{sec:auto_diff}

\MESA\ solves the equations of stellar evolution implicitly using a
Newton-Raphson method, which requires the partial derivatives of each
equation with respect to the basic structure variables in each cell (e.g., $\rho_k$, $T_k$).
These derivatives need to be computed accurately, typically to one part in $10^{6}$, often precluding use of finite differences.
These derivatives have historically been computed by hard-coding analytic expressions for each equation. This has accounted for a large fraction of the complexity and sources of error in \MESA.

We have largely eliminated this source of error and the associated complexity by using forward-mode operator-overloaded automatic differentiation~\citep{2000JCoAM.124..171B} in the new \texttt{auto\_diff} module.
This functionality provides partial derivatives of expressions automatically with respect to their input variables.
The \texttt{auto\_diff} module provides a number of Fortran derived types for this purpose.
For example, we define the type \code{auto\_diff\_real\_star\_order1}, which contains a floating-point number as well as its first partial derivative with respect to the basic stellar structure variables. The number of partial derivatives is specified at compile-time.
If \code{x} is a variable of this type, then it contains components \code{x\%val} representing the value of \code{x} and \code{x\%d1Array(j)} for the value of $\partial x/\partial \eta_j$, where $\eta_j$ is the $j$-th independent variable.

The \texttt{auto\_diff} types overload operators to implement the chain rule.
This means that a source code line such as \code{f = x * y} is equivalent to
\begin{verbatim}
f%val = x%val * y%val
f%d1Array(j) = x%d1Array(j) * y%val 
               + y%d1Array(j) * x%val
\end{verbatim}
Basic arithmetic and all special and trigonometric functions used in \MESA\, including functions such as \code{min}, \code{max}, and \code{abs}, are provided.
When these functions have discontinuities, we evaluate their derivatives as zero; and where they have discontinuous derivatives, we compute their derivatives as the average between the two sides of the discontinuity.

Using \code{auto\_diff}, expressions like
\begin{align}
	F &= \min\left(\rho_1 e^{T_1 / \sqrt{\rho_{0}}}, \cosh\left(r_{2}-r_1\right)\right)
\end{align}
can be written as
\begin{verbatim}
F = min(rho1*exp(T1/sqrt(rho0)),cosh(r2-r1)).
\end{verbatim}
Together with setup routines that link physical variables (e.g., \texttt{T1}) with the independent variables $\eta_j$, this code automatically provides correct partial derivatives of $F$.

By contrast, explicitly obtaining the partial derivatives of $F$ requires more complex and error-prone source code:
\begin{verbatim}
x0 = rho1*exp(T1/sqrt(rho0))
x1 = cosh(r2-r1)
F = min(x0,x1)
if (x0 < x1) then
   dF_drho1 = x0 / rho1
   dF_drho0 = -T1 * x0 / (2 * sqrt(rho0))
   dF_dT1 = x0/sqrt(rho0)
   dF_dr2 = 0
   dF_dr1 = 0
else
   dF_drho1 = 0
   dF_drho0 = 0
   dF_T1 = 0
   dF_dr2 = sinh(r2-r1)
   dF_dr1 = -sinh(r2-r1)
end if
\end{verbatim}

The \code{auto\_diff} module provides overloaded operators that were generated
using the \code{SymPy}~\citep{10.7717/peerj-cs.103} library in Python to compute power series  and extract chain-rule expressions.
We first optimized these expressions to eliminate common sub-expressions and minimize the number of division operators. We then translated these into Fortran.
This functionality is built on top of the CR-LIBM software package \citep{CR-LIBM}, which enables bit-for-bit identical results across all platforms (see \mesathree).

The \texttt{auto\_diff} module also provides additional \code{auto\_diff\_real} types for alternative use cases.
For convenience, types are provided to support the different hooks in \MESA.
For operations requiring higher-order derivatives, such as in the EOS (see \S\ref{sec:eos}), additional \code{auto\_diff\_real} types provide higher-order mixed partial derivatives.
The chain-rule expressions rapidly become more complicated for higher-order derivatives, but the basic principle is the same.
The \texttt{auto\_diff} machinery was used to benchmark the Skye EOS \citep{Jermyn2021}, with the result that the performance is similar to explicit expressions. Here we provide more detailed benchmarks.

\begin{deluxetable}{lccccc}[ht]
  \tablecolumns{6}
  \tablecaption{Ratio of runtimes for evaluating \code{auto\_diff} expressions relative to the same expressions with \code{real(dp)} (double precision \code{real}) variables. The \code{real(dp)} operations did not calculate any derivatives. Runtimes are averaged over $10^6$ trials, performed on the integers {$1-10^6$} (cast as \code{real(dp)}),
with intermediate results accumulated to prevent the compiler from optimizing away the operations. Label ``\code{*}'' refers to the multiplication operator \code{x*x}, labels ``Xvar\_orderY'' refer to the number of independent variables X and the maximum partial derivative order Y, label ``f'' refers to the function log(cosh(tanh))), and the label ``N'' refers to the number of partial derivatives computed. Timing data was obtained on a 2.4 GHz 8-Core Intel Core i9 running on a 2019 Macbook Pro. \label{tab:auto_diff_perf_op}}
  \tablehead{\colhead{} &\colhead{*}&\colhead{/}&\colhead{log}&\colhead{f}&\colhead{N}}
\startdata
real(dp) & 1  & 1  & 1  & 1 & 0\\
auto\_diff\_real\_1var\_order1 & 3.8 & 6.1 & 1.3 & 1.4 & 1\\
auto\_diff\_real\_2var\_order1 & 4.3 & 11 & 1.4 & 1.7 & 2\\
auto\_diff\_real\_2var\_order3 & 12 & 34 & 2.3 & 2.6 & 9\\
auto\_diff\_real\_star\_order1 & 35 & 77 & 4 & 2.7 & 33\\
\enddata

\end{deluxetable}

\begin{deluxetable}{lccccc}[ht]
  \tablecolumns{6}
  \tablecaption{Same as Table~\ref{tab:auto_diff_perf_op}, but comparing  \code{auto\_diff} performance against explicit \code{real(dp)} routines that include partial derivatives. Partial derivative expressions were constructed and simplified using Mathematica version 12, then implemented manually in Fortran.
\label{tab:auto_diff_perf_direct}}
  \tablehead{\colhead{} &\colhead{*}&\colhead{/}&\colhead{log}&\colhead{f}&\colhead{N}}
\startdata
real(dp) & 1  & 1  & 1  & 1 & 0 \\
auto\_diff\_real\_1var\_order1 & 2.3 & 1.9 & 1.2 & 0.75 & 1 \\
auto\_diff\_real\_2var\_order1 & 2.1 & 1.9 & 1.3 & 0.81 & 2 \\
auto\_diff\_real\_4var\_order1 & 1.9 & 1.3 & 1.3 & 0.74 & 4 \\
\enddata

\end{deluxetable}

Table~\ref{tab:auto_diff_perf_op} compares the runtime cost for several operations and several \code{auto\_diff} types to the cost of evaluating the same expressions in \code{real(dp)} types calculating no partial derivatives.
For operations like multiplication and division, this incurs an overhead of order the number of partial derivatives returned.
For more expensive operations, the overhead is much less, as the \code{auto\_diff} expressions are optimized to re-use intermediate results.

Table~\ref{tab:auto_diff_perf_direct} compares three first-order \code{auto\_diff} types and explicit \code{real(dp)} routines evaluating the same partial derivatives.
There is still overhead for simple operations, but the relative cost no longer scales with the number of derivatives.
For sufficiently complex operations, such as ${\rm f} = \log (\cosh(\tanh(x)))$, the optimized \code{auto\_diff} functions outperform our explicit routines.

For use in stellar evolution calculations, we find the runtime performance of hand-coded expressions are modestly better than those from \texttt{auto\_diff}, because most equations do not depend on all of the independent variables.
However, this overhead is small compared with the full cost of a timestep in \MESA.
Moreover, runtime is often significantly reduced by ensuring that all partial derivatives are correct, as inaccurate derivatives result in slow convergence and a larger number of small timesteps.
Some parts of the \MESA\ source code do not yet use \texttt{auto\_diff}, but this is gradually transitioning.

Four applications of the profound enabling capability of \texttt{auto\_diff} are shown in
\S\ref{sec:tdc} on time-dependent convection,
\S\ref{sec:eos} on \MESA's implementation of the EOS,
\S\ref{sec:starspots} on starspots, and
\S\ref{sec:superad} on superadiabatic convection.
The \texttt{auto\_diff} module can also be used in \texttt{run\_star\_extras}, as well as for software development outside of \MESA.


\section{Time-Dependent \revision{Local} Convection}
\label{sec:tdc}

The mixing length theory (MLT; \revision{\citealt{1932ZA......5..117B}}; \citealt{EBV}) has been used to parameterize convection in 1D stellar models for decades. It assumes that convective turbulence is in a steady state in which the energy input by the convective instability balances damping due to turbulent processes and radiative diffusion.
This is a good approximation when the composition and structure evolve on time-scales that are long compared to the characteristic time-scales of convection.

However, during particularly violent episodes of stellar evolution, it is possible for the structure to evolve faster than convection can reach a steady state.
This is the case in late-stage nuclear burning in massive stars (preceding core collapse), as well as during electron-degenerate ignition events (e.g., He shell flashes, Ne ignition, etc.).
In such cases, the dynamics of convective growth and decay must be incorporated.

To model this, we employ the time-dependent convection (TDC) formalism of \citet{1986A&A...160..116K} \revision{in the local limit}.
We build upon the implementation in \citet{2008AcA....58..193S}, 
introduced in \mesafive\ to model radial stellar pulsations in the \texttt{RSP} module.
\revision{More precisely, we use the one-equation version of the  \citet{1986A&A...160..116K} model,
both in the RSP module and now for general use in stellar evolution calculations.}
\revision{
We caution that combining different mixing models in a stellar evolution calculation
might lead to physically inconsistent solutions, because the different models 
have been developed separately and their underlying assumptions might not be compatible with each other.
Examples include combining the newly implemented time-dependent local limit convection model with
an overshooting model, or combining TDC with other models for chemical composition gradients, rotation, etc.
}

We describe the TDC formalism in \S\ref{sec:tdc_formal}.
In \S\ref{sec:tdc_stab} and \S\ref{sec:tdc_flip} we explain the modifications we have made relative to the implementation in \texttt{RSP} to make TDC numerically stable on long timescales.
Section~\ref{tdc:numerics} then details the TDC solver algorithm.
In \S\ref{sec:MLT_compare} we identify a change to the implementation which makes TDC agree with MLT in the limit of long timescales.
Finally, in \S\ref{tdc:wd} we explore the impact of TDC on models of white dwarfs (WDs) accreting He.

\subsection{Formalism}
\label{sec:tdc_formal}

Following the \citet{1986A&A...160..116K} model, TDC introduces a new variable, the specific kinetic energy in turbulence $e_t$,
which evolves according to
\begin{align}
	\DDt{e_t} + \alpha_{P_t} P_t \DDt{\rho^{-1}} = \epsilon_q + C - \frac{\partial L_t}{\partial m}~.
	\label{eq:et0}
\end{align}
Here $P_t \equiv (2/3) \rho e_t$ is the turbulent pressure, $\alpha_{P_t}$ is a dimensionless free parameter, $\epsilon_q$ is the viscous heating of bulk motion, and $L_t$ accounts for advection of kinetic energy between mass shells
\begin{equation}
L_t = -\area \alpha \alpha_t \rho h e_t^{1/2} \frac{\partial e_t}{\partial r} \ .
\label{eq:Lt}
\end{equation}
\revision{This expression is the same as the turbulent flux $F_t$ in \mesafive, but multiplied by \area\ to convert to a luminosity,
$\alpha_t$ is a convective flux parameter, and  ${h \equiv P/(\rho g)}$ is the pressure scale-height.}
Furthermore,
\begin{equation}
\begin{aligned}
	C \equiv \alpha e_t^{1/2} \frac{T P Q}{h \sqrt{6}} \mathcal{Y} - \alpha_D\left(\frac{8}{3}\sqrt{\frac{2}{3}}\right) \frac{e_t^{3/2}}{\alpha h} \\
                 - \frac{48\sigma \alpha_r}{\alpha^2}\left(\frac{T^3}{\rho^2 c_P \kappa h^2}\right) e_t
\end{aligned}
\label{eq:C}
\end{equation}
groups together sources and sinks of turbulent kinetic energy, including a source/sink from the superadiabaticity
\begin{align}
\mathcal{Y} \equiv \nabla - \nabla_{\rm ad}.
\end{align}
The coefficients $\alpha$, $\alpha_D$, and $\alpha_r$ are free parameters, 
and ${Q \equiv \partial \rho^{-1}/\partial T|_P}$ is the thermal expansion coefficient.
By default $\alpha_r = 0$, which means that TDC neglects radiative damping of convective motions.
\revision{We caution that using $\alpha_r = 0$ is an approximation that changes the physical contents 
and the physical completeness of the model \citep{1986A&A...160..116K, 1987PhDT.......162K, 1998A&A...340..419W}.}
This choice enables subsequent modifications (\S\ref{sec:MLT_compare}) that make TDC reduce to MLT in the limit of long timesteps.
Other defaults are $\alpha$\,=\,2, $\alpha_D$\,=\,1, and $\alpha_{P_{t}}$\,=\,0.
\revision{The choice $\alpha_D$\,=\,1 is equivalent to the \cite{1986A&A...160..116K}
choice of $C_D=(8/3)\sqrt{2/3}$ for compatibility with MLT, and $\alpha_{P_{t}}$\,=\,0 implies $P_t$=0;
see Table 3 of \mesafive \ and \cite{1998A&A...340..419W}.}

The turbulent energy is incorporated into the other equations of stellar structure via heat and momentum transport.
Specifically, in the momentum equation we include a turbulent pressure term $P_t$.
In the luminosity equation we incorporate
\begin{align}
	L = L_{\rm rad} + L_{\rm conv}~,
	\label{eq:eL0}
\end{align}
where $L_{\rm rad}$ is the radiative luminosity and
\begin{align}
	L_{\rm conv} = 4\pi r^2 \alpha \rho c_P T \frac{w}{\sqrt{6}} \mathcal{Y}
	\label{eq:Lconv}
\end{align}
is the convective luminosity.
Here $w \equiv \sqrt{e_t}$ is the turbulent velocity, \revision{and the factor of $\sqrt{6}$ arises from a choice of closure constants.}
\revision{Note $L_t$ is set to zero in Equation \ref{eq:eL0}, that is, the local limit solution is assumed.}
Finally, the luminosity enters the total energy equation, which sets the time evolution of the specific internal energy $e$ in each cell.

To implement TDC in \mesa{}, we drop the term $\epsilon_q$ from the energy equation,
simplifying our implementation. We do not expect this term to matter in most cases,
because bulk velocities are typically much smaller than convective velocities.
When using the Ledoux criterion for convective stability, we further modify TDC in \mesa\ relative to \code{RSP} to set
$\mathcal{Y}$ with the Ledoux gradient $\nabla_{\rm L}$ rather than the adiabatic
gradient $\nabla_{\rm ad}$, as in \cite{1986A&A...160..116K}.

\subsection{Numerical Stability}
\label{sec:tdc_stab}

In \mesa, \rsp\ solves equations~\eqref{eq:et0},~\eqref{eq:eL0} and~\eqref{eq:Lconv} implicitly along with other structure equations to evolve $e_t$ and $e$.
This approach works well on short (convective/pulsational) timescales, but it is numerically unstable on long (evolutionary) timescales.
This poses a challenge, as we want a method that can be used in both limits and smoothly transitions between them.

We conjecture that this numerical instability arises when $e_t$ is a solver variable.
The superadiabaticity $\mathcal{Y}$ sets the time evolution of $e_t$, and hence $L_{\rm conv}$ and $L$.
As $L$ is very sensitive to $\mathcal{Y}$, small errors in $\mathcal{Y}$ result in large errors in $L$.
These errors are not important over time steps shorter than the thermal timescale of a cell (as is the case in \rsp),
as an excess luminosity through one face heats one adjacent cell and cools the other, restoring thermal equilibrium.
With much longer time steps, errors in $L$ significantly alter the entropy profile,
propagating into $\mathcal{Y}$ and producing even larger errors in $L$ with each iteration.

An alternative approach, taken by the MLT implementation in \mesa, is to treat the luminosity as a solver variable determined implicitly by the energy equation (e.g., Equation~\ref{eq:dedt_form}).
From this, \mesa\ derives the temperature gradient needed to produce that luminosity, and requires that the temperature gradient between cells match that computed by MLT.
In effect, this flips the logic around, so that \mesa\ MLT solves for $\mathcal{Y}$ given $L$ whereas \rsp\ solves for $L$ given $\mathcal{Y}$.
Because $L$ is very sensitive to $\mathcal{Y}$, $\mathcal{Y}$ is relatively insensitive to $L$; thus, errors in $L$ produce \emph{smaller} errors in $\mathcal{Y}$, making this approach numerically stable.

\subsection{Flipped Equations}
\label{sec:tdc_flip}

To ensure numerical stability over long timescales, we implement TDC in \mesa\ in the same way as MLT, with $L$ as a solver variable.
We flip Equation~\eqref{eq:Lconv} to solve for $\mathcal{Y}$, accounting for the fact that $w$ depends on $\mathcal{Y}$ via Equation~\eqref{eq:et0}.
Doing so requires a few simplifications and a number of new approaches.

We numerically invert Equations~\eqref{eq:et0} and~\eqref{eq:Lconv} to solve for $\mathcal{Y}$ given $L$.
To do this, we note that the time evolution of $e_t$ in a single cell is nearly independent of $e_t$ in adjacent cells (see Equation~\ref{eq:et0}).
The only direct (rather than implicit) coupling between $e_{t,k}$ and $e_{t,k\pm 1}$ arises through $L_t$.
For simplicity, we currently set ${L_t = 0}$.
This makes $\mathrm{D}e_{t,k}/\mathrm{D}t$ independent of $e_{t,k\pm 1}$ except implicitly via the other structure variables. This in turn makes $\mathcal{Y}_k$ a function only of the local luminosity $L_k$ and solver variables in the adjacent cells.
We then solve for each $\mathcal{Y}_k$ using only local information, and preserve the basic structure of the Jacobian in \mesa\ as well as the runtime performance.

\subsection{Numerical Method}\label{tdc:numerics}

Our goal is to numerically solve Equations~\eqref{eq:et0} and~\eqref{eq:Lconv} for $\mathcal{Y}$ given $L$ with ${L_t = 0}$.
We first construct machinery to evaluate $L$ given $\mathcal{Y}$, and then perform a numerical root-find to obtain $\mathcal{Y}$ given $L$.

\subsubsection{$L$ given $\mathcal{Y}$}

We use ${e_t = w^2}$ to rewrite Equation~\eqref{eq:et0} as
\begin{align}
	2\DDt{w} = \xi_0 + \xi_1 w  + \xi_2 w^2 ,
	\label{eq:ew0}
\end{align}
where $\xi_0$, $\xi_1$, and $\xi_2$ are coefficients that we obtain by expanding the definitions of $C$ and $P_t$ in Equation~\eqref{eq:et0}.
We have divided through by $w$, and so have implicitly excluded one solution branch ($w=0$).
We will return to that branch shortly.

The coefficients are given by
\begin{align}
	\label{eq:xi0}
	\xi_0 &= \frac{\alpha}{h \sqrt{6}} c_P T \nabla_{\rm ad} \mathcal{Y}~,\\
	\label{eq:xi1}
	\xi_1 &= -\left(\frac{4\sigma T^3}{\rho^2 c_P \kappa}\left(\frac{2\alpha_r \sqrt{3}}{\alpha h}\right)^2 + \frac{2}{3}\alpha_{P_t} \rho \DDt{\rho^{-1}}\right)~, \\
	\label{eq:xi2}
	\xi_2 &= -\left(\frac{8}{3}\sqrt{\frac{2}{3}}\right)\frac{\alpha_D}{\alpha h}~.
\end{align}
We fix these to their end-of-step values
and solve for $w$ at the end of the time step given the initial value at the start of the time step.
This implicit approach is numerically stable, and the required  end-of-step values are readily available.
The form of the solution to Equation~\eqref{eq:ew0} depends on the sign of the discriminant $J^2 \equiv \xi_1^2 - 4 \xi_0 \xi_2$.

When $J^2 > 0$, the system is convectively unstable, with the solution
\begin{align}
	w = -\frac{1}{2\xi_2}\left(J \tanh\frac{\lambda + J \delta t}{4} + \xi_1\right),
	\label{eq:Wconv}
\end{align}
after a time step $\delta t$, where $\lambda$ is a constant depending on the initial value of $w$.
With long time steps, the solution grows to a plateau $w \rightarrow -(J+\xi_1)/(2\xi_2)$ independent of this initial condition. We show below that this is consistent with MLT.

When $J^2 < 0$, the system is convectively stable, with the solution
\begin{align}
	w = \frac{1}{2\xi_2}\left(|J| \tan\frac{\lambda+ |J| \delta t}{4} - \xi_1\right)~.
	\label{eq:wtan}
\end{align}
This solution eventually reaches $w(\delta t^*) = 0$ at some time $\delta t^* \sim 1/J$.
Beyond that point the system remains fixed at $w=0$, which is a valid solution to Equation~\eqref{eq:et0} but which was excluded in the form Equation~\eqref{eq:ew0} by dividing through by $w$.
When $J^2 < 0$ we must additionally check to see if the first root of $w$ occurs before the end of the time step and, if it occurs before, set ${w=0}$ at the end of the step rather than evaluating $w$ with Equation~\eqref{eq:wtan}.

Given $w$, we evaluate $L_{\rm conv}$ at the end of the step via Equation~\eqref{eq:Lconv}, and so now have $L$ given $\mathcal{Y}$ as desired.

\subsubsection{Numerical Inversion}

We now invert the relation between $L$ and $\mathcal{Y}$ by solving
\begin{align}
	L(\mathcal{Y}, \boldsymbol \aleph) = L_{\rm solver},
	\label{eq:Leq}
\end{align}
where $L_{\rm solver}$ is the desired luminosity produced by the Newton-Raphson solver, $L(\mathcal{Y},\boldsymbol \aleph)$ is the relation we constructed using Equations~\eqref{eq:Lconv} and~\eqref{eq:ew0}, and $\boldsymbol \aleph$ represents additional structure variables.

To solve equation~\eqref{eq:Leq}, we write it in the form
\begin{align}
	\mathfrak{R}(\mathcal{Y}) \equiv L(\mathcal{Y}, \boldsymbol \aleph) - L_{\rm solver} = 0~,
	\label{eq:Q0}
\end{align}
and then expand $L$ using Equations~\eqref{eq:eL0} and~\eqref{eq:Lconv} as
\begin{align}
	\mathfrak{R}(\mathcal{Y}) \equiv  (L_{\rm solver} - L_0 \nabla_{\rm L}) - (L_0 + c_0 w) \mathcal{Y}~.
	\label{eq:TDC_Q}
\end{align}
Here
\begin{align}
      L_0 \equiv \frac{16 \pi a c}{3} \left(\frac{G m T^4}{\kappa P}\right) \qquad
      c_0 \equiv 4\pi r^2 \frac{\alpha}{\sqrt{6}} \rho T c_P 
\end{align}
are positive quantities set by local properties and independent of $\mathcal{Y}$.
All quantities in Equation~\eqref{eq:TDC_Q} are evaluated at the end of the time step,
determining the sign of the solution for $\mathcal{Y}$ in advance.
The factor $L_0 + c_0 w$ is positive; hence, the sign of $\mathcal{Y}$ matches that of the first term, which is independent of $\mathcal{Y}$.

We evaluate $L_{\rm solver} - L_0 \nabla_{\rm L}$ to determine the sign of $\mathcal{Y}$, followed by a change of variables from $\mathcal{Y} \rightarrow \mathcal{Z} \equiv \ln |\mathcal{Y}|$.
This allows more resolution in $\mathcal{Y}$, which can vary by many orders of magnitude across a stellar model.
We restrict our search to $-100 \leq \mathcal{Z} \leq 100$, covering $10^{-43} \lesssim |\mathcal{Y}| \lesssim 10^{43}$.
We choose such a wide range because we have observed models that enter the extremes of this range, typically involving shocks where both TDC and MLT are suspect.
We have not encountered models with $|\mathcal{Y}|$ approaching $10^{43}$,
and those with $|\mathcal{Y}| < 10^{-43}$ are indistinguishable from $\mathcal{Y}=0$ for the purposes of calculations in \mesa, so this window should cover all cases of interest.

The TDC solver identifies and handles a variety of cases. It takes advantage of the fact that ${\dif w/\dif \mathcal{Y} \geq 0}$,
which follows because the convective velocity always increases as a region becomes more unstable.

We now discuss the different possible solutions.
When $\mathcal{Y} > 0$,
the root-finding problem is monotonic because $\dif w/\dif \mathcal{Y} > 0$ and $\dif \mathfrak{R}/\dif \mathcal{Y} < 0$.
We approach this by performing a bisection search in $\mathcal{Z}$ followed by a Newton-Raphson solve.
The bisection search ensures that the Newton-Raphson solve starts close to the true solution (we require the range $\Delta \mathcal{Z} \leq 1$ for termination).
The Newton-Raphson solve then rapidly refines the solution to near machine precision and, crucially, imbues the solution with a differentiable dependence on the solver variables, tracked by \code{auto\_diff} (\S\ref{sec:auto_diff}).
Even if the bisection search finds an adequate solution, we still require at least one Newton-Raphson iteration to ensure that the result contains the partial derivatives needed for the \mesa\ Jacobian.

When $\mathcal{Y} < 0$ and the initial $w = 0$, the entropy gradient is stable against convection. There is no pre-existing turbulence, and so $w=0$ for the entire step and $L=L_{\rm rad}$. This makes Equation~\eqref{eq:TDC_Q} linear in $\mathcal{Y}$.

Finally, when $\mathcal{Y} < 0$ and the initial $w > 0$, there can be up to three solutions to Equation~\eqref{eq:TDC_Q}:
\begin{itemize}
	\item In one solution, $\mathcal{Y}$ becomes large and negative. This forces $w \rightarrow 0$ before the end of the time step (e.g., the first root of Equation~\eqref{eq:wtan} occurs before time $dt$ passes), and $L=L_{\rm rad}$.
	\item In the other two solutions, $\mathcal{Y}$ becomes small and negative, and $w$ declines but does not reach zero by the end of the step. Here $L$ is carried by a mix of radiation and convection. There are two solutions because there is a tradeoff between the magnitude of $\mathcal{Y}$ and the decline of $w$, which compete in the $w\mathcal{Y}$ term in Equation~\eqref{eq:Lconv}.
\end{itemize}

Multiple solutions exist because, for long time steps, both $\mathcal{Y}$ and $w$ can evolve significantly in a single step.
One could force the time step to be smaller, such that there is just one solution.
However, a global time step limit is often undesirable, especially in cases where the precise means by which convection decays (e.g., for a retreating convective boundary on the main-sequence) is not usually of interest. Hence, it is often preferable to select one of the multiple solutions.

We disfavor the solution that decays fastest (e.g., $\mathcal{Y}$ is large and negative), as then convection decays on a dynamical timescale, which we suspect is unphysical.
Rather, we favor the slower-decaying (e.g., smaller-magnitude $\mathcal{Y}$) solution, which connects smoothly to the $\mathcal{Y}=0$ limit.
These preferences yield this rule: \emph{we always select the solution with the smallest $|\mathcal{Y}|$ and thus the slowest-decaying convection speed}.

\begin{figure}
\centering
\includegraphics[width=0.48\textwidth]{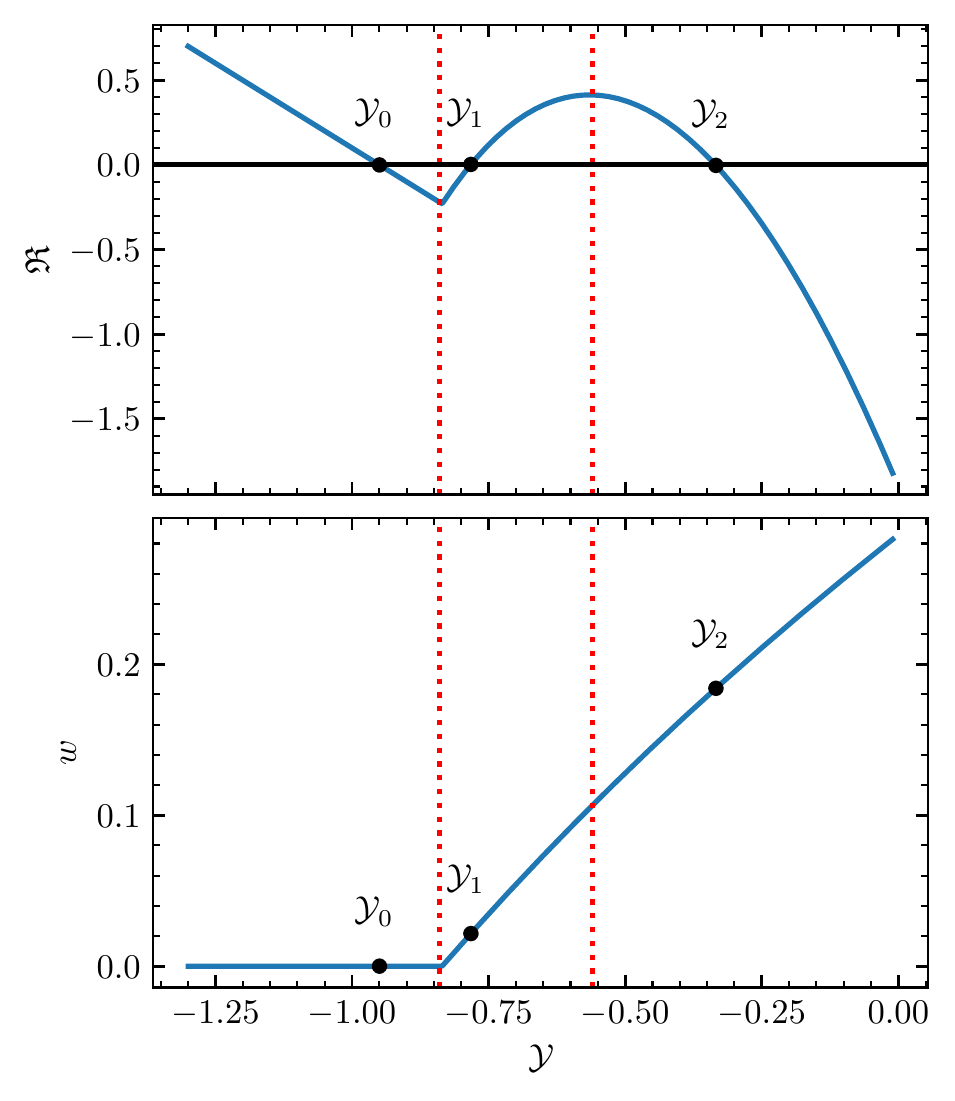}
\caption{The residual of the luminosity equation $\mathfrak{R}$ (upper) and $w$ at the end of the time step (lower), plotted as functions of $\mathcal{Y}$ in a case where the correct $\mathcal{Y} < 0$ and the initial $w > 0$. Vertical dashed red lines denote the two special points $\mathcal{Y}_0$ (left) and $\mathcal{Y}_1$ (right). Input parameters were chosen so that there are three solutions to $\mathfrak{R}(\mathcal{Y})=0$ (black dots), with $L_{\rm solver}=0.1$, $L_0=2$, $\nabla_{\rm ad} = 1$, $c_0 = 20$, $\xi_1 = 0$, $\xi_2 = -1$, $w_{\rm initial} = 1$, $|J| = \sqrt{-\mathcal{Y}}$, and $\delta t = t-t_0 = 5$, all in arbitrary units.}
\label{fig:tdc_three_sol}
\end{figure}

To find this solution, it is useful to examine $\mathfrak{R}(\mathcal{Y})$ in a representative case, shown in Figure~\ref{fig:tdc_three_sol}.
Each solution is a choice of $\mathcal{Y}$ such that $\mathfrak{R}(\mathcal{Y})=0$ (Equation~\ref{eq:Q0}).
Solutions are convective when $w > 0$ and radiative otherwise.

The first (slowest-decaying) solution is convective, with $w \approx 0.2$ and $\mathcal{Y} \approx -0.35$.
The second solution is \emph{also} convective, with $w \approx 0.05$ and $\mathcal{Y} \approx -0.75$.
Finally, the third solution is purely radiative, with $w = 0$ and $\mathcal{Y} \approx -0.95$.
The local maximum in $\mathfrak{R}$ is due to the fact that as $\mathcal{Y}$ becomes more negative, $w$ falls but $|\mathcal{Y}|$ rises, so the product $w\mathcal{Y}$ is not monotonic.

However, we do not know a priori how many solutions there are.
There can be no more than three, but by changing $L_{\rm solver}$ we can make the example shown in Figure~\ref{fig:tdc_three_sol} have just one (convective) solution.
Our approach is to first detect the number of solutions and isolate the one of physical interest.

The three solutions must be separated by two special points.
The first ($\mathcal{Y}_0$) is the smallest-magnitude $\mathcal{Y}$ with $w=0$, and the second ($\mathcal{Y}_1$) is the $\mathcal{Y}$ such that $w > 0$ and $\dif\mathfrak{R}/\dif\mathcal{Y} = 0$.
These are highlighted in Figure~\ref{fig:tdc_three_sol}.
Solutions of the first kind must occur at $\mathcal{Y} < \mathcal{Y}_0$, solutions of the second kind must occur at $\mathcal{Y} > \mathcal{Y}_0$, and at most one solution of the second kind occurs on either side of $\mathcal{Y}_1$.

Because ${\dif w/\dif\mathcal{Y} > 0}$, we search for $\mathcal{Y}_0$ using bisection in the interval ${-100 \leq \mathcal{Z} \leq 100}$.
We likewise identify $\mathcal{Y}_1$ by a bisection search over $-100 \leq \mathcal{Z} \leq \ln|\mathcal{Y}_0|$.

We use $\mathcal{Y}_0$ and $\mathcal{Y}_1$ to divide the interval ${-100 \leq \mathcal{Z} \leq 100}$.
The discriminant $\mathfrak{R}$ is monotonic over each subinterval by construction, so in each case we can search for a root using a combination of bisection search and Newton-Raphson refinement.
We check the intervals in order, from nearest to $\mathcal{Y}=0$ to furthest, and terminate the search as soon as a root is found.

\newpage
\subsubsection{Relation to \code{auto\_diff}}

TDC returns $\mathcal{Y}$ given $L$ and the other solver variables.
It additionally returns the partial derivatives of $\mathcal{Y}$ with respect to each of those variables.
This relies, fundamentally, on the new automatic differentiation feature (see \S\ref{sec:auto_diff}).
In particular, we used \texttt{auto\_diff} to calculate and propagate partial derivatives with respect to 33 variables of stellar structure \emph{through a Newton-Raphson solver}, producing the partial derivatives of a root-finding procedure with respect to its inputs.
The \texttt{auto\_diff} functionality enables the implementation of TDC.

\subsection{Reduction to Cox MLT}\label{sec:MLT_compare}

We now derive the modifications needed to ensure that TDC in \mesa\ agrees with MLT in the limit of long time steps.
\revision{While we use the $\alpha_r$\,=\,0 approximation in this section for clarity, the need for the correction is not removed by setting $\alpha_r > 0$.}

In TDC, the convective luminosity is given by Equation~\eqref{eq:Lconv}.
In MLT, the convective luminosity is
\begin{align}
	L_{\rm conv} = 4\pi r^2 f_2 \rho c_P T v_c \Lambda (\nabla - \nabla_e) h^{-1}
\end{align}
\citep{1999A&A...346..111L}, where $v_c$ is the convective velocity and $f_2$ is a parameter dependent
on the choice of MLT prescription.
Finally, $\nabla_e$ is the temperature gradient of a convective eddy, which is related to the efficiency parameter
\begin{align}
	\Gamma \equiv \frac{\nabla-\nabla_e}{\nabla_e-\nabla_{\rm L}}.
  \label{eq:tdc_Gamma}
\end{align}
We may write
\begin{align}
	\nabla - \nabla_e = \frac{\Gamma}{1+\Gamma}(\nabla-\nabla_{\rm L}) =  \frac{\Gamma}{1+\Gamma} \mathcal{Y},
	\label{eq:YGamma}
\end{align}
so
\begin{align}
	L_{\rm conv} = \frac{\Gamma}{1+\Gamma} 4\pi r^2 f_2 \rho c_P T v_c \Lambda \mathcal{Y} h^{-1}.
\end{align}
Next we identify $w = \sqrt{3/2} v_c$ (in steady state) and $\Lambda = \alpha h$, so
\begin{align}
	L_{\rm conv} = 4\pi r^2 \alpha \frac{\Gamma}{1+\Gamma}\left(\frac{2}{3} f_2^2\right)^{1/2} \rho c_P T w \mathcal{Y}.
\end{align}
This is nearly the same as Equation~\eqref{eq:Lconv}. In particular, in Cox%
\footnote{If desired, TDC may be modified to match other variants of MLT or other choices of $f_2$.}
MLT \citep{Cox:1968} $f_2 = 1/2$ and
\begin{align}
	L_{\rm conv} = 4\pi r^2 \alpha \frac{\Gamma}{1+\Gamma} \rho c_P T \frac{w}{\sqrt{6}} \mathcal{Y},
\end{align}
so the only difference is the term involving $\Gamma$.

That term, which controls the convective efficiency, is a genuine difference between TDC-in-\rsp\ and MLT.
We want TDC in \mesa\ to match the outputs of MLT in the steady state limit, in agreement with \citet{1986A&A...160..116K}, we modify Equation~\eqref{eq:Lconv} to include the factor $\Gamma/(1+\Gamma)$, giving
\begin{align}
	L_{\rm conv} = 4\pi r^2 \alpha \frac{\Gamma}{1+\Gamma} \rho c_P T  \frac{w}{\sqrt{6}} \mathcal{Y}.
	\label{eq:Lconv2}
\end{align}
We evaluate $\Gamma$ by calling MLT with the same inputs as TDC. We then treat this as fixed during the TDC iterations, which allows us to still use the algorithm described in \S\ref{tdc:numerics}.

With these modifications, the luminosity equations now agree, subject to $w = \sqrt{3/2}  v_c$ in steady state.
We now derive the conditions required to make this hold.

In MLT, the convective velocity is given by
\begin{align}
	v_c^2 = f_1 \Lambda^2 g \delta (\nabla - \nabla_e) h^{-1},
  \label{eq:vc2}
\end{align}
where $f_1$ is a parameter determined by the choice of MLT and
\begin{align}
	\delta \equiv -\left.\frac{\partial \ln \rho}{\partial \ln T}\right|_P = \frac{\chi_T}{\chi_\rho}.
\end{align}
Using $\Lambda \equiv \alpha h$ we can write Equation~\eqref{eq:vc2} as
\begin{align}
	v_c^2 = \alpha^2 f_1 h g \frac{\chi_T}{\chi_\rho} (\nabla - \nabla_e).
\end{align}
Next, with Equation~\eqref{eq:YGamma} we find
\begin{align}
	v_c^2 = \alpha^2 f_1 h g \frac{\chi_T}{\chi_\rho} \frac{\Gamma}{1+\Gamma} \mathcal{Y}.
	\label{eq:MLT_vc}
\end{align}

In TDC, we have identified the convection speed with
$v_c \equiv \sqrt{2/3} w$,
so we now proceed to prove that this is equivalent to that given by Equation~\eqref{eq:MLT_vc}.
When the TDC discriminant $J^2 < 0$, then $\mathcal{Y} < 0$ so the system is subadiabatic. Hence, at long times $w=0$ and therefore $v_c=0$, which matches the MLT answer.
When the discriminant is positive the system is convectively unstable, so we use Equation~\eqref{eq:Wconv} and find
\begin{align}
	v_c = -\sqrt{\frac{2}{3}} \frac{1}{2\xi_2} \left(J \tanh \frac{\lambda+\delta t J}{4} + \xi_1\right).
\end{align}
This solution was constructed assuming not only that $J^2 > 0$ but also that the relevant root is $J > 0$. As $\delta t \rightarrow \infty$ the $\tanh$ term approaches unity.
In this limit
\begin{align}
	v_c = -\sqrt{\frac{1}{6}} \frac{1}{\xi_2} \left(J + \xi_1\right)~,
\end{align}
and we also have $\xi_1 = 0$ because $\mathrm{D}\rho/\mathrm{D}t = 0$ and $\alpha_r = 0$ (Equation~\ref{eq:xi1}). Inserting the definition of $J$ and expanding with Equations~\eqref{eq:xi0} and~\eqref{eq:xi2} we find
\begin{align}
	v_c^2 = \frac{1}{6}\frac{J^2}{\xi_2^2} = -\frac{2\xi_0}{3 \xi_2} = \frac{\alpha^2 c_P T \nabla_{\rm ad} \mathcal{Y}}{8\alpha_D}~.
\end{align}
Comparing this with Equation~\eqref{eq:MLT_vc}, for equality
\begin{align}
	\frac{\alpha^2 c_P T \nabla_{\rm ad} \mathcal{Y}_{\rm MLT}}{8 \alpha_D} = \alpha^2 f_1 h g \frac{\chi_T}{\chi_\rho} \frac{\Gamma}{1+ \Gamma} \mathcal{Y}_{\rm TDC}.
	\label{eq:compare_TDC}
\end{align}

As before, to obtain equivalence between MLT and TDC we need to substitute $\mathcal{Y} \Gamma/(1+ \Gamma)$ for $\mathcal{Y}$ in the velocity equation.
In addition, we need to have
\begin{align}
	\frac{c_P T \nabla_{\rm ad}}{8\alpha_D} =  f_1 h g \frac{\chi_T}{\chi_\rho}.
\end{align}
With some rearranging, and using $h = P / \rho g$, we find
\begin{align}
	\frac{1}{8\alpha_D f_1} = \frac{P \chi_T}{\chi_\rho \rho c_P T \nabla_{\rm ad}}.
\end{align}
With $\Gamma_3 = 1 + (P / \rho c_V T) \chi_T$,
\begin{align}
	\frac{1}{8\alpha_D f_1} = (\Gamma_3 - 1)\frac{c_V}{\chi_{\rho} c_P \nabla_{\rm ad}} = \frac{\Gamma_3 - 1}{\Gamma_1 \nabla_{\rm ad}} = 1.
\end{align}
In Cox MLT $f_1 = 1/8$, and in TDC by default $\alpha_D = 1$, so the two sides are equal.

The net result is that in the limit of long time steps, TDC and Cox MLT solve the same luminosity equation with the same inputs and so are mathematically identical.
We find they agree numerically to around seven decimal places in $\mathcal{Y}$, even when $\mathcal{Y} \ll 1$.
\revision{The need for the correction is not removed by setting $\alpha_r > 0$. 
The asymptotic scaling in the inefficient limit is qualitatively different between 
(TDC with no correction and $\alpha_r > 0$) and (TDC with correction and $\alpha_r$\,=\,0).}
We further implement the calculation of convective mixing diffusivity and all other derived quantities using $v_c$ in the same way in both TDC and Cox MLT.

\begin{figure}
\centering
\includegraphics[width=0.48\textwidth]{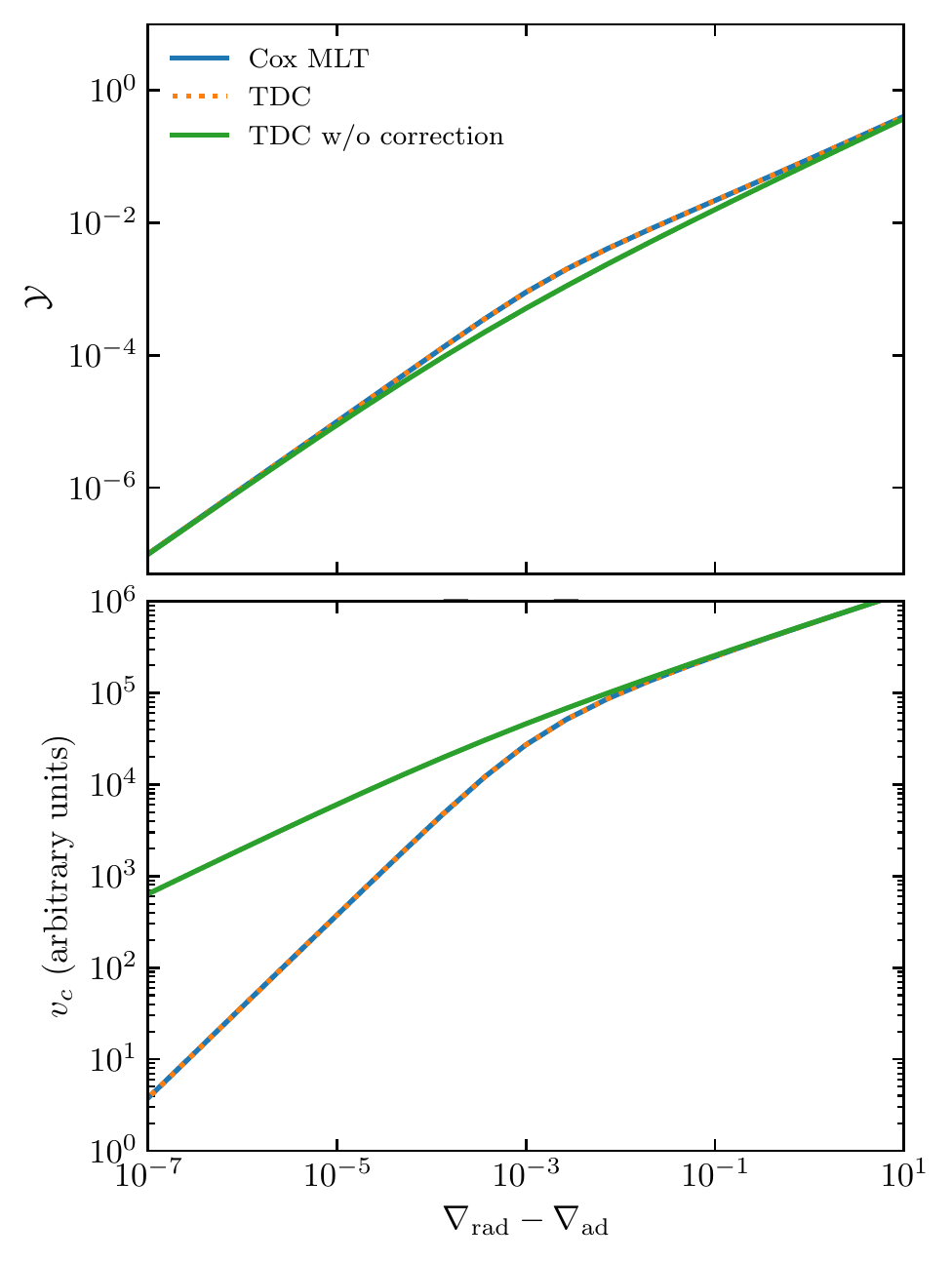}
\caption{The values of $\mathcal{Y}$ (upper) and $v_c$ (lower) plotted as functions of $\nabla_{\rm rad} - \nabla_{\rm ad}$ for MLT, TDC, and TDC \emph{without} the correction  $\mathcal{Y} \rightarrow \mathcal{Y} \Gamma/(1+ \Gamma)$ in the convective velocity and luminosity equations. The time step was chosen to be long enough that TDC reaches equilibrium.}
\label{fig:TDC_gamma}
\end{figure}

Figure~\ref{fig:TDC_gamma} shows the importance of the $\mathcal{Y} \rightarrow \mathcal{Y} \Gamma/(1+ \Gamma)$ correction.
In both panels, the solutions for TDC and MLT lie on top of each other.
The solution for TDC without the correction of $\mathcal{Y}$ in the equations, by contrast, deviates significantly in both panels.
This deviation is starkest in the lower panel, which shows a different $v_c$ scaling in the inefficient ($\nabla_{\rm rad} - \nabla_{\rm ad} \ll 1$) limit.

\subsection{Accreting White Dwarfs}\label{tdc:wd}

WDs accreting He at rates $\lesssim 10^{-6} \, \Msunyr$ undergo He shell flashes \citep{1989ApJ...342..430I}. These flashes can lead to He nova \citep[e.g., V445 Puppis;][]{2003A&A...409.1007A}, or even double-detonation type Ia supernovae \citep[e.g.,][]{2009ApJ...699.1365S,2011ApJ...734...38W,Kupfer2022}.
The time-dependent burning is controlled by three timescales: the local nuclear heating time,
\begin{align}
\theat \equiv \frac{ c_P T }{ \epsnuc } ,
\end{align}
being the characteristic timescale for temperature changes due to nuclear burning; the convective acceleration time,
\begin{align}
\taccel \equiv \frac{4}{J}
= \frac{3 h}{\sqrt{2 \alpha_{D} c_P T \grada \mathcal{Y}}},
\end{align}
being the timescale over which convection varies (see Equation \ref{eq:Wconv}); and the local dynamical time,
\begin{align}
\tdyn \equiv \frac{ h }{ c_{\rm s} } .
\end{align}
In steady state $\taccel$ is proportional to the eddy turnover time
\begin{align}
	t_{\rm eddy} \equiv \frac{\alpha h}{v_c},
\end{align}
but in cases of interest $\taccel$ and $t_{\rm eddy}$ can be quite different.

\cite{2009ApJ...699.1365S} showed that He shell masses of $\gtrsim 0.03 \, \Msun$ on a $\approx 1 \, \Msun$ WD can yield a $\theat$ comparable to or shorter than $\taccel$ or even $\tdyn$ near the base of the convection zone (BCZ). TDC will yield different results than MLT in this limit.

We construct these He flash models by accreting material comprising 99\% $^4{\rm He}$ and 1\% $^{14}{\rm N}$ by mass (as expected for solar metallicity stars that have undergone CNO burning) onto a $1 \, \Msun$ carbon-oxygen WD at constant $ \log ( \dot{M} / \Msunyr )$ between $-7.1$ and $-7.4$ in steps of $0.1$~dex.
Compressional heating results in a local temperature increase until the He shell ignites.
Lower $\dot{M}$ results in weaker compressional heating and a more massive He shell at ignition. Because heat is transported from the temperature peak towards both the core and the surface, ignition occurs above the base of the freshly accreted layer. We stop the accretion once a convective zone appears at the ignition site, and continue evolving through the He flash. Both the total accumulated He shell mass and location of ignition are impacted by the included reaction chain $^{14}{\rm N}(e^-, \nu) {^{14}{\rm C}} ( \alpha, \gamma) {^{18}{\rm O}}$ (NCO, \citealt{1986ApJ...307..687H,2017ApJ...845...97B}).

\begin{figure}
\centering
\includegraphics[width=0.48\textwidth]{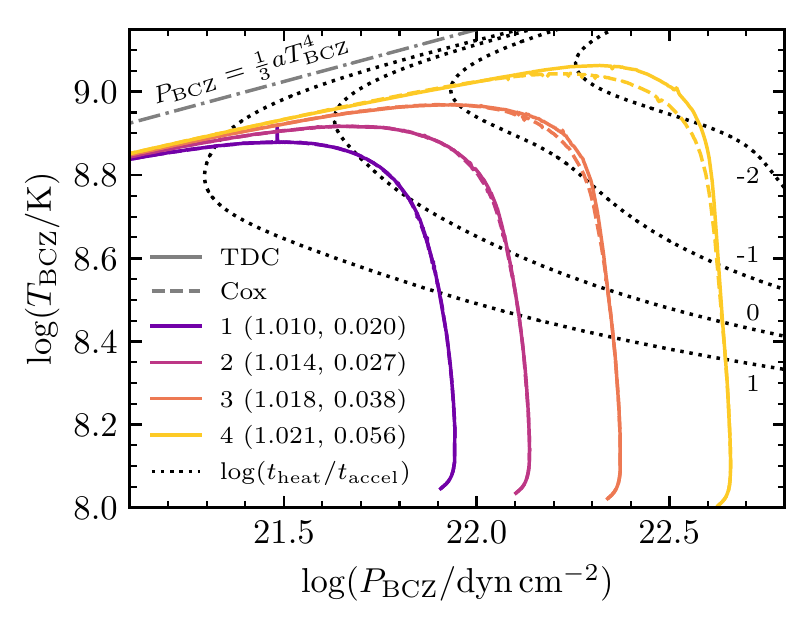}
\caption{Evolution of $\Tbcz$ and $\Pbcz$ during a He flash on a $ 1 \, \Msun$ WD with various He shell masses. Solid and dashed lines correspond to models run with TDC and Cox MLT respectively. The values in parentheses indicate the masses in solar units enclosed within and exterior to the BCZ, respectively. Dotted lines are contours of fixed logarithmic ratio between $\theat$ and $\taccel$\revision{, from $-2$ to $1$. Dot-dashed line gives the radiation pressure.} }
\label{fig:MLTDC_PT}
\end{figure}

In Figure \ref{fig:MLTDC_PT}, we label our models 1--4 at different $\dot{M}$ (with 1 corresponding to $-7.1$, 2 to $-7.2$, etc.) and note the masses enclosed by and exterior to the base of the convection zone (BCZ), which set the pressure $\Pbcz$ at ignition. The total accumulated He shell mass ranges from $0.03$ to $0.08 \, \Msun$.

Figure \ref{fig:MLTDC_PT} shows the evolution of $\Tbcz$ and $\Pbcz$ for models 1--4 with both TDC and Cox MLT. All models initially evolve at nearly constant $\Pbcz$, which increases with He shell mass. As temperature increases in the convection zone, the envelope expands and reduces $\Pbcz$. Concurrently, $\Tbcz$ reaches a maximum \citep{2009ApJ...699.1365S}.
Thicker He shells reach higher peak $\Tbcz$ and larger ratios between $\theat$ and $\taccel$.
\revision{For the contours here, $\taccel \approx 10 \, \tdyn$ (e.g., model 4 reaches $ \theat / \taccel \approx 0.01$, and correspondingly $ \theat / \tdyn \approx 0.1$.)}
When models show $\theat \lesssim \taccel$ (models 3 \& 4) and start expanding, TDC starts to deviate from Cox MLT, with greater deviations for thicker He shells. TDC shows higher $\Tbcz$ than Cox MLT at fixed $\Pbcz$ because TDC results in more superadiabatic convection. In contrast, when $\theat \gtrsim \taccel$ (models 1 \& 2) TDC and Cox MLT show good agreement in the evolution of $\Pbcz$ and $\Tbcz$.

\begin{figure}
\centering
\includegraphics[width=0.48\textwidth]{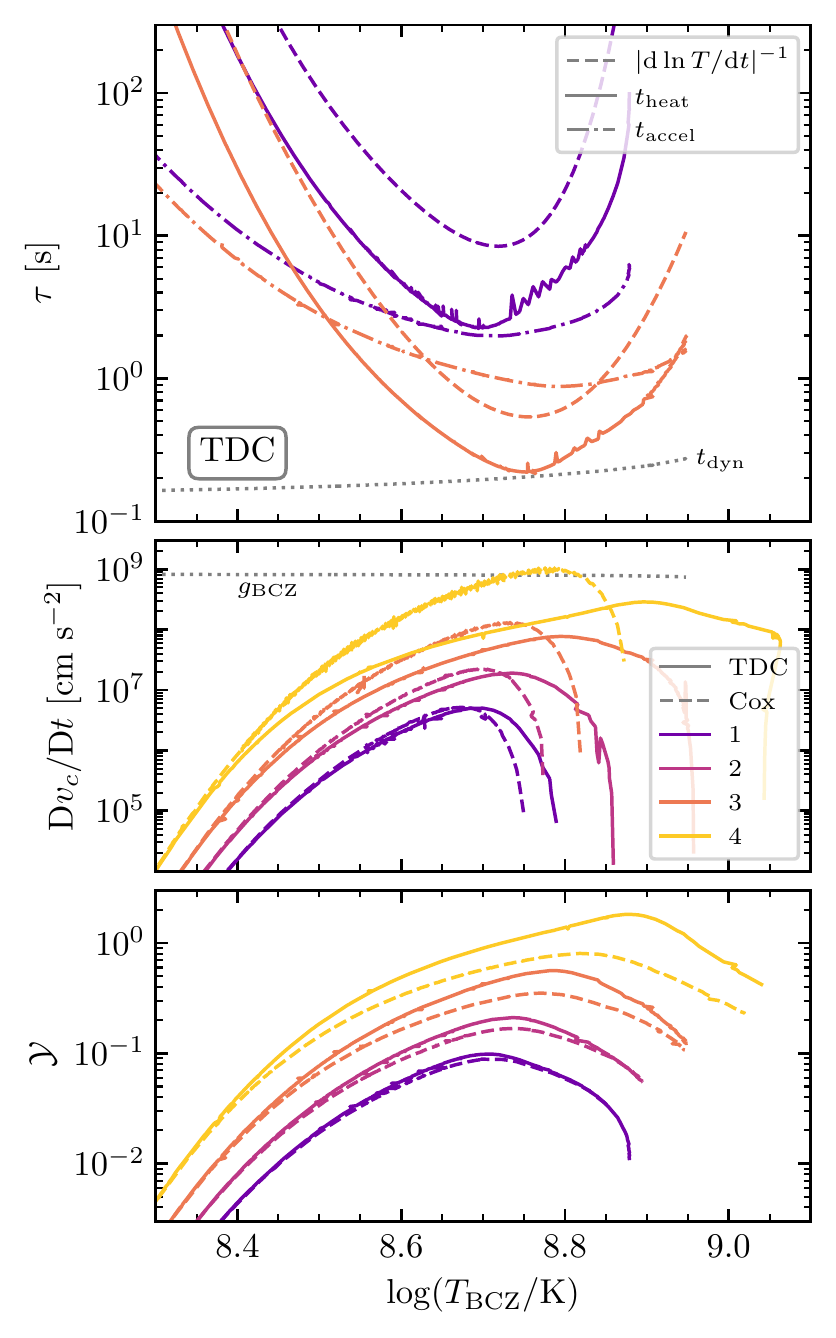}
\caption{Evolution of various time scales in the convection zone (upper), the time derivative of convective velocity at a fixed mass coordinate (middle), and the superadiabaticity at the same location (lower). The upper panel compares the heating timescale (solid line), actual timescale for temperature change (dashed line), convective acceleration timescale (dot-dashed line), and dynamical timescale (dotted line), for models 1 and 3 (TDC only). The first three timescales are evaluated at the BCZ, and the last at maximum convective velocity. Line styles in the middle and lower panels follow that of Figure \ref{fig:MLTDC_PT}, and in the middle panel we show the gravitational acceleration at the BCZ ($g_{\rm BCZ}$, dotted line) for comparison.}
\label{fig:MLTDC_tau}
\end{figure}

The upper panel of Figure~\ref{fig:MLTDC_tau} compares several timescales for TDC models 1 \& 3. The heating timescale, $\theat$, trends similarly with $ \dlnTdt $, but
the latter is larger than $\theat$ by factors of a few, because heat released by nuclear burning is distributed throughout some portion of the convection zone.
Due to the sharp dependence of $\epsnuc$ on $T$, both timescales decrease sharply with $\Tbcz$ until the WD starts to expand.
To reduce the noise in $ \dlnTdt $ displayed in Figure~\ref{fig:MLTDC_tau}, we fit it with a polylogarithmic function. The difference between $ \theat $ and $ \dlnTdt $ decreases with thicker He shells, as heat released by nuclear burning is increasingly trapped locally.

Another relevant timescale is $\taccel$, evaluated at maximum $v_c$. At $\logTbcz \lesssim  8.4$, $\taccel \approx (3/4)t_{\rm eddy}$. At $\logTbcz \gtrsim  8.4$, $\taccel$ evolves more quickly than $t_{\rm eddy}$, becoming up to 3 (6) times smaller than $t_{\rm eddy}$ in model 3 (4). This is because convection is no longer in steady-state, as $\theat \lesssim \taccel$ for $\logTbcz \gtrsim 8.4 - 8.5$ \citep[see also][]{2018MNRAS.476.2238G}. At minimum $\theat$, the hierarchy of timescales changes from
$$ \dlnTdt \gtrsim \theat \gtrsim \taccel \gg \tdyn $$
to
$$ \taccel \gtrsim \dlnTdt \gtrsim \theat \gtrsim  \tdyn $$
from model 1 to model 3, and ultimately to
$$ \taccel \gtrsim  \tdyn \gtrsim \dlnTdt \gtrsim \theat$$
in model 4. The fact that convection is not able to reach a steady state on the evolutionary timescale of the He flash
explains the difference between TDC and Cox MLT in models 3 \& 4 (see Figure \ref{fig:MLTDC_PT}).

We illustrate the difference in the evolution of $v_c$ and $\mathcal{Y}$ between TDC and Cox MLT in the middle and lower panels of Figure \ref{fig:MLTDC_tau}. For each TDC and Cox MLT pair, we locate the mass coordinate at which $v_c$ peaks when $\log ( L_{\mathrm{nuc}} / \Lsun ) = 9$ (arbitrarily chosen), and evaluate $\Dif v_{c} / \Dif t$ and $\mathcal{Y}$ during the initial acceleration phase. Initially, TDC and Cox MLT show good agreement when $ t_{\mathrm{heat}} \gg t_{\mathrm{accel}} $ ($ \log ( T_{\mathrm{bcz}} / \mathrm{K} ) \lesssim 8.4 $). When $ t_{\mathrm{heat}} \lesssim t_{\mathrm{accel}}$ ($ \log ( T_{\mathrm{bcz}} / \mathrm{K} ) \gtrsim 8.5 $), TDC shows slower evolution in $v_c$ and larger $\mathcal{Y}$ than Cox MLT. As $v_c$ is lower in TDC, heat is less efficiently transported out of the BCZ, resulting in higher $ \Tbcz $ and $\mathcal{Y}$ near maximum. With a thicker He shell, $ \Dif v_{c} / \Dif t $ may become comparable to $g$ (especially for Cox MLT model 4).

\begin{figure}
\centering
\includegraphics[width=0.48\textwidth]{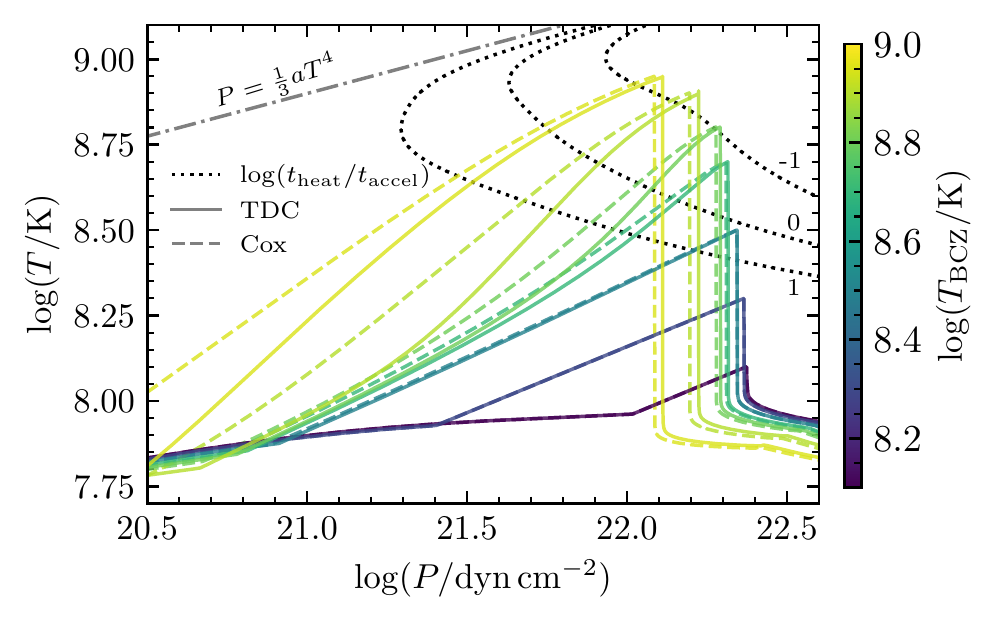}
\caption{Profiles of $T$ and $P$ within model 3 at different moments, when both TDC (solid) and Cox MLT (dashed) reach the same $ \Tbcz $. Curves of same color correspond to identical $\Tbcz$. \revision{Dotted and dot-dashed lines take the same meaning as in Figure \ref{fig:MLTDC_PT}.}}
\label{fig:MLTDC_profile_PT}
\end{figure}

\begin{figure}
\centering
\includegraphics[width=0.48\textwidth]{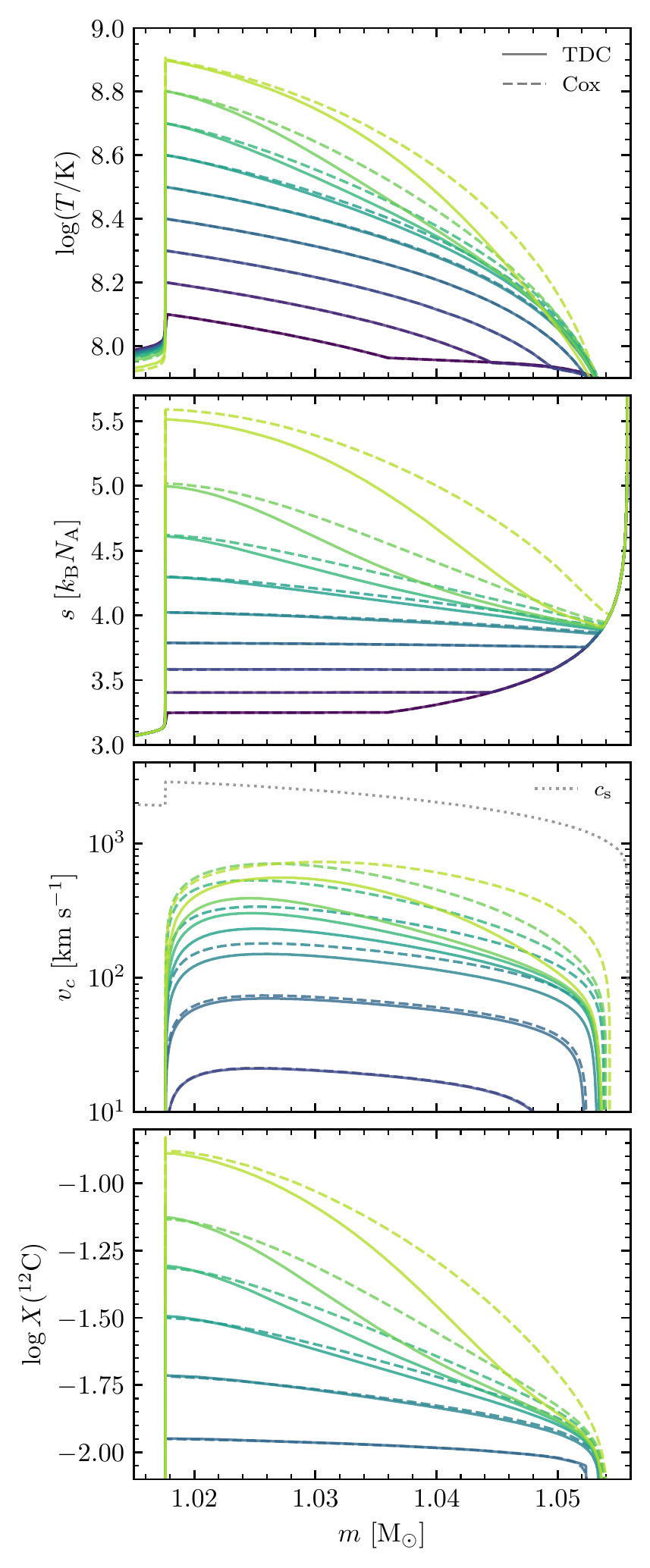}
\caption{Same as Figure \ref{fig:MLTDC_profile_PT}, but instead showing $T$, $s$, $v_c$, and $X(^{12}{\rm C})$ as a function of mass coordinate from top to bottom. In the third panel, $c_{\rm s}$ is shown for comparison. \revision{The colors match the colorbar in Figure \ref{fig:MLTDC_profile_PT}. }
}
\label{fig:MLTDC_profile_mass}
\end{figure}

We now study the evolution of model 3 in detail. In Figure \ref{fig:MLTDC_profile_PT}, we compare seven snapshots of the TDC and Cox MLT models in $T-P$ space, when both reach the same $\Tbcz$. The three coolest pairs of curves show good agreement and little superadiabaticity (third panel of Figure~\ref{fig:MLTDC_tau}). For the subsequent three hotter pairs, $\mathcal{Y}$ grows up to order unity near peak $\Tbcz$. Once $t_{\mathrm{heat}} \lesssim t_{\mathrm{accel}}$, heat is trapped more locally in TDC compared to Cox MLT. Therefore, TDC reaches the same $\Tbcz$ earlier in the evolution, and has a higher $\Pbcz$ due to comparably colder outer layers. Likewise, TDC shows less evolution in $T-P$ near the top of the convection zone and more superadiabaticity near the BCZ, again because of stronger heat-trapping near the BCZ.

Figure \ref{fig:MLTDC_profile_mass} compares TDC and Cox MLT in model 3 as a function of mass coordinate. The two show reasonable agreement in $v_{c}$ when $ \logTbcz \lesssim 8.5 $. At this point, $ \theat $ drops below $ \taccel $ (see Figure \ref{fig:MLTDC_tau}), which leads to TDC yielding lower $v_c$ than Cox MLT. For the same reason, $v_c$ near the top of the convection zone appears frozen in TDC for $ 8.6 \lesssim \logTbcz \lesssim 8.8 $.

At fixed $\Tbcz$, TDC shows lower $T$ throughout the convection zone, reflecting a local buildup of heat at the BCZ. Since TDC carries heat out of the BCZ less efficiently, it also shows a stronger entropy gradient, and for ${8.6 \lesssim \logTbcz \lesssim 8.8}$, $T$ and $s$ show little evolution near the top of the convection zone.

Both TDC and Cox MLT show appreciable abundance gradients, as $\carbon$ is produced near the BCZ but there is insufficient time for it to be transported outwards. Cox MLT shows higher $\carbon$ abundance overall, as it has more time to reach the same $\Tbcz$ and larger $v_{c}$. As TDC modifies both the $T$ and $X_i$ profiles, it may impact the potential for the ignition to develop into a detonation that would result in a thermonuclear transient.

In summary, we see that convection in TDC adjusts more slowly to changes in heating than in Cox MLT.
This results in slower, more superadiabatic convection during rapidly-burning phases of evolution.
The incorporation of the dynamics of how convection grows and decays is now possible and enabled by default in \MESA\ via TDC.


\begin{figure*}[!thb]
  \centering
  \includegraphics[width=1.0\columnwidth]{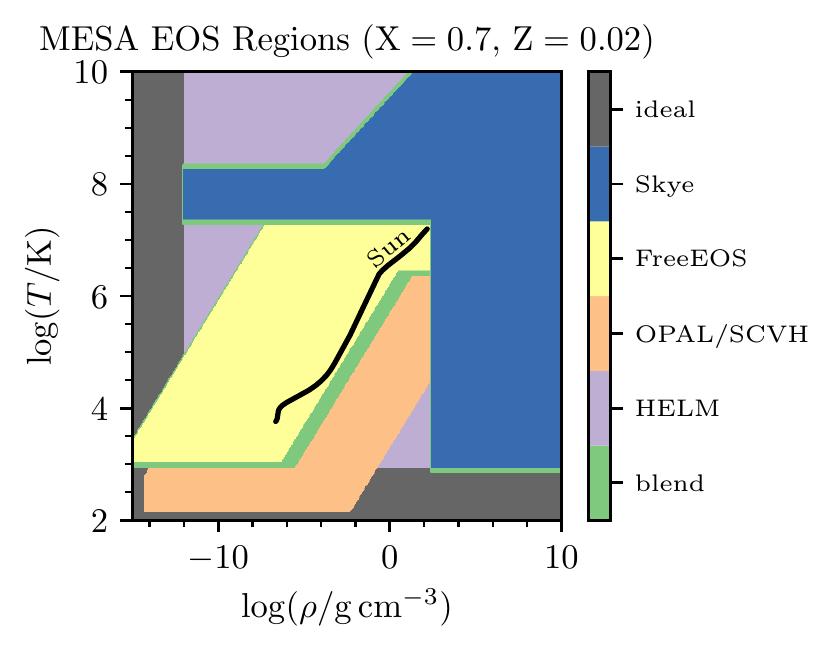}
  \includegraphics[width=1.0\columnwidth]{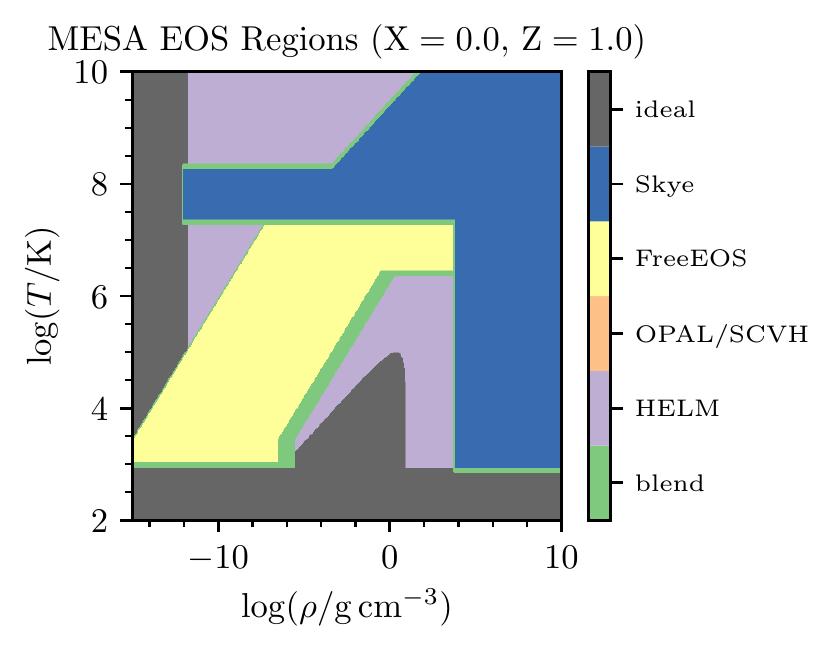}
  \caption{Default EOS boundaries for a solar-like composition (Z = 0.02, left)
           and for a pure-metal composition (Z = 1, 50\% $^{12}$C and 50\% $^{16}$O by mass, right).
           The colors and labels indicate an ideal gas approximation, Skye,
           FreeEOS, OPAL, SCVH, HELM,
           and blends between two EOSs. The black curve in the left panel shows a model for the present day Sun.
           \label{fig:eosbounds}}
\end{figure*}

\section{Equation of State}
\label{sec:eos}

\MESA\ models require thermodynamic quantities over a large span of $T$, $\rho$, and $X_{i}$.
This involves calling the \MESA\ EOS $\sim10^{4}$--$10^{10}$ times, depending on the chosen local physics and the number of iterations, cells, and time steps.
It would be ideal to have a single EOS that accurately represents the relevant physics in all regimes,
obeys all thermodynamic consistency relations to the limits of the arithmetic, and is as 
efficient in storage and execution as possible.
Below we report progress towards this ideal.

Figure \ref{fig:eosbounds} shows the default \MESA\ EOS boundaries
for solar and pure-metal (50\% $^{12}$C, 50\% $^{16}$O by mass) compositions.
Broadly, we prioritize HELM \citep{Timmes2000} at high $T$ and low $\rho$ for handling of the electron-positron plasma.
Elsewhere we prioritize Skye \citep{Jermyn2021}, limited by partial ionization at lower $T$ and $\rho$.
We then prioritize FreeEOS \citep{Irwin2004}, then OPAL \citep{Rogers2002} and SCVH \citep{Saumon1995},
and finally, when there are no other options, we use an ideal gas with radiation.
Blending boundaries between the different EOS prescriptions are set to defaults that have been motivated by specific use cases.
For example,
we have chosen the boundaries between FreeEOS and Skye such that
solar models at the age of the Sun stay fully on FreeEOS and do not encounter the
FreeEOS-Skye blend.

\subsection{Skye}
\label{sec:Skye}

Skye is an EOS for fully ionized matter \citep{Jermyn2021}.
A motivation for developing Skye was eliminating the blend between HELM and PC \citep{Potekhin2010}.
There is a blend between HELM and Skye that occurs at much higher $T$ and lower $\rho$ (see Figure~\ref{fig:eosbounds}),
where the two agree.
Skye includes the effects of positrons, relativity, and electron degeneracy~\citep{Timmes2000,baturin_2019_aa}, Coulomb interactions~\citep{ICHIMARU198791,Potekhin2010,PhysRevE.62.8554,1999CoPP...39...97D,PhysRevE.64.057402,PhysRevE.47.4330}, non-linear mixing effects~\citep{1999JChPh.111.9695C,2009PhRvE..80d7401P,2013A&A...550A..43P,1993PhRvE..48.1344O,2010PhRvE..81c6107M}, and quantum corrections~\citep{HANSEN1975187,PhysRevA.36.1859,1961PhRv..124..747C,PhysRevE.62.8554,Potekhin2010,2019MNRAS.488.5042B,2022MNRAS.510.2628B}.
Skye determines the point of Coulomb crystallization in a self-consistent manner, accounting for mixing and composition effects.
A defining feature of Skye is the use of analytic Helmholtz free energy terms and automatic differentiation (see \S\ref{sec:auto_diff})
to provide thermodynamic quantities.
Skye is thus readily extendable to new physics by including additional terms in the free energy \citep{Jermyn_2022}.

Skye is both a standalone software instrument and integrated into \mesa.
The two implement the same input physics and options.
At times this has required modifications of other parts of \mesa.
Here we describe the most important of these modifications.

\subsubsection{Crystallization}

Skye determines the crystallization phase transition by minimizing the Helmholtz free energy,
which permits derivatives to be discontinuous at the transition.
For instance, the entropy discontinuity reflects the latent heat of crystallization.
This posed a challenge in \mesa.
Consider the expression
\begin{align}
  \epsilon_{\rm grav} \equiv -T \DDt{s}~.
  \label{eq:TDsDt}
\end{align}
The entropy $s \equiv -\partial F/\partial T|_\rho$ undergoes a discontinuity at the phase transition.
If D$s$/D$t$ is evaluated by finite differences, then no time step will be small enough to produce a converged result for $\epsilon_{\rm grav}$.
We could write
\begin{align}
    \DDt{s} = \left.\frac{\partial s}{\partial T}\right|_\rho \DDt{T} + \left.\frac{\partial s}{\partial \rho}\right|_T \DDt{\rho},
    \label{eq:skye:partials}
\end{align}
but this form misses the latent heat of the phase transition because, except for the infinitesimal vicinity of crystallization,
thermodynamic derivatives of $s$ contain no information about the transition.
At the phase transition, derivatives of $s$ contain a Dirac delta contribution, which cannot be directly implemented in numerical calculations.
The choice is between poor convergence (finite differences of $s$) or neglecting the latent heat (Equation~\ref{eq:skye:partials}).

To address crystallization, Skye returns a parameter $\phi$ that provides a smoothed representation of the phase.
Specifically, $\phi = 1$ in the solid phase, $\phi = 0$ in the liquid phase, and near the phase transition $\phi$ smoothly interpolates between these limits.
The transition in $\phi$ is tuned so that the crystallization boundary is numerically resolved and yet spans a small fraction of a stellar model.
Using $\phi$, Skye then constructs a smoothed version of the latent heat of crystallization, which is only significant in the transition region.
This allows use of Equation~\eqref{eq:skye:partials} to avoid numerical issues near the phase transition,
but requires that we include an extra heat source in the energy equation to capture the latent heat:
\begin{align}
    \epsilon_{\rm latent} \equiv L_T \DDt{\ln T} + L_\rho \DDt{\ln \rho},
\end{align}
where $L_T$ and $L_\rho$ represent the differences between smoothed and original versions of the entropy derivatives $T \partial s/\partial \ln T$ and $T \partial s/\partial \ln \rho$. The original derivatives lack the latent heat, while the smoothed ones contain it, so $L_T$ and $L_\rho$ produce additional heating.
With this procedure, \mesa{} is able to model phase transitions, remain numerically converged, and accurately capture the latent heat of crystallization.
This procedure smears only the latent heat of crystallization and does not smear the thermodynamics of the phase transition, which would produce unphysical results such as negative sound speeds.

The Skye EOS approach
represents a significant improvement for the \mesa\ latent heat
treatment.
Previously, \mesa\ relied on a finite difference of the
entropy calculated in the PC EOS for solid and liquid phases so that
latent heat could be included in \epsgrav via
Equation~\eqref{eq:TDsDt}, smoothing this quantity near
the phase transition for numerical convergence (\mesafour). Another common approach is to include latent heat release with an
explicit heating term using
$l_{\rm cr} \approx 0.77 k_{\rm B} T /\langle A \rangle m_{\rm p}$
based on the calculation of \cite{Salaris2000}. Our new
approach based on Skye has the advantage that the phase diagram and
latent heat release are both calculated from first principles and are
self-consistent with the underlying thermodynamics of the EOS. 
\cite{Jermyn2021} showed that the net latent heat release is commensurate with the \cite{Salaris2000} value.

\subsection{FreeEOS}
\label{sec:FreeEOS}

We use FreeEOS version 2.2.1 \citep{Irwin2004} to expand the chemical composition parameter space covered
by partial ionization, as compared to the OPAL tables. This replaces the eosPTEH tables of \mesafive.
FreeEOS minimizes a Helmholtz free energy to span essentially the same thermodynamic range as OPAL.

The FreeEOS tables generated for \MESA\ use the `EOS1' mode, which is the highest level of physical accuracy provided by FreeEOS.
The tables are parameterized by the metal mass fraction $\rm{Z}=0$, 0.02, 0.04, 0.06, 0.08, 0.10, 0.20, 0.30, 0.40, 0.50, 0.60, 0.70, 0.80, 0.90, and 1.00.
All tables assumed a scaled-solar chemical composition based on \citet{GS98}.
For Z $\ge$\,0.80, there is also a set of tables with $X(^{12}$C) = $X(^{16}$O) for use with WD interiors.
For each Z a range of H mass fraction values between 0 and $1-\rm{Z}$ are provided, allowing for a complementary range of He mass fractions.
The tools to generate a new set of \MESA\ EOS tables for an arbitrary chemical composition using FreeEOS
are provided in \texttt{MESA\_DIR/eos/eosFreeEOS\_builder} with the exception of the FreeEOS library,
which can be downloaded from the FreeEOS repository.

\subsection{EOS Blends}
\label{sec:eos_blends}

The \MESA\ EOS blends several EOS prescriptions. 
Each EOS returns fundamental quantities and the partial derivatives of those quantities.
The blends of fundamental quantities and derivatives are treated differently because they are
used by \MESA\ for different purposes.
Fundamental quantities enter into physical equations, and so must be physical (e.g., positive sound speed), while their derivatives are used to construct the solver Jacobian, and so must represent accurate derivatives of the fundamental quantities.

The EOS returns a vector \texttt{res}  containing fundamental EOS quantities such as 
$e$, $s$, and $c_V$ (see \mesaone\ Table~3), as
well as blending fractions for the various EOS components. The EOS
also returns corresponding vectors \texttt{d\_dlnd} and \texttt{d\_dlnT}
of partial derivatives of each of the quantities in \texttt{res} with
respect to $\rho$ and $T$.

At the boundary between a pair of EOS prescriptions (EOS1 and EOS2) we calculate blends of 
$\texttt{res}$, $\texttt{d\_dlnd}$ and $\texttt{d\_dlnT}$ independently.
The EOS at a point in the blending region between EOS1
and EOS2 is evaluated with blending coefficient $\alpha \in [0,1]$
representing the fraction of EOS1, and $1-\alpha$ representing the
fraction of EOS2. We construct blending
coefficients using the quintic polynomial
\begin{equation}
  \alpha = 6x^5 - 15x^4 + 10x^3~,
\end{equation}
\revision{which maps the interval $x \in [0,1]$ (representing distance
  across a blend in $\rho$ or $T$) onto the interval $\alpha \in [0,1]$}
with zero slope at the blending boundaries.
The blending coefficients therefore
have non-zero derivatives with respect to $\rho$ and $T$ in blending
regions. Quantities in the resulting $\texttt{res}_{\rm blend}$ vector are
evaluated as a linear mix using the blending coefficient,
\begin{equation}
  \texttt{res}_{\rm blend} = \alpha\, \texttt{res}_{1} + (1 - \alpha)\, \texttt{res}_{2}~.
\end{equation}
\revision{Our choice of a quintic polynomial for the blending
  coefficient  ensures that both $\alpha$ and $(1-\alpha)$ are
  non-negative everywhere in the blending region, and therefore the
  EOS blending never introduces negative quantities into blends of
  non-negative values for the EOS $\texttt{res}$ vector.}
For the derivative vectors, we include additional terms to account for
the derivatives of the blending coefficients,
\begin{equation}
  \begin{split}
    \texttt{d\_dlnd}_{\rm blend} =&
    \alpha\, \texttt{d\_dlnd}_{1}
    +\frac{\partial \alpha}{\partial \ln \rho} \texttt{res}_{1} \\
    &+(1 - \alpha)\, \texttt{d\_dlnd}_{2}
    -\frac{\partial \alpha}{\partial \ln \rho}\texttt{res}_{2}~,
  \end{split}
\end{equation}
and similarly for \texttt{d\_dlnT}. Including these terms for the
blending coefficients in the derivative blends provides correct
derivatives for the solver, reducing the number of Newton iterations.

Some quantities in the fundamental EOS \texttt{res} vector are
themselves derivatives of other EOS quantities, such as
$c_V \equiv (\partial e/\partial T)_\rho$. The different blending
treatments for EOS quantities and their derivatives mean that
thermodynamic identities may be violated in blending regions.
Physical equations such as the energy equation must use quantities
such as $c_V$ from \texttt{res} rather than the theoretically equivalent but
numerically different derivative quantities from the \texttt{d\_dlnT}
vector.
The latter can lead to
unphysical results such as negative heat capacities or negative sound
speeds. This inconsistency is unavoidable so long as we must blend between different EOS prescriptions.

\subsection{Thermodynamic Consistency}
\label{sec:eos_consistency}

\begin{figure*}
\centering
\includegraphics[width=0.48\textwidth]{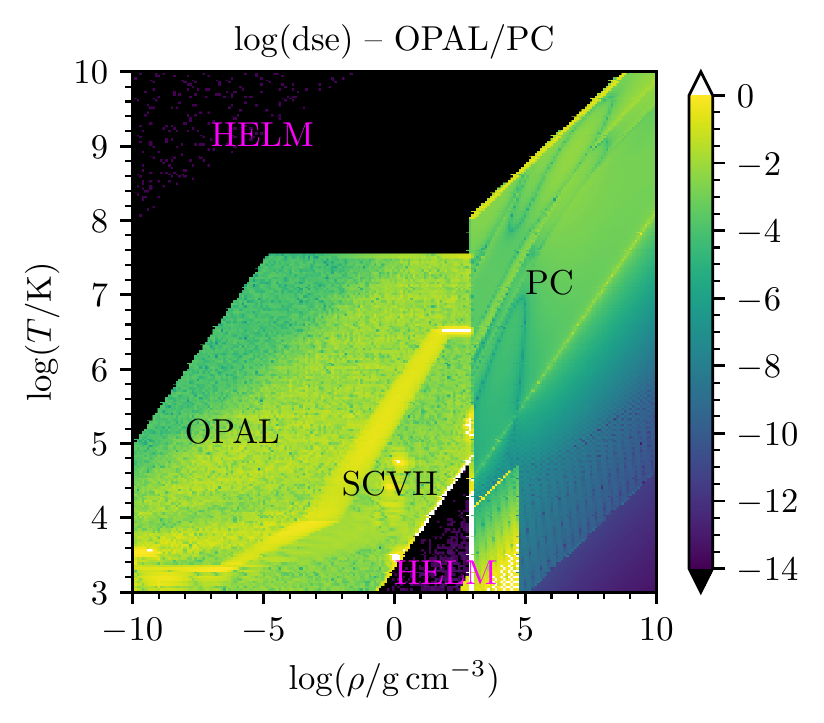}
\includegraphics[width=0.48\textwidth]{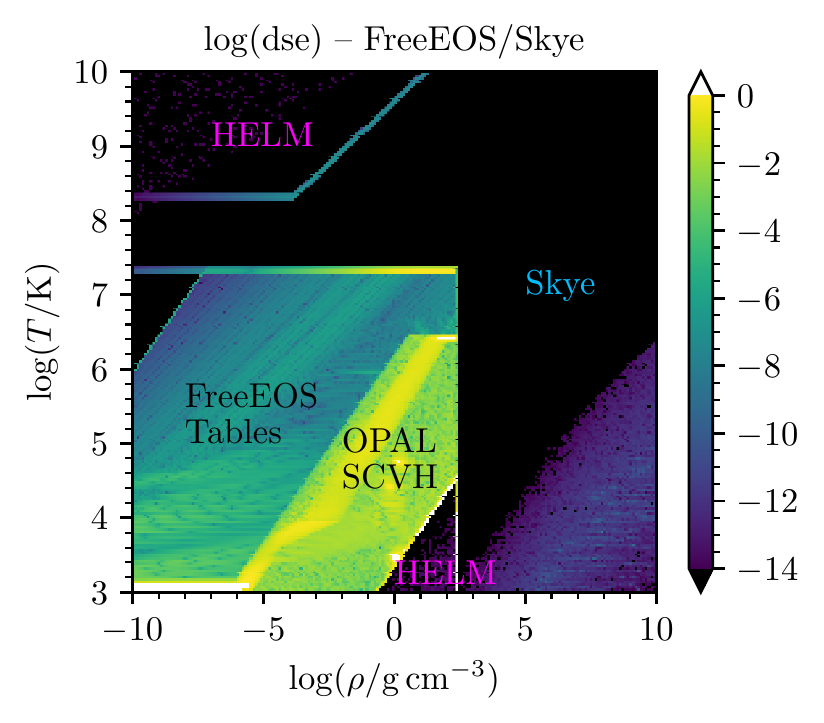}
\caption{The logarithm of the thermodynamic consistency measure $\mathrm{dse}$ for the \mesa\ EOS computed with PC and OPAL (left) and Skye and FreeEOS (right) with $\rm{X}=0.7$, $\rm{Y}=0.28$, and $\rm{Z}=0.02$, with $\rm{Z}$ comprising equal mass fractions of $^{12}\mathrm{C}$ and $^{16}\mathrm{O}$. 
}
\label{fig:eos_consistency}
\end{figure*}

One desirable feature in an EOS is thermodynamic consistency, which ensures that the Maxwell relations hold --- e.g., mathematically equivalent forms of the equations of stellar structure are also numerically equivalent within the floating point precision of the arithmetic.
Unfortunately, several of the EOS prescriptions in \mesa\ are not fully thermodynamically consistent. This can cause errors in energy conservation, 
making mathematically equivalent formulations of the structure equations behave differently.

Here we report on the current state of thermodynamic consistency in \mesa.
Figure~\ref{fig:eos_consistency} compares the consistency measure
\begin{equation}\label{eq:dse}
\mathrm{dse} \equiv T \frac{\partial s/\partial T |_{\rho}}{\partial e/\partial T |_{\rho}} - 1
\end{equation}
for the \MESA\ EOS with PC and OPAL (left, former default) and with Skye and FreeEOS (right, current default). The quantity $\mathrm{dse}$ is zero
in thermodynamically consistent systems.

As Skye derives all quantities from partial derivatives of a Helmholtz free energy, it is thermodynamically consistent to near machine precision.
Without Skye, the corresponding regions of the EOS are covered by PC and HELM.
The regions with Skye active show thermodynamic consistency to near machine precision,
representing a significant improvement for $\log(\rho/\mathrm{g\,cm^{-3}}) \gtrsim 3$.
The band at $\log \left(T/\mathrm{K}\right) \approx 9-10$ in the right panel is due to a blend in the EOS from Skye to HELM, which is required to remedy a
floating point loss-of-precision issue in Skye when electron-positron
pairs dominate the EOS. In the left panel the PC region shows a stripe
of high error due to Coulomb crystallization. 
FreeEOS is
thermodynamically consistent to near machine precision. Our current
method of interpolating the \mesa\ FreeEOS tables does not preserve this
property. Still, these tables show significant improvement relative to OPAL.


\section{Energy Equations}
\label{sec:energy_eq}

Section~3 of \mesafive\ highlighted the importance of numerical energy conservation in \mesa\ models, and introduced a new form of the energy equation aimed 
at improving energy conservation. This new form motivated several solver improvements, leading to
tighter tolerances for equation residuals and corrections.
We now advance that discussion by further explaining
the multiple formulations of the energy equations in \MESA{}. 
We contrast the advantages conferred by each formulation across different applications.
We also clarify the meaning of the
quantities \texttt{rel\_E\_err} and \texttt{rel\_run\_E\_err} reported
for \MESA\ runs, and elucidate what these quantities do and do not
tell us about the quality of the \MESA\ solutions.

After reviewing the energy-equation fundamentals and recent
implementation improvements in \S\ref{sec:energy_fundamentals},
we explore an idealized example problem without any composition
changes or EOS complexities in \S\ref{sec:carbon_kh}. This
example motivates a new time-centered approach for evaluating
the \texttt{eps\_grav} form of the energy equation, and demonstrates
that a lower value of \texttt{rel\_run\_E\_err} does not always
indicate evolution that is more physically accurate.
In \S\ref{sec:eosproblems} we describe the additional complexities
introduced by thermodynamic inconsistencies that can be present in the
EOS, especially in EOS blending regions, and how these manifest in
different ways for different energy equation implementations. 

Finally,
in \S\ref{sec:1M_pre_ms_to_1TP} we illustrate the various
contributions to energy error terms through the example of a
$1\,\Msun$ star including both composition changes due to nuclear
burning and EOS blends and inconsistencies.
This example demonstrates that the quantity
\texttt{rel\_run\_E\_err} must be interpreted differently for
different forms of the energy equation. When using the \texttt{dedt} form of
the energy equation, the energy error reflects the quality of the
residuals from the \MESA\ solver, even though larger energy errors
associated with the EOS are still present in the model. When using the
\texttt{eps\_grav} form of the energy equation, the energy error reports much
larger values that reflect the presence of these EOS errors, even when
the quality of solutions may be comparable to or better than the
\texttt{dedt} form.

Convergence tests and comparisons between multiple forms of the
energy equation remain vital for understanding the reliability and
accuracy of solutions in different regimes.
Significant progress has been made in ensuring that different
forms of the energy equation converge to the same result.  In
degenerate conditions, the \texttt{eps\_grav} forms generally perform better
(i.e., they are closer to the converged answer at a given time
resolution).
With the \texttt{dedt} form, the numerical energy conservation error often
measures the quality of the solution (i.e., the size of the
residuals). Focusing on improving that quantity has driven
significant solver improvements and motivated the development of an
accurate energy accounting infrastructure.
This energy accounting work has also motivated improving the \texttt{eps\_grav}
form to account for composition changes, as well as
an implicit trapezoidal time-centering scheme. \MESA\ now includes
these changes by default when using the \texttt{eps\_grav} form of the energy
equation. Further progress rests on improvements to the quality of the EOS.

\subsection{Fundamentals and Implementations}
\label{sec:energy_fundamentals}

\MESA\ has two primary energy equations.  One, called the
``\texttt{eps\_grav} form'',%
\footnote{In \mesafive, we referred to this equation as the
  ``\texttt{dLdm} form''.  That was an unfortunate choice as a $\partial L/\partial m$ term occurs in all versions of the equation.}
is the standard stellar structure energy
equation \citep[e.g.,][]{Kippenhahn2012} and is the equation introduced in \mesaone.
This equation is
\begin{equation}
  \ddm{L} = \epsilon + \epsgrav~,
  \label{eq:epsgrav_form}
\end{equation}
where $L$ is the luminosity, $\epsilon$ is a specific energy
generation source term (e.g., nuclear reactions, neutrinos), and
\begin{equation}
  \begin{split}
    \epsgrav & \equiv-\DDt{e} + P\DDt{}\left(\frac{1}{\rho}\right) \\
    & = -T \DDt{s} - \sum_i \left(\frac{\partial e}{\partial X_i}\right)_{s,\rho, \{X\ne X_i\}} \DDt{X_i}~.
  \end{split}
  \label{eq:epsgrav_defn}
\end{equation}
In practice, the total Lagrangian time derivative of $e$ is expanded and further manipulated to reach the
final form evaluated in \MESA\ (see \mesafour, \S8).

The other, called the ``\texttt{dedt} form'',
is an energy equation for
the time evolution of the total specific energy of a Lagrangian cell,
\begin{equation}
  \label{eq:dedt_form}
  \DDt{}\left(e + \frac{1}{2}u^2 - \frac{Gm}{r} \right)
  =
  \epsilon - \ddm{}\left(L + P \area\,u\right) ~,
\end{equation}
where $u$ is cell velocity and $\area = 4\pi r^2$ is the area of the
cell face.
The relationship between these two forms was derived in \mesafour,
Section 8.3 and the \texttt{dedt} form was introduced as a
powerful tool in support of improved numerical energy conservation in
\mesafive, \S3.

When solutions are numerically converged (i.e., have sufficient
space/time resolution to give resolution-independent results)
and the EOS is thermodynamically consistent and provides correct
partial derivatives (see \S\ref{sec:eosproblems}), these
two equations should give identical results.  Conversely, the
solutions may differ when unconverged.

The error in numerical energy conservation during a step,
$E_{\rm err}^{\rm step}$, is evaluated as the difference between the
change in total energy of the model across the time step and the
expected change in total energy due to known energy sources and sinks
(e.g., nuclear reactions, neutrinos, surface luminosity).
Total energy is defined as
\begin{equation}
  \label{eq:Etot}
  \begin{split}
    E_{\rm tot} &\equiv \int \left(e + \frac{1}{2}u^2 - \frac{Gm}{r} \right) \, \dif m \\
    &= \sum_k dm_k \left(e_k + \frac{1}{2}u_k^2 - \frac{Gm_k}{r_k} \right)~,
  \end{split}
\end{equation}
where $dm_k$ is the mass contained within cell $k$.
Additional terms for rotational kinetic energy can also be
  included in Equation~\eqref{eq:Etot} when rotation is enabled, and
  turbulent energy is included for RSP models.

A cumulative sum of
the per-step energy errors, $E_{\rm err}^{\rm run}$, is tracked
during a run.  When divided by the total energy at the end of the
step, $E_{\rm err}^{\rm step}$ and $E_{\rm err}^{\rm run}$ respectively become the quantities \texttt{rel\_E\_err}
and \texttt{rel\_run\_E\_err} that are reported by \MESA. As stated in \mesafive{}, these quantities are primarily meant
to represent a measure of the numerical reliability of solutions
accepted for \mesa\ evolution steps, rather than a measure of physical validity and
completeness of \mesa\ models. In \S\ref{sec:eosproblems}
and \S\ref{sec:1M_pre_ms_to_1TP}, we focus on further clarifying the meaning
of these energy error quantities, which require a different
interpretation when using the \texttt{eps\_grav} form of the energy
equation than when using the \texttt{dedt} form.

\MESA\ does not solve its discretized, finite-mass form of the stellar
structure equations exactly.  When a trial solution is accepted, the
residual difference between the left- and right-hand sides of the equation
becomes an error in numerical energy conservation.  Therefore, one
necessary step in ensuring good numerical energy conservation is to
select tight tolerances for the acceptance of a solution. This requires
sufficiently high quality derivatives in the Jacobian that the
solver can reach these tolerances in a reasonable number of
Newton iterations (see \S3 of \mesafive).

However, even achieving zero residuals is not sufficient to ensure
numerical energy conservation.  When \MESA\ modifies the stellar model
outside of the Newton solve, the resulting changes in total energy
must be correctly included in the accounting.  When physically
appropriate, compensating energy source terms must be included in the
equations that are solved during the Newton iterations.
For example, mass changes of the stellar model are one such process, and the
procedure that ensures numerical energy conservation is described in
\S3.3 of \mesafive.  At that time, this procedure was applied only
when using the \texttt{dedt} form of the equation.  Now, it is used
with all forms of the energy equation, and the less general approach
originally used with the \texttt{eps\_grav} form (\mesathree, \S7) has been removed from \MESA.

The composition changes associated with element diffusion (\S3 of \mesafour) 
and convective premixing (\S5~of \mesafive) are also
incorporated in an operator-split manner (i.e., adjustments to the
model made outside of the Newton iterations for the implicit structure solve during an evolutionary
step; see also \S\ref{sec:rates}), and so require special accounting.  The energy changes due
to these composition changes are now tracked and compensating source terms
are added to the equations, improving numerical energy
conservation.

Non-conservation of numerical energy can also occur when the equations
being solved are approximated in ways that do not conserve energy.
Historically, the default \MESA\ implementation of $\epsgrav$
(\mesaone, Equation 12) dropped the term associated with composition
changes. While the energy associated with composition changes is
dwarfed by the energy released by nuclear reactions (see \mesafour),
the integrated energy error introduced by dropping this term
is not negligible compared to the value of $E_{\rm tot}$ by the end of
the MS. 

In \mesafive, Figure~25, the
``\texttt{dLdm}-form'' calculation (right panel) did not include
composition changes in \epsgrav, and so the large values of the
relative energy error shown during core He burning effectively
quantify the impact of dropping the composition term
rather than characterizing the numerical quality of the \MESA\ solution.
In this case, the scale of the reported error appears significant
because \MESA\ adopts $E_{\rm tot}$ as the reference
value for checking cumulative numerical energy conservation. 
A larger reference value, like the time-integrated
radiated energy of the star, is typically used to justify dropping
the composition term from $\epsgrav$.

A continued focus on numerical energy conservation requires equations that are
energy conserving, so \MESA\ now includes the composition term in its
default implementation of \epsgrav. With $(\rho, T)$ as the
thermodynamic structure variables, we have
\begin{equation}
\label{eq:epsfifth}
\begin{aligned}
\epsgrav = &-c_V T \DDt{ \ln T}
- \left[ \rho \left(\frac{\partial e}{\partial \rho}\right)_T - \frac P \rho \right] \DDt{ \ln \rho} \\
&+ \epsgravX~,
\end{aligned}
\end{equation}
where $c_V \equiv (\partial e/\partial T)_\rho$.
As shown in \mesafour, Equation (65), \MESA\ implements the equivalent expression
\begin{equation}
  \label{eq:eps_grav_rewrite}
  \begin{split}
    \epsgrav =& -c_P T \left[(1 - \nabla_{\rm ad} \chi_T)\DDt{\ln T} - \nabla_{\rm ad} \chi_\rho \DDt{\ln \rho}\right] \\ & + \epsgravX~,
  \end{split}
\end{equation}
where ${c_P \equiv (\partial e/\partial T)_P - (P/\rho^2)(\partial \rho/\partial T)_P}$ and
${\grada \equiv (\partial \ln T /\partial \ln P)_s}$.
The composition term is
\begin{equation}
  \epsgravX \equiv - \sum_i \left(\frac{\partial e}{\partial X_i}\right)_{\rho,T, \{X\ne X_i\}} \DDt{X_i}~.
\label{eq:epsgravX}
\end{equation}
When implemented in \MESA, the quantity \epsgravX\ is evaluated as a
finite-difference approximation to the directional derivative along
the change in the composition vector over the time step:
\begin{equation}
  \epsgravX = -\frac{1}{\dt}\left[e(\rho, T, \{X_i\}) - e(\rho, T, \start{\{X_i\}})\right] ~.
\end{equation}
This is analogous to the approach used in evaluating the spatial
composition derivatives that enter into the \bvv\ frequency (\S3.3 of \mesatwo).  
In addition to being simpler to evaluate, this
approximation is numerically convenient because it only requires first
derivatives of $e$ with respect to composition in order to form the
Jacobian.  The \MESA\ \texttt{eos} module and its interface with
\MESAstar\ have been upgraded either to provide these partial
derivatives when available, or to construct approximations to these
partial derivatives for the Jacobian based on finite differences using
small variations of the composition when analytic derivatives are not
available.

The total derivatives of the structure variables in
Equation~\eqref{eq:eps_grav_rewrite} are evaluated as their
differences over the time step. In previous implementations of the
\texttt{eps\_grav} form of the energy equation, the thermodynamic
quantities that multiply the total derivative quantities were evaluated
at the end of the step (in the standard \MESA\ backwards-Euler approach).
As a means of further improving numerical energy conservation when using the
\texttt{eps\_grav} form, we have now introduced a higher-order (in
time) version of \epsgrav\ using the implicit trapezoidal rule. This replaces
end-of-step quantities with time-centered versions (i.e., averages of
the values at the start and end of the step).  We refer to this as
``\texttt{eps\_grav} (centered)'' in contrast to the previous implementation,
which we indicate as ``\texttt{eps\_grav} (end of step).''
As we shall show in the following sections, including  both composition
changes and time-centering in the \texttt{eps\_grav} implementation
greatly improves energy conservation, so we now include both of these
improvements by default in \mesa\ when using the \texttt{eps\_grav}
form.

In the following sections, we use test cases to demonstrate the
performance and physical meaning of numerical energy conservation
under the various forms of the energy equation in \mesa.
We also show that in some circumstances, such as degenerate stars, the
\texttt{eps\_grav} form of the energy equation converges to accurate
entropy and temperature evolution substantially faster than the
\texttt{dedt} form does, even while reporting larger errors in
numerical energy conservation.

\subsection{Results: carbon\_kh}
\label{sec:carbon_kh}

As an illustrative test case, we follow an initially low-density,
1.3\,\Msun\ sphere of pure carbon as it undergoes Kelvin-Helmholtz contraction.
The model begins at a central density of $\logrhoc = 1$ and we follow
the contraction over a factor
$\gtrsim10^7$ increase in $\rho_{\rm{c}}$.  Nuclear reactions are
not considered.  For simplicity, we assume that the radiative
opacities are given by electron scattering and include standard
thermal neutrino losses. This model does not experience convection.
We exclusively use the HELM EOS, as the use
of a single EOS that is formulated from the Helmholtz
free energy avoids most of the EOS inconsistencies that we
discuss in \S\ref{sec:eosproblems}.

This case is not meant to model a real object, but
provides a simple example problem that has neither mass changes nor
composition changes.  It is nonetheless demanding as the conditions in
the star vary tremendously during the evolution as material goes from
non-degenerate conditions to conditions of relativistic electron
degeneracy, and the dominant energy loss mechanism transitions from
radiative diffusion to optically-thin neutrino cooling.

We explore three different versions of the energy equation: the \texttt{dedt}
form, the \texttt{eps\_grav} form (end of step), and the \texttt{eps\_grav} form
(centered).  We use a temporal convergence study to illustrate the
performance of the different variants of the energy equation.  For
each equation, we show three time resolutions, and compare against a
family of ultra-resolution runs that serve as reference
solutions. These ultra-resolution reference runs still show
small differences depending on which form of the energy equation is
selected, so we also show that smaller level of disagreement to
indicate the level of differences that should be interpreted as
significant. We interpret the small magnitude of disagreement between
ultra-resolution runs as evidence that the different versions of the
energy equation are converging to the same result for sufficiently
high resolution.

\begin{figure}
  \centering
  \includegraphics[]{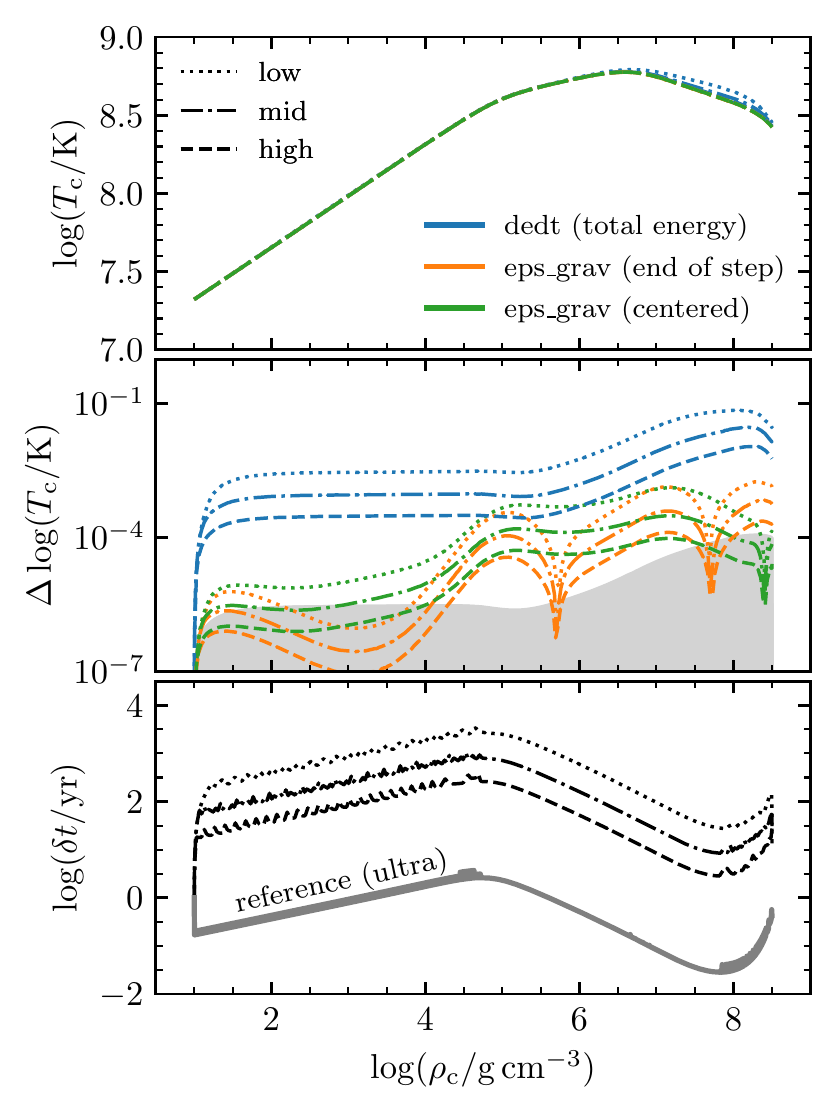}
  \caption{Convergence tests of the evolution of the central
    quantities for the contracting 1.3\,\Msun\ sphere of pure carbon using
    different forms of the energy equation. The central density is a
    proxy for time.  The top panel shows $T_{\rm c}$.
    The middle panel shows the $T_{\rm c}$ difference from a reference solution.
    The three line styles represent the three levels of resolution as
    indicated by the legend in the top left corner of the upper panel.
    The grey region marks the level of agreement between the set of
    reference solutions.  The lower panel shows the time step for each
    resolution, with the time step of the runs used in the
    ultra-resolution reference set indicated as the solid grey line. }
  \label{fig:ckh_center}
\end{figure}

Figure~\ref{fig:ckh_center} shows the trajectory of $T_{\rm{c}}$ versus $\rho_{\rm{c}}$.  
While this calculation does not
consider nuclear reactions, in calculations that do, the $T$
and $\rho$ sensitivity of the nuclear reaction rates motivates
obtaining solutions that are converged in these quantities
(e.g.\ \citealt{Schwab2015}). This example does include thermal
neutrinos, which lead to central cooling at high density.
The top panel shows that the two \texttt{eps\_grav} versions agree (to within
the line width) at all resolutions, while the \texttt{dedt} form shows visible
differences during the evolution after the model has reached its
maximum $T_{\rm{c}}$.  The level of difference from the
reference solution is shown in the middle panel.  All forms exhibit
first-order convergence, where a 1~dex reduction in the time step leads
to a 1~dex reduction in the error in $T_{\rm{c}}$.
However, at a fixed resolution, the \texttt{eps\_grav} forms show similar
performance to each other and superior performance relative to the
\texttt{dedt} form.

In order to understand why the \texttt{eps\_grav} forms perform better under
degenerate conditions, consider an adiabatic change,
${\dif e + P\dif (1/\rho) = T \dif s = 0}$ at fixed composition.
This expression is satisfied exactly for
infinitesimal changes and a perfect EOS. 
When we integrate across a time step, we know the integral
of total time derivatives (e.g.\ ${\Dif e /\Dif t}$ or ${\Dif \ln \rho /\Dif t}$)
exactly, but approximate the integral over the time step for quantities that
are not total time derivatives.  The extent to which our scheme will
fail to reproduce an adiabatic evolution is the error in approximating
these other integrated quantities appearing in the energy equation
(e.g., $P$ or $c_V$).  Recall that the usual backwards Euler approach in \MESA{} is
effectively like assuming that the non-total-time-derivative part is
constant and equal to the end-of-time step value (e.g., $P = P_{\rm end}$).

For nearly adiabatic evolution in electron degenerate conditions, we have a
cancellation between large $\dif e$ and ${P\dif (1/\rho)}$ terms, but this
cancellation ends up incomplete in \mesa\ because the evaluation of
the former term is exact while the latter has error.
The error is usually small compared to the order of the terms being
subtracted, and so imperfect cancellation often will not introduce
large errors.  But in degenerate material, the scale of the cancelling
terms is larger than the thermal energy by roughly the degeneracy
parameter ${\eta \equiv \mu_e / k_{\rm B} T}$, where $\mu_e$ is the electron
chemical potential. Therefore, otherwise small cancellation errors can
be amplified by a factor of $\eta$ for the temperature evolution.%
\footnote{Numerical cancellation errors are a common pitfall for
  evolution in electron degenerate material. See \cite{Brassard1991}
  for a detailed discussion of an analogous problem in evaluating the
  \BV\ frequency in WD interiors.}

By contrast, when we write the $\epsilon_{\rm grav}$ form, this
cancellation for adiabatic evolution instead occurs in ${[\rho
(\partial e/\partial \rho)_T - P/\rho]}$ (Equation~\ref{eq:epsfifth})
which is replaced with ${c_P T \nabla_{\rm ad} \chi_{\rho}}$ in
the form of Equation~\eqref{eq:eps_grav_rewrite} that \MESA\ uses for
its \texttt{eps\_grav} implementation. This captures adiabatic density
evolution in terms of EOS derivative quantities that are not subject
to cancellation errors. Instead, accuracy in this form is limited by
the accuracy of our approximations over finite time steps for thermodynamic quantities like
$\nabla_{\rm ad}$ appearing in the energy equation. The error
associated with time discretization,
$[c_P T (1 - \nabla_{\rm ad}\chi_{T})]_{\rm end}\Delta \ln T
+ [c_P T \nabla_{\rm ad} \chi_{\rho}]_{\rm end}\Delta \ln \rho$,
is at least a factor of $\sim \eta$ smaller than the
cancellation error, and in practice can be even better.

\begin{figure}
  \centering
  \includegraphics[]{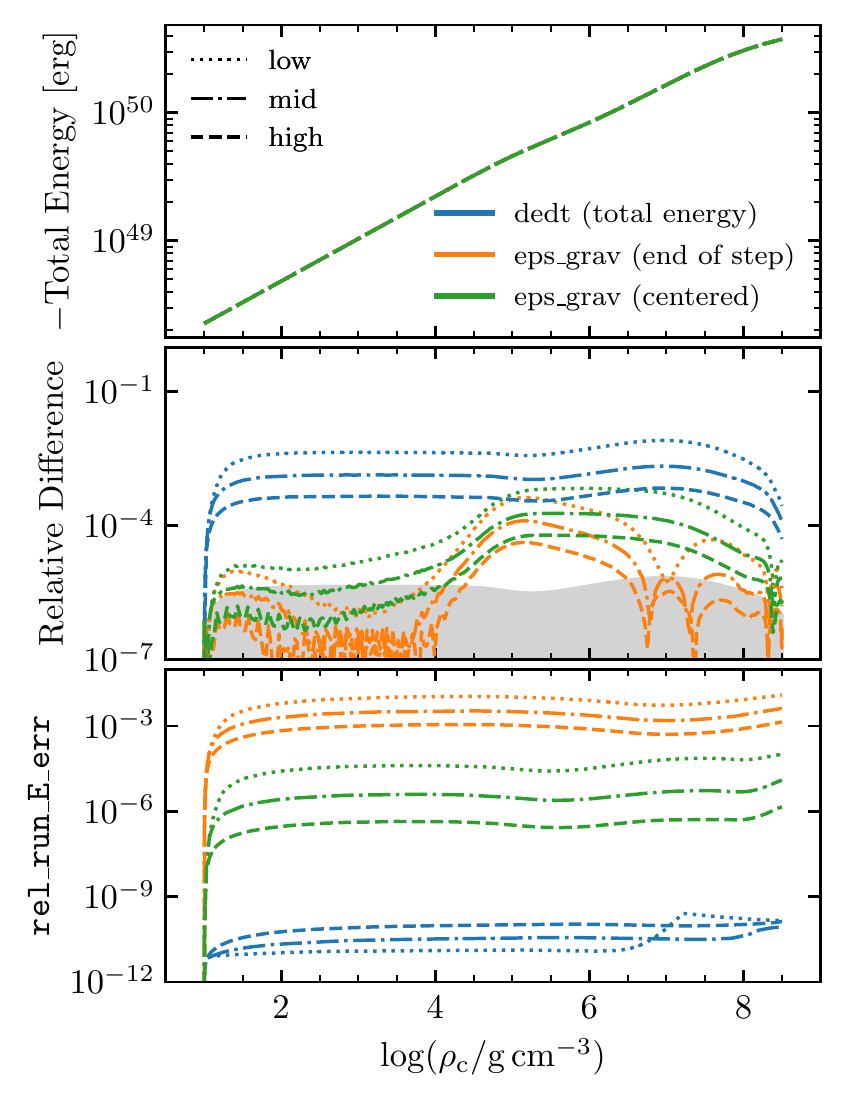}
  \caption{Convergence tests of the evolution of the total energy for
    the contracting 1.3\,\Msun\ sphere of pure carbon using different forms of
    the energy equation.
    The top panel shows the negative total energy.  The middle panel shows the
    difference from a reference solution.  The grey region marks the
    level of agreement between the set of reference solutions.  The
    lower panel shows the relative cumulative energy error for each
    run.}
  \label{fig:ckh_tote}
\end{figure}

Figure~\ref{fig:ckh_tote} shows the total energy of the model.  The
top panel shows that all runs agree in this quantity within the line
width, while the middle panel reveals the level of relative
difference.
We emphasize that even though the \texttt{dedt} runs report by far the best
cumulative energy error as a measure of step-to-step internal energy
consistency (lower panel of Figure~\ref{fig:ckh_tote}), they nevertheless
show less accurate evolution of the total energy and temperature
relative to the ultra high-resolution reference runs (middle panel of
Figure~\ref{fig:ckh_tote}). This is
because the cumulative energy error reports the degree to
which energy is conserved by evolution steps, while the total energy
is a function of the global stellar structure which can slowly diverge
even with zero reported energy error. This reflects the fact
that ``energy error'' as reported by \MESA\ is primarily a measure of
the internal consistency of the stellar structure solver, and should
not be construed as always reflecting globally accurate energy
evolution. This reported error is still a useful diagnostic for \MESA\
models, but must be interpreted with caution.

All forms of the energy equation in Figure~\ref{fig:ckh_tote} approximately show first-order convergence in
the total energy.  The lower panel shows that the different forms exhibit notably different
behaviors with increasing time resolution.  The \texttt{dedt} form has
excellent numerical energy conservation that does not depend on time
resolution. The error is roughly the error due to the non-zero
residuals in the solution of the energy equation.  The \texttt{eps\_grav} forms
display worse performance in this quantity, though the error shrinks
as the time step decreases.  The ``end of step'' form shows first-order
convergence, while the ``centered'' variant exhibits more rapid,
second-order convergence with smaller numerical energy conservation
errors at fixed resolution. We would expect these trends to continue until the
numerical energy conservation error is no longer dominated by errors
due to the temporal discretization, at which point it reaches the
floor set by non-zero residuals or imperfect EOS thermodynamics.

The pure carbon case shows that the time-centered \texttt{eps\_grav} form of the
energy equation is the best choice for models evolving under
degenerate conditions, with the best balance between accurate
temperature evolution and step-to-step energy conservation according
to Figures~\ref{fig:ckh_center} and~\ref{fig:ckh_tote}.
This case was idealized to focus on the effects of finite equation
residuals and time discretization. We now move on to discussing the
additional complexities introduced by EOS imperfections.

\subsection{Quantifying EOS shortcomings}
\label{sec:eosproblems}

The value of $e$ returned by the EOS is an essential ingredient in
evaluating the total energy of the model, and high-quality partial
derivatives of EOS quantities are critical for accurate and efficient
solver performance. We now discuss three primary EOS
issues that influence energy conservation and solver performance.
First, an EOS may return low quality partial derivatives that degrade
convergence of the implicit solver. We now mitigate this with more
careful derivative accounting described in \S\ref{sec:eosderivatives}.
Second, an EOS may have internal inconsistencies in its reported
thermodynamics. We have mitigated this by upgrading the \mesa\ EOS
patchwork with Skye \citep{Jermyn2021} and FreeEOS \citep{Irwin2004}
as described in \S\ref{sec:eos}.
Third, even when individual EOS components yield excellent
thermodynamic consistency, the necessity of blending between EOS
components to provide continuous coverage across different regimes
inevitably introduces additional thermodynamic inconsistency.
We have mitigated this last issue by minimizing the number and
severity of EOS blends as much as possible, but unavoidable energy
inconsistencies remain, and we discuss their implications for energy
conservation in \S\ref{sec:eosblend_error}.

\subsubsection{EOS Derivatives}
\label{sec:eosderivatives}

In \mesafive, we addressed the quality of the EOS derivatives by
introducing new options that used bicubic spline interpolation in
high-resolution tables of $P_{\rm{gas}}$, $s$, and $e$.  
This provided accurate first and second
partial derivatives by evaluating analytic derivatives of the
interpolating polynomials rather than by interpolating values of
tabulated derivatives.
While this approach successfully ensured that the partial derivatives
corresponded to how the interpolated EOS values actually changed in
response to small changes of the parameters, it inevitably led to
small, interpolation-related artifacts in partial derivative quantities
such as \grada\ or \gammaone.  In asteroseismic applications that require smooth profiles of the \bvv\ frequency, this
approach proved unsatisfactory.

\MESA\ now adopts an approach that separately treats quantities
that appear in the equations (and happen to be partial derivatives) and the
places where these theoretically equivalent, but numerically different
quantities appear in the Jacobian (as partial derivatives of other quantities
that appear in the equations).  That is, the Jacobian uses the
partial derivatives of \revision{bicubic spline} interpolants, while the equations use the \revision{bicubic spline}
interpolants of
partial derivatives.  This enables both efficient numerics and smoother solutions
at the cost of some additional bookkeeping.
\revision{A potential pitfall is that negative values for non-negative 
quantities can be encountered. In practice, we find that we do not
encounter negative interpolants for the physical quantities that enter
the equations. While we may encounter negative values from the
derivatives of the interpolants used for the Jacobian, these only
guide the Newton iterations in converging toward a solution.
In this scheme, negative
derivatives of interpolants cannot introduce physical errors into the
equations used for model solutions.
}

\subsubsection{Thermodynamic Consistency and EOS Blends}
\label{sec:eosblend_error}

In order to quantify how models employing the different forms of the energy equation
experience inconsistencies in the EOS differently,
we establish a measure of the quality of the \MESA\ EOS during the
evolution of a model as follows.
In the $(\rho, T, \{X_i\})$ basis, the total derivative of the
specific internal energy, $e$, mathematically satisfies
\begin{equation}
  \begin{split}
  \DDt{e} - &\left[\ddp{e}{\rho}_{T,\{X_i\}} \DDt{\rho} + \ddp{e}{T}_{\rho,\{X_i\}} \DDt{T} \right.\\
  & \left. \;\; + \sum_i \left(\frac{\partial e}{\partial X_i}\right)_{\rho,T, \{X\ne X_i\}} \DDt{X_i} \right] = 0.
\label{eq:dedt}
   \end{split}
\end{equation}

For a Lagrangian volume corresponding to a cell $k$, we evaluate the
time integral of the left hand side of Equation~(\ref{eq:dedt}) across
a time step.  We replace the sum over individual composition
derivatives with a single directional derivative along the direction
of composition change over the time step, as in the evaluation of
\epsgravX.  We approximate terms that are not the integral of total
derivatives using the implicit trapezoidal rule.  Weighting by $dm_k$ 
and summing over all cells this gives, for a single
step,
\begin{equation}
  \begin{split}
  E_{\rm err,eos}^{\rm step} =
  \sum_{k = 1}^n dm_k \Bigg\{\Delta e
    & - \overline{\left[\rho \left(\frac{\partial e}{\partial \rho}\right)_{T}\right]} \Delta \ln \rho \\
    & - \overline{[c_V T]} \Delta \ln T
    - \overline{\Delta e_{X_i}}
  \Bigg\}_k~,
  \label{eq:e_err_eos_step}
\end{split}
\end{equation}
where overline quantities are the trapezoidal rule estimates
corresponding to the average at the start and end of the time step.
The value $\Delta e_{X_i} = e(\{X_i\}) - e(\{X_i\}_{\rm start})$ is the change in
specific internal energy due to composition changes alone at a given
$\rho$ and $T$, so
\begin{equation}
  \begin{split}
  \overline{\Delta e_{X_i}}
= \frac{1}{2} \big[&e(\start{\rho}, \start{T}, \{X_i\}) - e(\start{\rho}, \start{T}, \start{\{X_i\}}) \\
& + e(\rho, T, \{X_i\}) - e(\rho, T, \start{\{X_i\}}) \big] ~.
\end{split}
\end{equation}
Summing the per-step errors over a \MESA\ run, then at the $n$-th
time step, we have
\begin{eqnarray}
  E_{\rm err,eos}^{\rm run} = \sum_{i =1}^n E_{\rm err,eos}^{{\rm step},i}.
\end{eqnarray}
As the time step is reduced, the error from the temporal discretization
shrinks and $E_{\rm err,eos}^{\rm run}$ converges to a measure
of the energy error incurred as a result of EOS shortcomings.

We define two other energy errors.  By limiting the sum in
Equation~\eqref{eq:e_err_eos_step} to those zones that are in an EOS
blend during a particular time step, we can isolate the per-step energy error
due to the blend and so analogously define
$E_{\rm err, blend}^{\rm step}$ and $E_{\rm err, blend}^{\rm run}$.
We define the per-step residual energy error
$E_{\rm err, res}^{\rm step}$ as the mass-weighted sum of the energy
equation residuals over the model and also track its cumulative value
$E_{\rm err, res}^{\rm run}$.

To understand how these different forms of error might manifest under
different treatments of the energy equation, it is helpful to consider
the idealized case of an EOS that is a blend of EOS1 and EOS2, identical 
except for their definition of where the energy zero
point lies, so that ${e_{\rm EOS1} = e_{\rm EOS2} + e_{\rm offset}}$,
where $e_{\rm offset}$ is some constant. Physically, either EOS should
produce the same evolution, and all EOS derivatives will be the
same. However, evolution through the blend between EOS1 and EOS2 will
not satisfy Equation~\eqref{eq:dedt}, and therefore must lead to
non-zero values of $E_{\rm err,eos}^{\rm step}$ regardless of which
form of the energy equation is used. In particular, the thermodynamic
identity ${c_V = (\partial e/\partial T)_\rho}$ is violated in the
blending region because the blending coefficient derivatives are not
included as part of the blended value of $c_V$
(see \S\ref{sec:eos_blends}).

In practice, such energy offsets (in addition to other
inconsistencies) always occur at the locations of \MESA\ EOS blends
because it is impossible to construct a blending region in
which two distinct EOS treatments agree exactly. Due to different
input physics assumptions in different EOS components, energy offsets
have more complexity than simple constant zero point differences
across the range of parameters where blending is necessary. As far as
possible, we have chosen blending locations to minimize the
differences between EOS components and to minimize the residual amount
of unavoidable offset (e.g., by adjusting the definitions of internal
energy to be as consistent as possible about where the zero point
lies). However, no general solution is currently available to
completely eliminate inconsistencies for blends between our current EOS
components, and EOS blends therefore remain one of the largest
potential sources of energy error when present in \mesa\ models.

When evolving using the \texttt{dedt} form
of the energy equation, Equation~\eqref{eq:dedt_form} will lead to
$e_{\rm offset}$ being folded into the $\Dif e/\Dif t$ term of the energy
equation for regions of the model evolving through the blend,
injecting spurious heating/cooling into those regions. However, the
energy error reported by \MESA\ in these regions may still be 0, because
\MESA\ evaluates energy error according to the blended $e$ from the
EOS. It is therefore possible to have models for which $E_{\rm
  err,eos}^{\rm step}$ is significantly larger than \texttt{E\_err}
reported by \MESA\ when using the \texttt{dedt} form of the energy equation, as
we shall see in the following subsection. On the other hand, in this
idealized scenario the \texttt{eps\_grav} form of the energy equation
(Equations~\ref{eq:epsfifth} and~\ref{eq:eps_grav_rewrite}) would give
the physically correct evolution since it is evaluated in terms of
derivatives that are unaffected by the energy offset in the blending
region.
However, both $E_{\rm err,eos}^{\rm step}$ and \texttt{rel\_run\_E\_err}
will report large values under the \texttt{eps\_grav} form in
this scenario, reflecting the thermodynamic inconsistency of $e$ in
the blended EOS rather than inaccuracy in the evolution.

\subsection{Results: 1M\_pre\_ms\_to\_1TP}
\label{sec:1M_pre_ms_to_1TP}

Having demonstrated the performance of the various forms of the energy
equation in idealized circumstances in \S\ref{sec:carbon_kh}, we
now model the evolution of a 1\,\Msun\ star from the pre-main sequence
to its first He thermal pulse on the AGB.
This example includes composition changes due to nuclear reactions,
and it uses the current default \MESA\ EOS (blending together FreeEOS
and Skye for the $(\rho,T)$ regions encountered by this model).
A small portion of the envelope of this model encounters the
FreeEOS-OPAL/SCVH blend near ZAMS, then the model lies entirely on
FreeEOS for most of the first $\approx$5\,Gyr of MS evolution (see
\S\ref{sec:eos} and Figure~\ref{fig:eosbounds}), after which the
core evolves toward higher density and encounters the FreeEOS-Skye blend.
The approach described in \S\ref{sec:eosproblems} allows us to
quantify the various sources of energy errors, identifying how much
error comes from EOS inconsistencies and blends, and how much is due
to residuals of the equation solutions.

\begin{figure}
  \centering
  \includegraphics[width=\apjcolwidth]{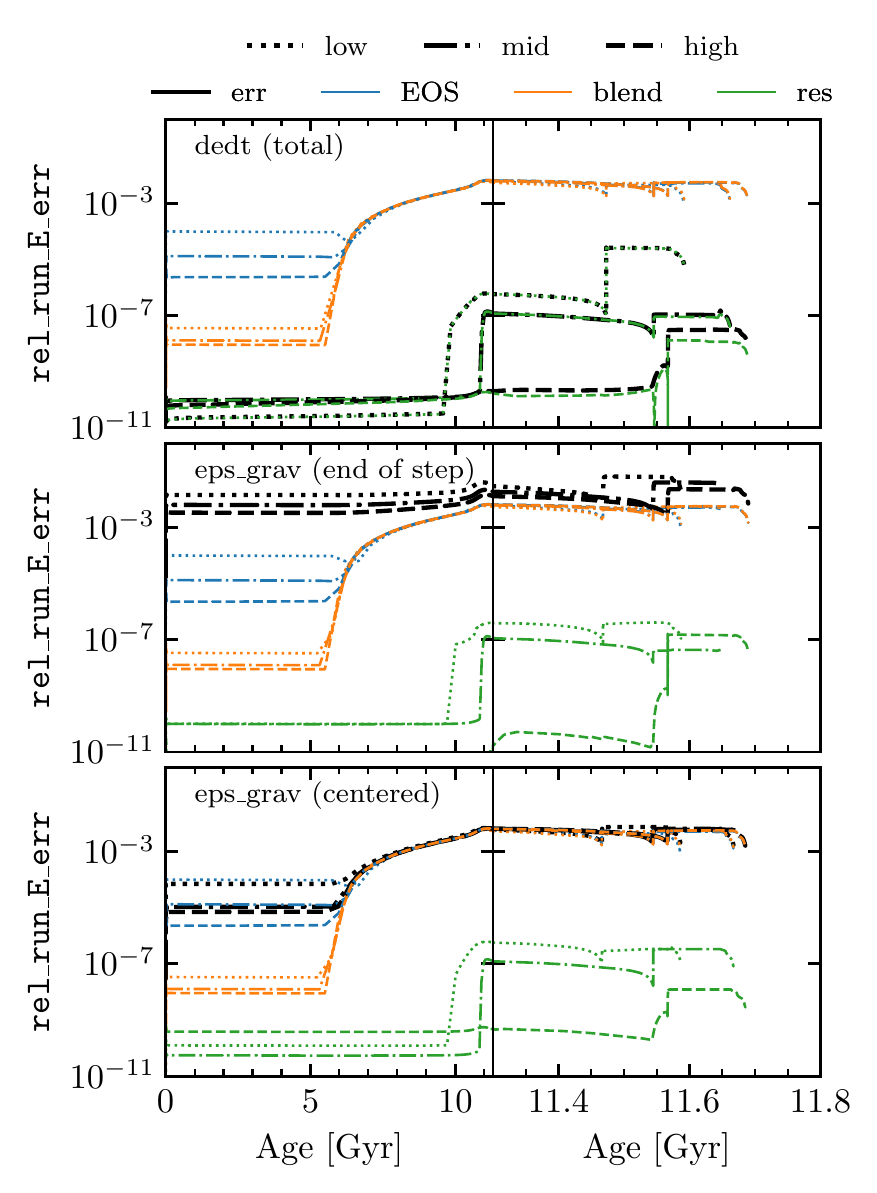}
  \caption{Comparison of types of energy errors for a 1\,\Msun\
    model.  Each panel shows a different form of the energy equation.
    Each line color indicates a different type of energy error.  Each
    line style indicates a different time resolution. The thick black lines show the error in numerical energy
conservation that \mesa\ reports as \texttt{rel\_run\_E\_err}.
The thinner colored lines show the errors due to the
EOS, EOS blends, and equation residuals.  The composition
    term \epsgravX\ is included for both of the \texttt{eps\_grav}
    panels in this figure.}
  \label{fig:1M_err_res}
\end{figure}

Figure~\ref{fig:1M_err_res} summarizes the energy errors defined in
\S\ref{sec:eosproblems} when using different forms of the energy
equation. Because the models are
approximately converged and the different forms of the energy equation
converge to the same solution, the colored lines at a given time
resolution are similar in all panels.
\revision{In particular, the measure of the energy error
associated with the EOS blend has approximately converged
(agreement among orange lines in the middle of each panel).
However, the measure of the EOS inconsistency in places where this quantity is
not dominated by the blend (i.e., where the blue lines are above the orange lines) 
continues to shrink with increasing
time resolution, showing that this measure is not converged and that the
true inconsistency is less than that indicated by the highest resolution line.
}

In the top panel, the \texttt{dedt} form shows a numerical energy error (black
lines) that is roughly the energy error associated with the equation
residuals (green lines). The EOS energy errors are present in the
model, but do not show up in the reported \texttt{rel\_run\_E\_err} for this
form of the energy equation, as explained in
\S\ref{sec:eosproblems}.
In the middle panel, the end-of-step \texttt{eps\_grav} form shows that the
numerical energy conservation error remains above the energy errors
associated with the EOS, while these in turn are generally well above
the errors associated with the equation residuals.
In the bottom panel, the centered \texttt{eps\_grav} form shows a numerical
energy error (black lines) that is roughly the energy error associated
with the EOS inconsistencies.  In particular, by the end of the main
sequence, the numerical energy error is dominated by the blend error.
Further resolution increases or improvements to the energy equation
will not improve the numerical energy conservation.
Progress can only come through improvements to the EOS.


\section{Atmosphere}
\lSect{atmosphere}

\newcommand{\dqdtau}{\mathrm{d}q/\mathrm{d}\tau}

The \MESA\ \texttt{atm} module uses an atmosphere model to evaluate
the pressure $\Psurf$ and temperature $\Tsurf$ at the outermost
($k=1$) cell boundary. These `model surface' values are in turn
incorporated in the outer boundary conditions applied to the stellar
model, as specified in Equation~(10) of \mesaone. Here we
describe various improvements and fixes in the \texttt{atm} module
since \mesaone\ and \mesatwo.

In older \MESA\ releases, the choice of atmosphere model was
controlled by the \texttt{which\_atm\_option} inlist parameter. In
recent releases this parameter is renamed \texttt{atm\_option}, with
three possible choices:
\begin{itemize}
  \item \texttt{atm\_option = \tqs T\_tau\tqs}: atmosphere based on $T(\tau)$ relations, as discussed in
    \Sect{atmosphere-T-tau}.
  \item \texttt{atm\_option = \tqs irradiated\tqs}: irradiated atmosphere, as discussed in
    \Sect{atmosphere-irrad}.
  \item \texttt{atm\_option = \tqs table\tqs}: tabulated atmosphere, as discussed in
    \Sect{atmosphere-table}.
\end{itemize}
In addition, \MESA\ now offers two complementary approaches to
including the atmosphere structure in model data passed as input into
pulsation codes. These are described in \S\ref{sec:atm-recon} and
\S\ref{sec:atm_as_int}.

\subsection{Atmospheric $T(\tau)$ Relations} \lSect{atmosphere-T-tau}

Setting \texttt{atm\_option = \tqs T\_tau\tqs} builds an atmosphere in
which temperature at each optical depth $\tau$ is specified by a
function $T(\tau)$; $\Psurf$ and $\Tsurf$ are obtained by evaluating
the atmospheric thermodynamic state at an optical depth $\tausurf$
corresponding to the nominal model surface. This optical
depth can be much smaller or much greater than the optical depth
$\tau \approx 2/3$ typically associated with stellar photospheres;
that is, the model surface need not correspond to the photosphere.

The $T(\tau)$ functions have the form
\begin{equation}
  \label{eq:Ttau}
  T^{4}(\tau) =\frac{3}{4} \Teff^{4} \left[ \tau +  q(\tau) \right].
\end{equation}
Selection of $T(\tau)$ is set by the \texttt{atm\_T\_tau\_relation}
inlist parameter, with four possible choices:
\begin{itemize}
\item \texttt{atm\_T\_tau\_relation = \tqs Eddington\tqs}: the gray,
  Eddington-approximation relation, where $q(\tau)=2/3$.
\item \texttt{\atm\_T\_tau\_relation = \tqs solar\_Hopf\tqs}: the $q(\tau)$
  function described in Equation~(A9) of \mesatwo, which is a fit to
  Model C of the solar atmosphere by \citet{Vernazza:1981},
  often referred to as VAL C.
\item \texttt{\atm\_T\_tau\_relation = \tqs Krishna\_Swamy\tqs}: the
  relation from Equation~(33) of \citet{Krishna-Swamy:1966}.
\item \texttt{\atm\_T\_tau\_relation = \tqs Trampedach\_solar\tqs}: the
  relation from \citet{Ball:2021}, itself a fit to the solar
  atmosphere simulation by \citet{Trampedach:2014}.
\end{itemize}
For a given $T(\tau)$ relation, the corresponding
$P(\tau)$ throughout the atmosphere is obtained by integrating the
hydrostatic balance equation
\begin{equation} \label{atm-hse}
  \dd{P}{\tau} = \frac{g}{\kappa},
\end{equation}
from $\tau \ll 1$ inward to $\tau=\tausurf$. In this integration the
gravity is set to the constant value $g=GM/R^{2}$, in accordance with the
assumption that the atmosphere is geometrically thin and contains
negligible mass. The opacity evaluation is controlled by the
\texttt{atm\_T\_tau\_opacity} inlist parameter, with three possible
choices:
\begin{itemize}
  \item \texttt{\atm\_T\_tau\_opacity = \tqs fixed\tqs} --- uniform $\kappa$ throughout
    the atmosphere, with a value set by the current opacity
    $\kappa_{1}$ in the outermost cell.
  \item \texttt{\atm\_T\_tau\_opacity = \tqs iterated\tqs} --- uniform
    $\kappa$ throughout the atmosphere, with a value obtained from the
    \kap\ module for $(\Psurf,\Tsurf)$. As indicated by its name, this choice requires
    iteration because $\Psurf$ is not known a priori.
  \item \texttt{\atm\_T\_tau\_opacity = \tqs varying\tqs} --- varying
    $\kappa$ throughout the atmosphere, with a value obtained from the
    \kap\ module for $(P,T)$ at the local $\tau$.
\end{itemize}
With the last choice, the \texttt{dopri5} ($5^{\rm th}$-order
Dormand-Prince) differential equation integrator from the \num\ module
is employed with specifiable error tolerance and maximum number
of steps. With the first and second choices, however, the fact that
$\kappa$ does not depend on $\tau$ means that Equation \eqref{atm-hse} can be
integrated analytically to yield 
\begin{equation} \label{atm-P}
P(\tau) = \frac{g}{\kappa} \tau \left( 1 +  \texttt{Pextra\_factor} \times \frac{\kappa L}{6 \pi G M c \tau} \right).
\end{equation} 
The second term in the parentheses arises as a constant of
integration, and accounts for non-zero radiation pressure in the limit
of small $\tau$. The term is obtained from Equation~(20.16) of
\citet{Cox:1968}. The parameter \texttt{Pextra\_factor} depends
on the assumed angular dependence of the radiation specific intensity,
with the default value of unity corresponding to isotropic-outward,
and a value of $1.5$ corresponding to radial-outward. While
unphysical, setting $\texttt{Pextra\_factor} > 1.5$ can sometimes be a
useful numerical strategy to improve convergence in models that are
close to the Eddington limit. However, caution is warranted as this
strategy can produce incorrect stellar radii.

\begin{table*}[ht!]
\centering
  \begin{tabular}{ccc}
    Release r11701 and earlier &
    \multicolumn{2}{c}{Release r12115 and later} \\
    \texttt{which\_atm\_option} & \texttt{atm\_T\_tau\_relation} & \texttt{atm\_T\_tau\_opacity} \\ \hline
    \texttt{\tqs simple\_photosphere\tqs} & \texttt{\tqs Eddington\tqs} & \texttt{\tqs fixed\tqs} \\
    \texttt{\tqs gray\_and\_kap\tqs} & \texttt{\tqs Eddington\tqs} & \texttt{\tqs iterated\tqs} \\
    \texttt{\tqs Eddington\_gray\tqs} & \texttt{\tqs Eddington\tqs} & \texttt{\tqs varying\tqs} \\
    \texttt{\tqs Krishna\_Swamy\tqs} & \texttt{\tqs Krishna\_Swamy\tqs} & \texttt{\tqs varying\tqs} \\
    \texttt{\tqs solar\_Hopf\tqs} & \texttt{\tqs solar\_Hopf\tqs} & \texttt{\tqs varying\tqs}
  \end{tabular}
  \caption{Mapping between \texttt{which\_atm\_option} parameter
    choices in releases of \MESA\ up to r11701 (see \S\ref{sec:nudocker}), and the corresponding
    parameter choices in releases since r12115 (to be used in tandem with
    \texttt{atm\_option = \tqs T\_tau\tqs}).}
  \lTab{atm-t-tau}
\end{table*}

For $T(\tau)$ atmospheres, \Tab{atm-t-tau} summarizes the
mapping between the \texttt{which\_atm\_option} parameter choices
supported in older \mesa\ releases, and the combinations of
\texttt{atm\_T\_tau\_relation} and \texttt{atm\_T\_tau\_opacity}
parameter choices that provide the replacement functionality.
 
In implementing the changes described here, we uncovered two issues
that impacted the accuracy and performance of the \texttt{atm} module
in older releases.
 
First, calls to the \eos\ and \kap\ modules to evaluate $\rho(P,T)$
and $\kappa(\rho,T)$ did not use the same tables and/or configuration
options as the interior model, leading to possible inconsistencies at
the surface where the atmosphere and the interior join. To fix this
problem, we implemented a callback system so that the \texttt{star}
module can pass appropriately configured EOS and opacity wrapper
routines to the \texttt{atm} module.

Second, in cases where \texttt{atm\_T\_tau\_opacity = \tqs
  iterated\tqs}, the partial derivatives of $\Psurf$ and $\Tsurf$ with
respect to dependent variables ($L_{1},r_{1},\rho_{1},T_{1}$) in the
outermost cell were incorrectly evaluated; in some cases, this caused
the global Newton solver to converge slowly or not at all. To fix
this problem, we implemented the correct expressions, which follow
from application of the chain rule to the $T(\tau)$ relation and the
hydrostatic solution (Equation~\ref{atm-P}).

\subsection{Irradiated Atmospheres} \lSect{atmosphere-irrad}

Setting \texttt{atm\_option = \tqs irradiated\tqs} provides functionality
similar to \texttt{\tqs T\_tau\tqs}, but adopting the
$T(\tau)$ relation for an externally irradiated atmosphere
given in Equation~(6) of \citet{Guillot:2011}.
Equation \eqref{atm-hse} is integrated analytically, with
opacity evaluation controlled by the \texttt{atm\_irradiated\_opacity}
inlist parameter; the possible choices \texttt{\tqs fixed\tqs} and
\texttt{\tqs iterated\tqs} behave the same as described in
\Sect{atmosphere-T-tau}. The \texttt{\tqs iterated\tqs}
case replaces the \texttt{which\_atm\_option = \tqs
  gray\_irradiated\tqs} choice described in \mesatwo, and fixes a bug
related to incorrect $\Teff$\ evaluation.

\subsection{Tabulated White Dwarf Atmospheres} \lSect{atmosphere-table}

When \texttt{atm\_option = \tqs table\tqs}, $\Psurf$ and $\Tsurf$ are
obtained by interpolating in pre-computed atmosphere tables. In
addition to the options described in \mesaone\ and \mesatwo,
\MESA\ now provides a new set of atmosphere tables for WDs with
He-dominated surfaces (DB WDs). These tables provide
$\Psurf$ and $\Tsurf$ data over the ranges $5,000\,\Kelvin < \Teff
< 40,000\,\Kelvin$ and $6.0 < \log(g/\cmpersecondSc)< 9.4$. They assume a
He dominated composition of $n_{\rm{H}} = 10^{-5}n_{\rm{He}}$, and are based on model
atmospheres calculated using the \citet{Koester10} code and evaluated
at $\tau=25$.  The limits of this grid are now explained.

At $\Teff \gtrsim 40,000$\,K
He undergoes its second ionization, and non-LTE effects that are not included in the
atmosphere code become important.  The lower limit
of $\Teff = 5,000\,\Kelvin$ is imposed by the \citet{Koester10} code's EOS.  Studies of spectroscopic samples of
DBs \citep[e.g.,][]{Eisenstein2006, Genest-Beaulieu2019} have shown
that DBs are well described for surface gravities lower than
$\log(g/\cmpersecondSc) = 9.5$.  The lower limit of
$\log(g/\cmpersecondSc) = 6.0$ is imposed by the convergence of the
models at $\Teff=40,000\,\Kelvin$, since radiation pressure competes
with gravity in this regime.

\begin{figure}
\includegraphics[width=\apjcolwidth]{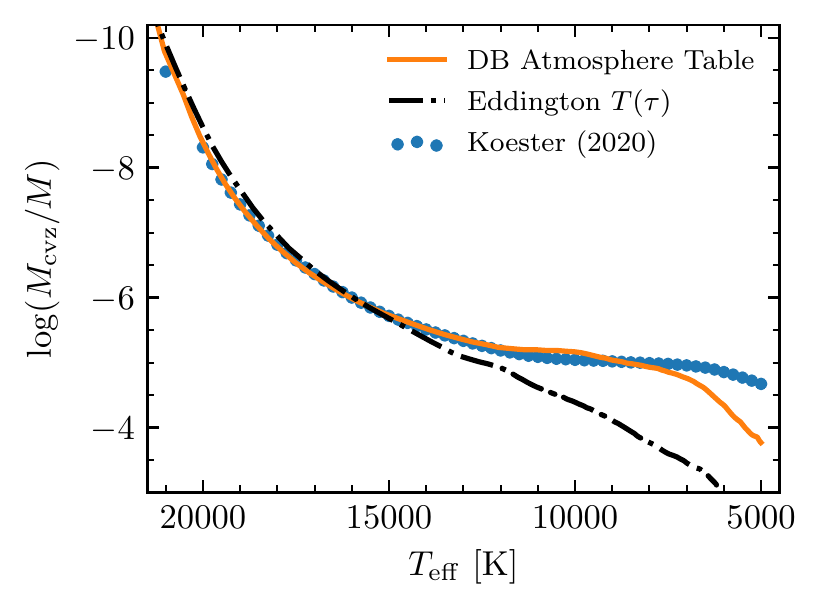}
\caption{Depth of the surface convection zone (measured as fractional convection
  zone mass $M_{\rm cvz}/M$) for a
  $0.57\,\Msun$, $\log(g/{\rm cm\,s^{-2}}) \approx 8.0$ WD with a He
  atmosphere. The curves indicate \MESA\ results with the
  DB tables and the Eddington $T(\tau)$ relation, while the points are
  based on \citet{Koester2020}.}
\label{fig:db-wd}
\end{figure}

As a diagnostic for these new tables, we use the depth of the surface
convection zone in cooling DB WD models, which is primarily sensitive
to the atmospheric boundary condition and the stellar model EOS in the convective
region.  \Figure{db-wd} demonstrates the improvements brought by these
new tables, plotting the mass $M_{\rm cvz}$ of the surface convection
zone versus \Teff.  Also shown are
results from the atmosphere and envelope models of
\citet{Koester2020}, as well as the outcome from using
\texttt{atm\_option = \tqs T\_tau\tqs} with
\texttt{atm\_T\_tau\_relation = \tqs Eddington\tqs} and
\texttt{atm\_T\_tau\_opacity = \tqs fixed\tqs}. 

The new
tables reproduce the \citet{Koester2020} calculations much better at
low $\Teff$ than the Eddington $T(\tau)$ atmospheres. The latter
diverge for $\Teff \lesssim 15,000\,\Kelvin$
because the conditions at the surface cross the boundaries of the He opacity table coverage in \mesa.
The tabulated atmospheres agree with \citet{Koester2020} to
$\Teff \approx 7,000\,\Kelvin$. 
For cooler temperatures, the
uncertainty in $M_{\rm cvz}$ is due to the uncertain EOS at the base of the convection zone \citep{Saumon1995}.
\footnote{See discussion at
  \url{http://www1.astrophysik.uni-kiel.de/~koester/astrophysics/astrophysics.html},
  where the full tables of convection zone depths based on
  \cite{Koester2020} are hosted.}

\subsection{Atmosphere Reconstruction for Pulsation Codes}
\label{sec:atm-recon}

\begin{figure}
  \includegraphics{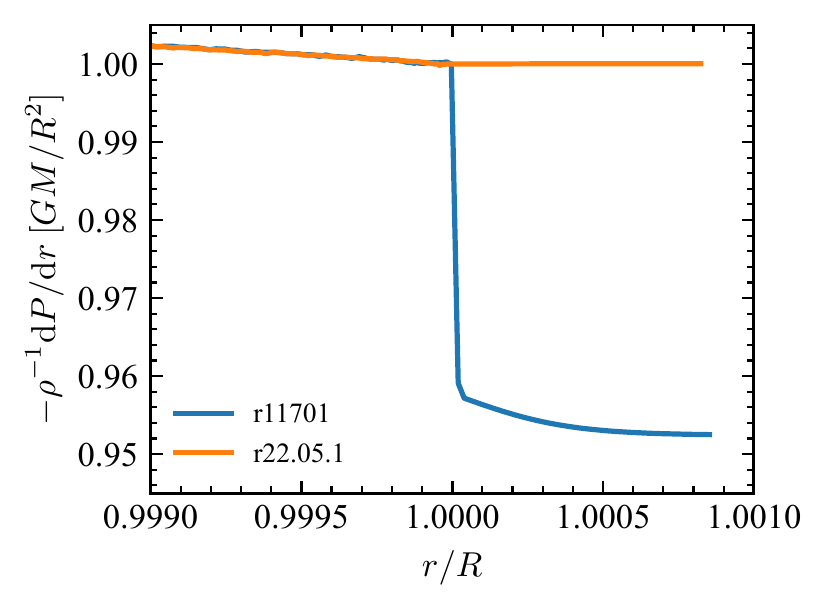}
  \caption{The hydrostatic term $\rho^{-1} \D P/\D r$ (in units of the
    surface gravity $GM/R^{2}$) as a function of $r/R$ for \GYRE-format models of the present-day
    Sun that include reconstructed atmospheres
    with \texttt{atm\_option = \tqs T\_tau\tqs},
    \texttt{atm\_T\_tau\_relation = \tqs Eddington\tqs} and
    \texttt{atm\_T\_tau\_opacity = \tqs varying\tqs}. The two curves
    show releases r11701 (prior to the reconstruction fix) and
    r22.05.1; only the latter reproduces the correct behavior
    $\rho^{-1} \D P/\D r = -GM/R^{2}$ throughout the reconstructed
    atmosphere $r/R \geq 1$.} \label{fig:atm-hse}
\end{figure}

As discussed in \mesatwo\ and \mesathree, \MESA\ can pass models to
the \ADIPLS\ \citep{Christensen-Dalsgaard:2008} or
\GYRE\ \citep{Townsend:2013} linear pulsation codes, either in-memory
during an \astero-module optimization or via files written to disk in
a variety of formats. Often, it is desirable to reconstruct the
atmosphere structure from $\tau=\tausurf$ out to $\tau
\ll 1$ before passing them to the pulsation codes. This has no impact on the interior
model, but can improve asteroseismic modeling.

For $T(\tau)$ atmospheres (\Sect{atmosphere-T-tau}), setting
\texttt{add\_atmosphere\_to\_pulse\_data = .true.} enables this
reconstruction. The radial coordinate $r$ throughout the atmosphere is
then determined by integrating the $\tau$ equation
\begin{equation} \lEq{atm-tau}
  \dd{r}{\tau} = - \frac{1}{\kappa \rho}
\end{equation}
outward from $\tau = \tausurf$ to $\tau \ll 1$. For this integration,
$\kappa$ is evaluated in accordance with the
\texttt{atm\_T\_tau\_opacity} parameter discussed previously, while
$\rho(P,T)$ is obtained from the \eos\ module for the
local $\tau$.

In releases of \MESA\ prior to r12115, the outward integration used an explicit Euler scheme with a default step-size too large to
accurately follow $r(\tau)$. Together with the
\eos/\kap\ table inconsistency highlighted in \Sect{atmosphere-T-tau},
this led to departures from hydrostatic balance in the
reconstructed atmosphere, as highlighted in Figure~D.1 of
\citet{Christensen-Dalsgaard:2020} for $r/R > 1$. To fix this issue,
release r12115 and later use the \texttt{dopri5} integrator for the
outward integration, with a specifiable error tolerance, step-size
and outermost optical depth. \revision{Figure \ref{fig:atm-hse}} demonstrates these
improvements by showing hydrostatic balance for
\GYRE-format models of the present-day Sun calculated using releases
r11701 (pre-fix) and r22.05.1.

\subsection{Atmospheres as Part of the Interior}
\label{sec:atm_as_int}

\begin{figure}
  \includegraphics{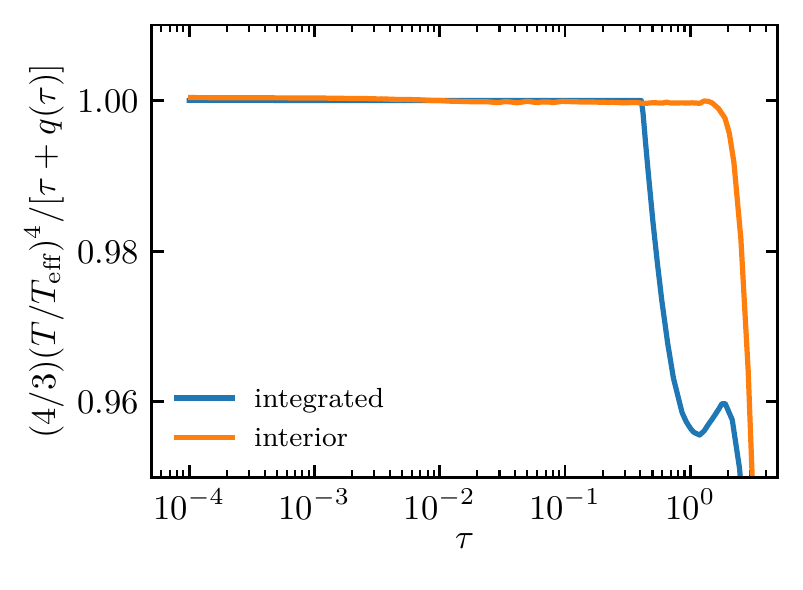}
  \caption{Ratio of the left- and right-hand sides of Equation~\eqref{eq:Ttau},
    by which we compare the temperature structure of two models
    in FGONG format, one with the atmosphere reconstructed as in
    \S\ref{sec:atm-recon} (integrated) and the other with the atmosphere
    modeled with the interior as in \S\ref{sec:atm_as_int} (interior).}
  \label{fig:atm-Ttau}
\end{figure}

The optically-thin outer layers of a star are usually treated separately
from the interiors of stars because they do not satisfy
the assumptions under which the interior structure equations
are derived.  
Given a $T(\tau)$ relation, however, it is possible to correct
the equation for radiative heat transport so that the temperature stratification produced by
solving the stellar structure equations matches the desired $T(\tau)$ \citep{Trampedach:2014,Mosumgaard:2018}.
The radiative temperature
gradient $\nabla_\mathrm{atm}$ of an atmosphere that follows a given
$T(\tau)$ relation is
\begin{equation}
  \nabla_\mathrm{atm}=\nabla_\mathrm{rad}\left(1+\frac{\mathrm{d}q}{\mathrm{d}\tau}\right).
\end{equation}
We can therefore recover any $T(\tau)$ relation by scaling the canonical $\nabla_\mathrm{rad}$ by $1+\dqdtau$.
This procedure is enabled with the new option \texttt{use\_T\_tau\_gradr\_factor}. 
For the gray, Eddington-approximation relation, $q(\tau)$ is constant
and no correction is necessary.  

To include the optically thin layers
in the interior model, the surface boundary should be set or
relaxed to a smaller $\tau$ using the flags
\texttt{set\_tau\_factor} or \texttt{relax\_tau\_factor} and their
associated controls.
The surface boundary conditions, now at smaller
$\tau$, still follow \S\ref{sec:atmosphere-T-tau}.  They are evaluated at
$\tausurf=$~\texttt{tau\_factor}$\times\tau_\mathrm{eff}$ rather than
$\tau_\mathrm{eff}$, where $\tau_\mathrm{eff}$ is the optical depth
at which the $T(\tau)$ relation is equal to $\Teff$.

This approach has several advantages.  First, like atmospheres
reconstructed with \texttt{atm\_T\_tau\_opacity = \tqs varying\tqs},
the atmospheric structure is kept consistent with \MESA's EOS and
opacity routines.
Second, the computation can leverage the parallelization of \MESA. 
Finally, this approach accounts for the fact that $g$ is not exactly constant throughout the atmosphere (\S\ref{sec:atm-recon}), which is assumed by $T(\tau)$ atmospheres. 

Figure~\ref{fig:atm-Ttau} shows the ratio of the left- and right-hand
sides of Equation~(\ref{eq:Ttau}) in a $1\Msun$ ZAMS model with the
\texttt{\tqs solar\_Hopf\tqs} $T(\tau)$ relation when the atmosphere
is either reconstructed as in \S\ref{sec:atm-recon} or included as
part of the interior. Both models deviate at
$\tau\gtrsim2$ because convection starts to transport heat.

The reconstructed atmosphere deviates in $\tau_\mathrm{eff}<\tau\lesssim2$
because $\dqdtau \neq 0$ just below the photosphere,
which is treated as part of the interior without correcting
$\nabla_{\rm{rad}}$.  When using analytic $T(\tau)$
relations, this can be corrected by using the \texttt{use\_T\_tau\_gradr\_factor}
option but not when using tabulated atmospheres, for which the
correction factors $1+\dqdtau$ have not been provided.
Though inconsequential for tables at $\tausurf\gg1$,
where $\dqdtau\to0$ anyway, it introduces inaccuracy in the
temperature stratification when using tables at small $\tau$.
This inaccuracy may be acceptable, depending on the
scientific problem. 
This inaccuracy is not present if $\dqdtau=0$ in the affected regions, which includes the Eddington
$T(\tau)$ relation.

\subsection{Choosing atmosphere options}

There is limited consensus on when to use which atmosphere options, but
we nevertheless offer a few guiding remarks.

The most commonly used and current default in \MESA\ is a gray Eddington atmosphere with
the surface boundary at $\tausurf=2/3$, with $\kappa$ fixed throughout the atmosphere. 
If the precise behavior of the stellar atmosphere is
not important, this should suffice.

There is a hierarchy of accuracy at the expense of greater computational cost among the choices for \texttt{atm\_T\_tau\_opacity}. 
On the basis of the self-consistency of $\kappa$,
\texttt{\tqs varying\tqs} is more accurate than \texttt{\tqs iterated\tqs}, which is
in turn more accurate than \texttt{\tqs fixed\tqs}. 

For the calculation of pulsations that have significant amplitude near
the surface (e.g., solar p modes), it is
important to choose an option that allows the atmosphere to be
reconstructed for the equilibrium stellar model (\S\ref{sec:atm-recon}).

Tabulated atmospheres provide boundary conditions that are typically 
computed using more complete physics (e.g., non-LTE) than can be described 
by the stellar structure equations.  As shown in Figure~\ref{fig:atm-Ttau}, 
tabulated atmospheres at $\tau\lesssim2$ lead to a stellar model in which the
near-surface temperature stratification is equivalent to an Eddington atmosphere. 
The models in Figure~\ref{fig:atm-Ttau} differ by $\sim 50$~K at $\tau_{\rm{eff}}$. 
This inaccuracy might be outweighed by the benefits of a complete atmosphere model.  
The correct structure could in principle be recovered by extracting the appropriate $T(\tau)$ relations
\citep{Trampedach:2014} from the detailed atmosphere models, but these
are not generally available. 

Ideally, we would have access to grids of $T(\tau)$ relations and
corresponding bolometric corrections extracted from advanced
simulations of stellar atmospheres, with parameters that cover the HR
diagram.  Until this ideal is realized, stellar modellers must decide
which aspects of the atmospheric boundary condition are most important
for their calculations and choose appropriate options.


\section{Convection in the Outer Layers of Stars}\label{sec:mltconv}

\def\B{\rm{B}}

\subsection{Starspots}
\label{sec:starspots}

Starspots are common
for stars with $M \lesssim \Msun$.
Models of M dwarfs that include starspots and surface magnetism
have inflated radii close to those inferred by observation
(e.g., \citealt{Feiden2013, Mann2015}).

We thus implement in \mesa\
the treatment of starspots introduced in the \texttt{SPOTS} models of \citet{Somers2020} (also see \citealt{cao_2022_aa}), which are based on the Yale Rotating Stellar Evolution Code (\texttt{YREC};
\citealt{Demarque2008,YY2013}) and described in detail in \citet{Somers2015} and \citet{Somers2020}.

\subsubsection{Starspots formalism}
\citet{Somers2020} parameterize the variance of the surface flux due to magnetic pressure from starspots by modifying the atmospheric boundary condition.
\citet{Somers2015} characterized the degree of ``spottiness'' on the stellar surface by two parameters:
\begin{enumerate}
\item[$\bullet$] \texttt{SPOTF} (hereafter $f_\text{spot}$), a coverage fraction, or ``spot filling factor''

\item[$\bullet$] \texttt{SPOTX} (hereafter $x_{\text{spot}}$), representing the temperature contrast between the spotted and unspotted regions at $r=R$: $x_\text{spot}= T_\text{spot}/T_\text{photosphere}$.
\end{enumerate}
Numerically, values from $0.0$ to $1.0$ are permitted for both parameters.
 Observationally constrained coverage fractions $f_\text{spot}$ 
are described in \citet{Cao2022}, who find that a value $f_\text{spot}=0.34$ is a reasonable fit to observations of sub-solar-mass stars in the $\lambda$ Ori cluster. 

The spot-induced temperature contrast, $x_\text{spot}$, is restricted to physically meaningful values of $0.5 - 1.0$. A value of $x_\text{spot} = 1.0$ indicates that the effective temperature in the spotted region, $T_\text{spot}$, does not differ from the surrounding, unspotted effective temperature, $T_\text{amb}(r)$ (the ``ambient temperature,'' or \texttt{ATEFF} in \texttt{YREC}). At the surface, $T_\text{amb}(r=R) = T_\text{photosphere}$.
A value of $x_\text{spot}=0.5$, on the other hand, corresponds to the statement that $T_\text{spot}$ differs from $T_\text{amb}$ by the maximum degree permitted by magnetic equipartition: namely, when the magnetic pressure contribution constitutes half of the total pressure. 

The temperature contrast $x_\text{spot}$ perturbs the radiative gradient,
$\nabla_{\text{rad, spot}}$,
in the surface convection zone. This effect can be made depth-dependent via
\begin{equation}
x_\text{spot}(r) = 1 - (1 - x_{\text{spot}}) \frac{T_\text{amb}(r)}{T(r)},
\label{eq:depth_dependent_spot}
\end{equation}
where $T(r)$ is the temperature at $r$ and the quantity $T_\text{photosphere} - T_\text{spot} = T_\text{amb}(R) - T_\text{spot}$ is held constant as a function of $r$.
In Equation \eqref{eq:depth_dependent_spot}, $x_\text{spot}$ is a scalar 
parameter.

Per equations (1) through (4) in \citet{Somers2015}, the ``spotted'' luminosity is set to 
\begin{equation}
L_\text{spotted} \equiv L/\left[ f_\text{spot}\,x_\text{spot}^4 + (1-f_\text{spot}) \right].
\end{equation}
\texttt{YREC} models the suppression of convective flux at the stellar surface via a two--part lookup in its atmospheric boundary tables, invoking ``ambient'' versus ``spotted'' effective temperatures that differ by a factor of
\begin{equation}
\alpha_\text{spot} = 1 + f_\text{spot}(x_\text{spot}^4 - 1),
\label{eq:alpha_spot}
\end{equation}
where $x_\text{spot}$ can optionally be a function of the local temperature at a given depth within the convective envelope. 

\subsubsection{\MESA\ implementation}
\MESA's implementation is equivalent, but modifies the surface boundary conditions in terms of pressure rather than \Teff, as \Teff\ is strictly an output in \MESA. Through the specification of the magnetic pressure term, $P_{\B}$, the temperature contrast is given by
\begin{equation}
x_\text{spot}(r) = \frac{P(r) - P_{\B}}{P(r)},
\label{eq:xspot_pressure}
\end{equation}
where $P_{\B}$ is assigned at the start of the step via
\begin{equation}
    P_{\B} = (k_B N_A \rho / \mu) (1 - x_\text{spot}(R))\Teff.
\label{eq:PB_defn}
\end{equation}
Choosing $x_\text{spot} = 1$ corresponds to $P_{\B}=0$, hence there is no magnetic pressure and no perturbation to 
$P$. Choosing $x_\text{spot} = 0$ yields the other extreme: $P_{\B} = P$. Once again, a practical limit on $P_{\B}$ is set by the assignment $x_\text{spot} = 0.5$, which corresponds to magnetic equipartition.

Using \texttt{auto\_diff} (\S\ref{sec:auto_diff}) we account for the modified pressure term and its partial derivatives at every mass shell. In Equations \eqref{eq:xspot_pressure} and \eqref{eq:PB_defn}, $P$ and $P_{\B}$ are \texttt{auto\_diff} variables.
Likewise, the radiative temperature gradient due to the presence of spots (i.e., magnetic inhibition of the convective flux),
\begin{equation}
\nabla_{\text{rad, spot}} =  \nabla_\text{rad}/( f_\text{spot}\, x_\text{spot}(r)^4 + 1 - f_\text{spot}),
\label{eq:perturbed_Trad}
\end{equation}
is an \texttt{auto\_diff} variable, since it is a function of the \texttt{auto\_diff} quantities $\nabla_\text{rad}$, $P$, $x_\text{spot}(r)$, and scalars.

The quantity $P_{\B}$ is assigned once at the beginning of every evolutionary step and held constant over all Newton iterations within that step.
The use of \texttt{auto\_diff} data types ensures that the Newton solver automatically receives correct partial derivatives of the modified radiative temperature gradient with respect to, e.g., depth and other stellar structure variables.
The modification to $\gradr$ is evaluated once per Newton iteration. To obtain the spotted luminosity, $L$
is adjusted by a factor of $\alpha_\text{spot}$ (Equation~\ref{eq:alpha_spot}).

\begin{figure}
\begin{center}
\includegraphics[width=\linewidth]{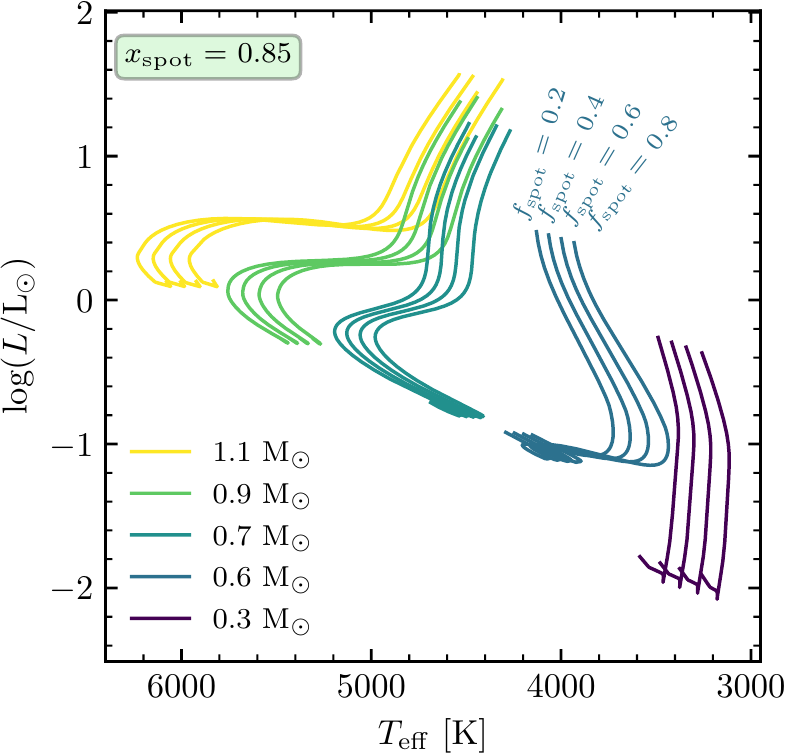}

\caption{Evolutionary tracks showing the effects of star spots for $x_\text{spot} = 0.85$.
Color indicates mass. Within clusters of identical mass,  $f_\text{spot}$ values are ordered lowest (left-most) to highest (right-most).
}
\label{fig:starspots}
\end{center}
\end{figure}

\subsubsection{\MESA\ Models}
We demonstrate the \texttt{starspots} functionality by computing a grid of spotted evolutionary tracks for $M = 0.2 - 1.3\, \Msun$ and $\rm{Z} = 0.014$. We use the \texttt{photosphere} table option for atmospheric boundary conditions across all tracks, though for the lowest--mass stars (e.g., \ $M \lesssim 0.5\, \Msun$), a choice of \texttt{tau\_1m1} would be more appropriate. We use the \texttt{Henyey} MLT prescription with $\alpha_{\text{MLT}} = 1.95$.

Figure~\ref{fig:starspots} shows tracks using $f_\text{spot} = \{0.2, 0.4, 0.6, 0.8\}$ and 
$x_\text{spot} = 0.85$. 
When comparing \MESA\ models to the \texttt{YREC}-based \texttt{SPOTS} models of \citet{Somers2020}, we observe the following features: 
\begin{itemize}

    \item Tracks computed with \MESA\ \texttt{starspots} shift smoothly by the same magnitude and in the same direction 
    as a function of $f_\text{spot}$. 
    
    \item We also find that the lower the initial value of $x_\text{spot}$ (i.e., more extreme in terms of magnetism), the greater the impact of an increased coverage fraction.

    \item Below $\approx0.6\,\rm{M}_{\odot}$, the \texttt{SPOTS} models are cooler than the \MESA\ \texttt{starspots} models. This is due to differences in our choice of the atmospheric boundary condition.

    \item The degree of radial inflation for an $0.3\,\rm{M}_{\odot}$ M dwarf  predicted by \MESA\
    \texttt{starspots}, using $f_\text{spot} = 0.34$ and $x_\text{spot} = 0.85$, is of the order 3\%-5\%, in agreement with the   \texttt{SPOTS} models.

\end{itemize}

The \MESA\ \texttt{starspots} test case can be found in \verb|MESA_DIR/star/test_suite/starspots|.

\subsection{Superadiabatic Convection in Radiation-dominated Regions}
\label{sec:superad}

Modeling stars near the Eddington limit is a complex numerical problem. Under
such conditions, convective regions with density inversions are expected
\citep{Joss+1973,paxton_2013_aa} and 1D models using standard
MLT can develop extended low-density envelopes, becoming red supergiants
before finishing their MS evolution (e.g.,
\citealt{Sanyal+2015,Szecsi+2015}). Three-dimensional radiative
hydrodynamic simulations are just starting to explore the physics of energy
transport near the Eddington limit \revision{\citep{Jiang+2015, TsangMilosavljevic2015, Schultz+2020, Goldberg2022, Moens+2022}}, and will hopefully provide
a way to accurately model these regions in 1D stellar evolution
instruments. In the meantime, 1D simulations using MLT near the Eddington limit are expensive, requiring small timesteps down to
the point that some calculations become impractical. Enhanced convective energy
transport in these regions can inhibit the formation of density inversions and
facilitate calculations. One such approach is the use of a density scale height
rather than a pressure scale height in MLT
\citep{NishidaSchindler1967,StothersChin1973,Maeder1987}.

The MLT++ formalism is a stellar-engineering approach that has been commonly used in \MESA\ to reduce superadiabaticity in regions nearing the Eddington limit (\mesathree{}). \revision{Although convection is expected to operate in regions of the star approaching the Eddington limit, the efficiency of convective energy transport is uncertain. By reducing the expected superadiabaticity, MLT++ provides an ad-hoc enhancement. Such an enhancement is supported by results from 3D simulations \citep{Jiang+2015,Schultz+2020}, but the method is not calibrated to detailed simulations or observations. The main motivation for MLT++ is to enable computations of massive star evolution up to late stages, and users need to assess if the deviations from a more physical model such as MLT are relevant to their results.
One important limitation of MLT++ is that it is a non-local explicit method}, which can lead to large step-to-step variations that produce unphysical results and prevent the
solver from finding a valid solution.
By making use of \texttt{auto\_diff} (\S\ref{sec:auto_diff}), we have implemented a fully implicit and local alternative to
MLT++, which allows the modelling of a larger range of masses and
metallicities. Setting the option
\texttt{use\_superad\_reduction} activates this method.

In hydrostatic equilibrium, the ratio between the radiative luminosity and the local Eddington luminosity is
\begin{align}
    \Gamma_\mathrm{Edd}\equiv\frac{L_\mathrm{rad}}{L_\mathrm{edd}}=\frac{4aT^4}{3P}\nabla.
\end{align}
For a particular model of energy transfer (e.g., TDC as discussed in \S\ref{sec:tdc}, or MLT),
the expected Eddington factor is
\begin{align}
  \Gamma_\mathrm{Edd,exp}\equiv\frac{4aT^4}{3P}\nabla_\mathrm{exp},
\end{align}
where $\nabla_\mathrm{exp}$ is the temperature gradient predicted by the energy transfer model. As in MLT++,
we artificially enhance energy transport in convective regions where the expected Eddington factor is high by adjusting $\nabla_\mathrm{rad}$. The difference between the
radiative and the Ledoux gradient is reduced to
\begin{align}
    \nabla_\mathrm{rad,new}-\nabla_\mathrm{L}=\frac{\nabla_\mathrm{rad}-\nabla_\mathrm{L}}{f_{\Gamma}},
\label{eq:gradr_new}
\end{align}
where $\nabla_\mathrm{rad,new}$ is the adjusted radiative temperature gradient
and $f_\Gamma\geq 1$ determines the reduction of
${\nabla_\mathrm{rad}-\nabla_\mathrm{L}}$ in the convective region. Such a
scaling of $\nabla_\mathrm{rad}$
can be interpreted as an effective lowering of $\kappa$ in regions near the
Eddington limit. Results from \cite{Schultz+2020} suggest that the impact on
radiative transfer of a vigorously convecting region supports this choice. The adjusted $\nabla_\mathrm{rad,new}$
is then used instead of $\nabla_\mathrm{rad}$ to recompute $\nabla$ according to the convection model.

The functional form of $f_\Gamma$ is arbitrary, and was determined empirically so that Equation~\eqref{eq:gradr_new} can
be applied in a large number of cases while minimizing adjustments in the limit
$\Gamma_{\rm Edd, exp} \rightarrow 0$. Just as with MLT++, it serves as a stellar-engineering method
to circumvent complex evolutionary stages, rather than a specific physical model
that accounts for how convection is modified near the Eddington limit. While exploring different options for $f_\Gamma$,
\texttt{auto\_diff} played a critical role by removing the need to directly specify partial derivatives.

\begin{figure}
  \centering
  \includegraphics[width=\columnwidth]{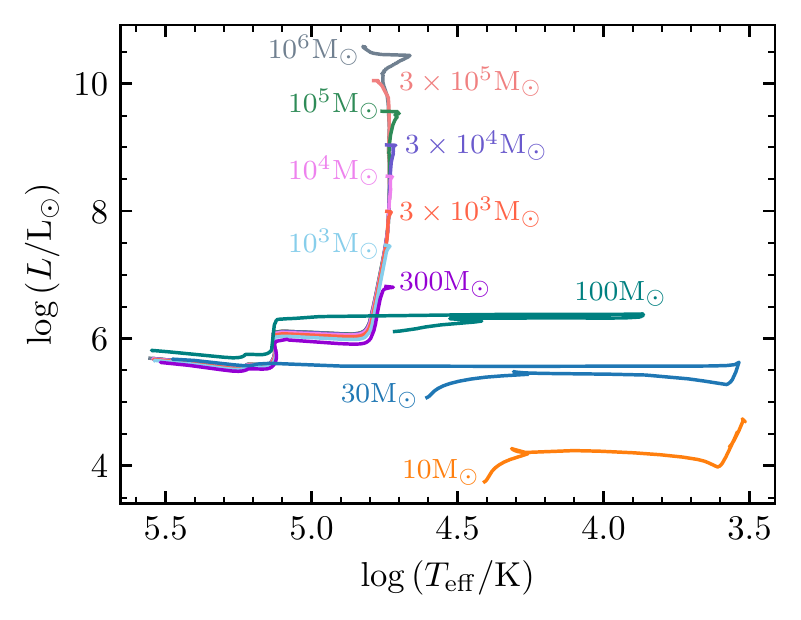}
  \caption{Evolution of stars ranging from 10 to 10$^6$\,\Msun\ at solar
    metallicity, using the
    implicit method to increase the efficiency of energy transport in regions
    approaching the Eddington limit.
    }
  \label{fig:hr_vms}
\end{figure}

Our choice for $f_\Gamma$ is the following:
\begin{eqnarray}
\begin{aligned}
f_\Gamma&=1\\&+\frac{\alpha_1 g(\Gamma_\mathrm{Edd,exp}/\Gamma_\mathrm{c}-1)+\alpha_2 g(\Gamma_\mathrm{exp}/\Gamma_\mathrm{inv}-1)}{\sqrt{\beta}}\\
&\times h((\nabla_\mathrm{exp}-\nabla_\mathrm{L})/\delta_{\rm{c}}),
\end{aligned}
\end{eqnarray}
where $\beta$ is the ratio of gas to total pressure and $\Gamma_\mathrm{inv}\equiv 4(1-\beta)/(4-3\beta)$ is the Eddington factor at which
an ideal gas with radiation develops a density inversion
(\citealt{Joss+1973}, \mesathree). The parameters $\Gamma_\mathrm{c}$, $\alpha_1$, $\alpha_2$ and $\delta_{\rm{c}}$ regulate
the enhancement of energy transport. The function
\begin{eqnarray}
  g(x)\equiv
  \begin{cases}
    0 & x<0 \\
    x^2/2 & 0<x<1 \\
    x-1/2 & x>1
  \end{cases}
\end{eqnarray}
is continuous with a continuous first derivative. No correction will be applied if
$\Gamma_\mathrm{exp}<\Gamma_\mathrm{c}$ and
$\Gamma_\mathrm{exp}<\Gamma_\mathrm{inv}$. If either of those
thresholds is exceeded, $\alpha_1$ and $\alpha_2$ set the strength of the
enhancement in energy transport for each. The $1/\sqrt{\beta}$ term further
enhances the effect in regions dominated by radiation pressure. The function $h(x)$ is \revision{chosen as
\begin{eqnarray}
  h(x)=\begin{cases}
    0 & x\le0 \\
    6x^5-15x^4+10x^3 &  0<x\le1 \\
    1 & x>1
  \end{cases},
\end{eqnarray}
such} that it is equal to zero for $x<0$ and equal to one for $x>1$, while monotonically increasing in between with a zero derivative
at $x=0$ and $x=1$. This choice ensures significant corrections are
only applied in cases where a superadiabaticity comparable to
$\delta_{\rm{c}}$ would be expected.

\begin{figure}
  \centering
  \includegraphics[width=\columnwidth]{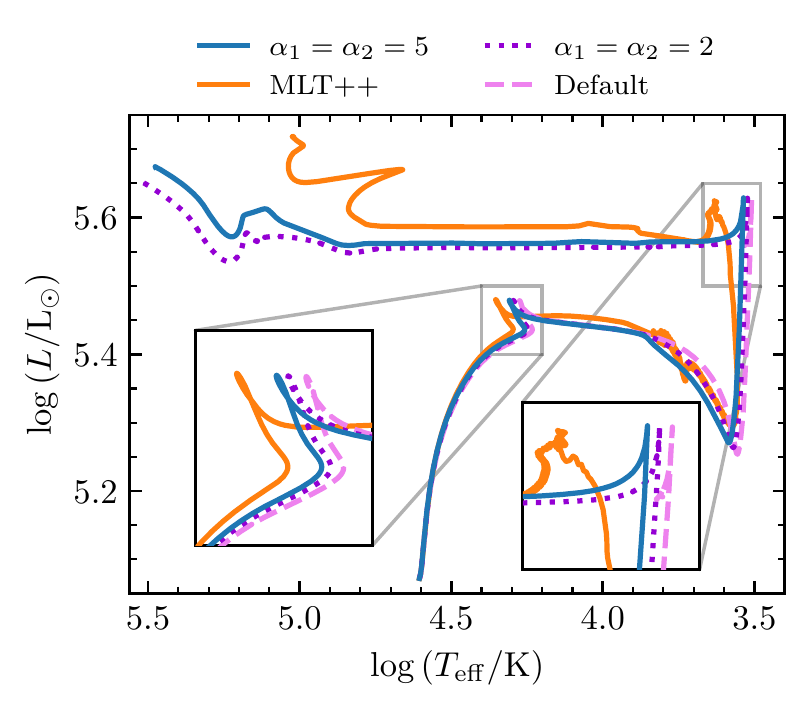}
  \caption{Evolution of a 30\,\Msun\ star at solar metallicity. The different tracks use
    the implicit method to enhance energy transport near the Eddington limit (with
    different choices for $\alpha_1$ and $\alpha_2$), MLT++,
    and the \texttt{MESA} default which includes no energy-transport enhancement. Both
    simulations with the new method use $\Gamma_\mathrm{c}=0.5$
    and $\delta_{\rm{c}}=10^{-2}$. The \texttt{MESA} default simulation stalls when evolving towards the blue after a
    RSG phase. Inset plots are made to highlight variations between
    the runs at TAMS and at the end of the
    RSG phase.
    }
  \label{fig:hr_comp}
\end{figure}
Figure~\ref{fig:hr_vms} illustrates the evolution of stars \revision{up to} 10$^6$\,\Msun\ at high metallicities ($\rm{Z}=0.0142$) using this new approach. 
The very high mass models are not necessarily meant to
represent real stars, but serve as an extreme test of this new
approach. \revision{In particular, realistic models of supermassive stars need to also take into
account general relativistic effects \citep{Chandrasekhar1964,Fricke1973}, which are not
included in these simulations.}
The calculations all used $\Gamma_\mathrm{c}=0.5$, $\alpha_1=\alpha_2=5$ and
$\delta_{\rm{c}}=10^{-2}$, which we found to perform consistently across a broad
range of masses.

A comparison between the new method, MLT++ and a simulation without
any enhancement of energy transport is shown for a 30\,\Msun\ model in Figure~\ref{fig:hr_comp}.
Overall, the new method provides smoother evolution while remaining closer to the
result obtained without enhancing energy transport. MLT++ introduces undesirable numerical
variations that are particularly visible when the model moves from the blue to the red
at TAMS, and when it evolves from the red to the
blue after being stripped of most of its H~envelope. Lowering $\alpha_1$
or $\alpha_2$, or increasing $\Gamma_\mathrm{c}$ or $\delta_{\rm{c}}$, will produce
results closer to those without energy-transport enhancement. Figure~\ref{fig:hr_comp}
also shows two different simulations with the new method using
$\alpha_1=\alpha_2=5$ and $2$. The simulations performed with
$\alpha_1=\alpha_2=2$ take almost triple the number of steps and
computation wall time. Selection of these parameters requires balancing performance and similarity to the unenhanced behavior. 
\revision{
  Users need to carefully assess if such variations have a meaningful impact in their conclusions.
}

\begin{figure}
  \centering
  \includegraphics[width=\columnwidth]{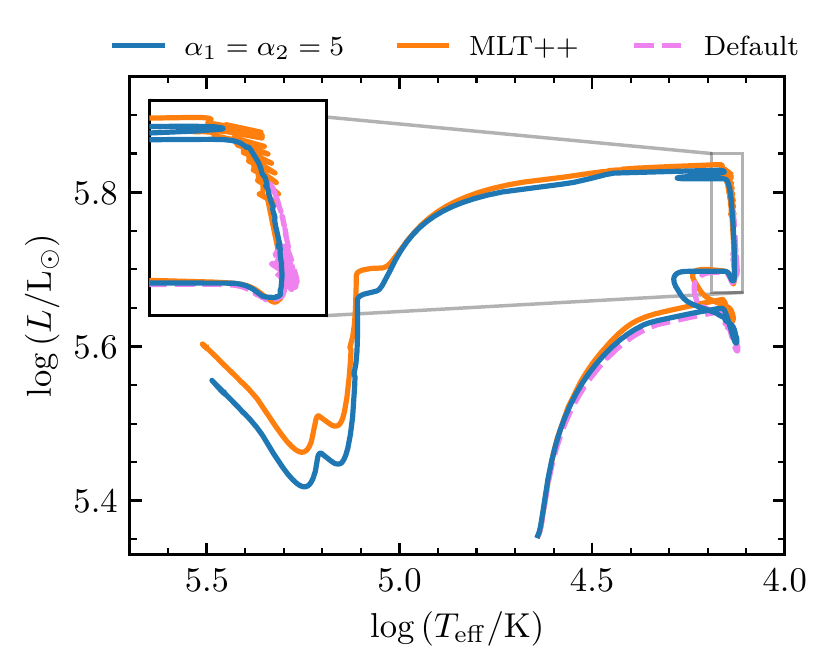}
  \caption{
    Evolution of a 40\,\Msun\ star with a 30\,\Msun\ point mass companion with an initial
    orbital period of 50 days. Models are computed for solar metallicity and using
    different methods to enhance energy transport in regions near the Eddington limit.
    }
  \label{fig:mt_history_HR}
\end{figure}

\begin{figure}
  \centering
  \includegraphics[width=\columnwidth]{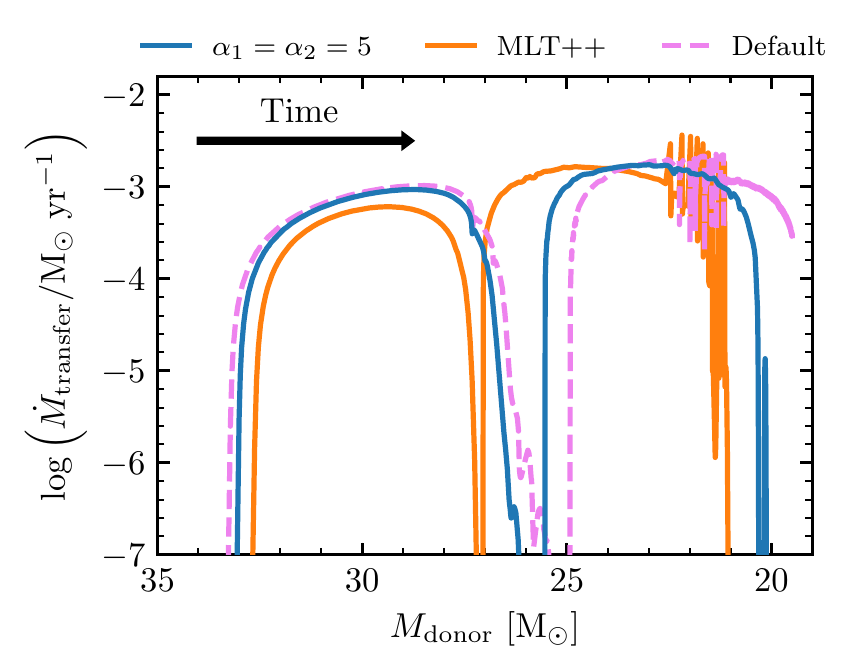}
  \caption{
    Mass transfer rate $\dot{M}_{\rm{transfer}}$ as a function of donor mass for the binary system shown
    in Figure~\ref{fig:mt_history_HR}. All simulations have two significant
    phases of mass transfer corresponding to interaction during the
    MS (Case~A) and after TAMS (Case~AB).
    }
  \label{fig:mt_history}
\end{figure}

The stability of this implicit method is particularly useful in simulations of binary systems, where small step-to-step variations in $R$ can result in large changes to $\dot M$ during Roche lobe overflow. This makes the previous MLT++ method inappropriate. 

Figure~\ref{fig:mt_history_HR} shows the evolution of a
40\,\Msun\ model with a 30\,\Msun\ point mass companion at an initial orbital period of $50$ days. 
Simulations are performed using no enhancement of energy transfer, MLT++, and the implicit method. In all cases, the models experience an initial mass transfer phase during the MS, and a second mass transfer phase right after TAMS. The simulation without enhanced energy transfer stalls during this second mass transfer phase, exemplifying the computational complexity of modelling stars near the Eddington limit. 

Figure~\ref{fig:mt_history} demonstrates that although the MLT++ model evolves beyond detachment, large step-to-step
variations lead to large changes in mass transfer rates. In contrast, the new implicit method provides a smooth solution throughout
the mass transfer phase. \revision{An early example of this implicit method enabling the succesful computation of massive binary evolution are the models of $\zeta$ Ophiuchi computed by \cite{renzo_2021_ab}.}


\section{Opacity}
\label{sec:opacity}

\subsection{Molecular Opacities}

Molecules contribute significantly to stellar opacity for
$T \lesssim 5000$K \citep{Alexander94low,Ferguson05low}. 
The \MESA{} \texttt{kap} module has been expanded to include
low-temperature molecular opacities from \citet{Lederer09low} and \AE SOPUS \citep{Marigo09}.
Both opacity sets allow for varying CNO levels. 
CNO-enhanced molecular opacities find applications in models of 
red giants, 
AGB stars, and
R Coronae Borealis stars \citep{Schwab19}.

\citet{Lederer09low} provide Rosseland mean
opacity tables for 14 metallicities ranging from Z\,=\,$10^{-5}$ to Z\,=\,4$\times$10$^{-2}$, and three
H mass fractions X = 0.5, 0.7, or 0.8. One can
specify seven Z-dependent levels of C-enhancement and three Z-dependent
levels of N-enhancement, all with \citet{Lodders03} solar abundances. 
The tables span $3.2 \leq \log(T/\rm{K}) \leq 4.05$ and 
$-7 \leq \log(\rho T_6^{-3}/ \grampercc) \leq 1$. These opacities
are enabled by setting \texttt{kappa\_low\_T\_prefix = \textsc{\char13}kapCN\textsc{\char13}}.

\AE SOPUS \citep{Marigo09} allows computation of opacity tables for a
variety of solar compositions with the optional inclusion of enhancements (and depletions) to C, N and C/O on top of the basic mixture. 
We provide a set of 
 \AE SOPUS opacity tables. Additional \AE SOPUS opacity tables can be generated through the \AE SOPUS web-interface\footnote{\url{http://stev.oapd.inaf.it/cgi-bin/aesopus}},
and \texttt{MESA\_DIR/kap/preprocessor/AESOPUS} contains information on preparing the tables for \MESA.

To compare \mesa\ and the Monash stellar evolution code \citep{Lattanzio86, Frost96, Karakas07},
custom \AE SOPUS tables were generated with $3.2 \leq \log(T/\rm{K}) \leq 4.5$ in steps of 0.01 dex and $-7 \leq \log(\rho T_6^{-3} / \grampercc) \leq 1$ in steps of 0.05 dex. We use steps of 0.5 dex at higher $T$, the \citet{Lodders03} solar composition, reference metallicities of Z\,=\,0.01 to 0.10 in steps of 0.01 dex and $0.5 \leq \rm{X} \leq 0.8$ in steps of 0.1. We set the CNO abundance variation factors \verb|fc| = $-1$, 0.2, 0.4, 0.6, 1, 1.5, \verb|fco| = $-1$, $-0.5$, 0, 0.5, 1, 1.5 and \verb|fn| = 0, 0.4, 0.7, 1. 

These tables were installed in \MESA\ and the Monash code. The physics used in the
Monash code is reconstructed as closely as possible in
\MESA\ \citep{cinquegrana_2022_aa, Cinquegrana22solarcal}. 
This includes the \texttt{basic.net} reaction network,
treatment of MLT convection (with independently calibrated $\alpha_{\rm MLT}$ parameters 
of $\alpha_{\rm MLT}$\,=\,1.931 in \MESA, $\alpha_{\rm MLT}$\,=\,1.86 in Monash), 
high temperature and molecular opacities \citep{Iglesias96, Marigo09},
an Eddington gray atmosphere, and
mass-loss approximations \citep{Reimers75, Blocker95a}. We use 
an $\eta _{\rm Reimers}=0.477$ efficiency factor, and
$\eta _{\text{Bl\"{o}cker}}$\,=\,0.01 in \MESA\ and
$\eta _{\text{Bl\"{o}cker}}\,=\,0.02$ in the Monash code \citep{cinquegrana_2022_aa}. 
Functionality that is not available in both includes the process of defining the border between convective
and radiative regions. Here, a relaxation method is used in the Monash code \citep{Lattanzio86}, and 
the predictive mixing algorithm in \mesa\ (\mesafour).

\begin{figure}[!htb]
\includegraphics[width=\columnwidth]{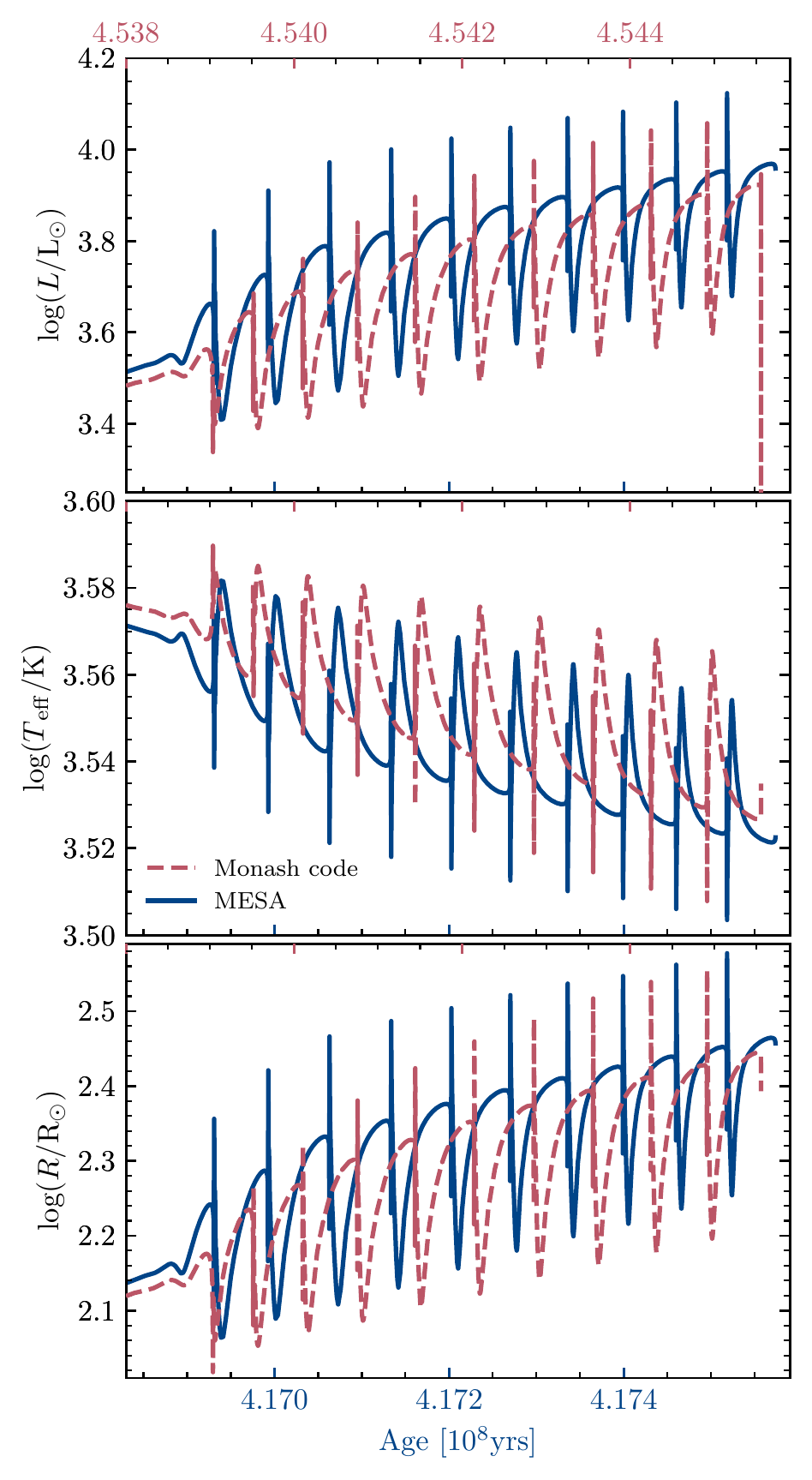}
\caption{Evolution of $L$, \revision{\Teff\ and $R$, respectively,} during the TP-AGB phase of 3\,\Msun, Z\,=\,0.014 models from \MESA\ (solid) and the Monash code (dashed).
The age of the \MESA\ (Monash) stellar model is shown on the bottom (top) x-axis.
\label{fig:LTR}}
\end{figure}

We evolve a $3\,\Msun$, Z\,=\,0.014 model with \MESA\ and the Monash code.
Both models use the same initial conditions and are terminated at the 11th thermal pulse.
Figure \ref{fig:LTR} compares the $L$, \revision{\Teff, and $R$} histories of the two models during the TP-AGB phase. 
The \MESA\ model reaches higher peak $L$, smaller peak \Teff, and larger peak $R$ than the Monash model.
The differences decrease as the evolution proceeds. Both have very similar interpulse periods, $\simeq$\,7.2$\times$10$^4$ yr, after the first few pulses.

\subsection{Compton Opacities}

At temperatures $\log(T/\rm K) \gtrsim 8.7$, beyond the boundary of
the standard opacity tables, the radiative opacity is set by Compton
scattering (\mesaone, \S4.3).  Such conditions are often
realized in simulations of massive stars and thermonuclear bursts
on neutron stars.
\MESA{} now calculates the Compton scattering opacity using the
prescription of \citet{Poutanen2017}, which improves upon the previous
approach from \citet{Buchler1976}.  See \S4 in
\citet{Poutanen2017} for a detailed comparison of the two approaches.

\subsection{Conductive Opacities}

Energy transport via electron conduction plays an important role,
especially in degenerate stellar interiors.  The conductive opacities
in \MESA\ are expanded versions of the tables from
\citet{Cassisi2007}, see \mesatwo, Appendix A.3.
\citet{Blouin2020_kap} improved the conductive
opacities for H and He in the regime of moderate Coulomb coupling
and moderate degeneracy, primarily relevant for the envelopes of DA
and DB WDs.
We implement their analytical fits for conductive opacities for H and He in \MESA.

\begin{figure}
   \centering
   \includegraphics[width=\columnwidth]{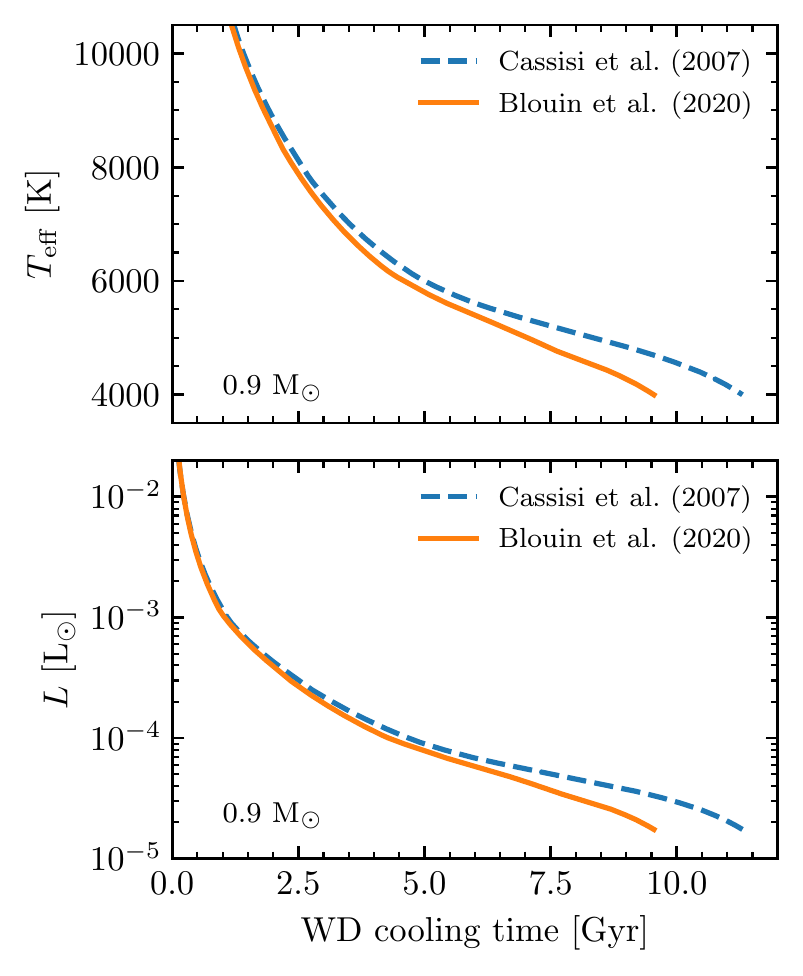}
   \caption{Cooling tracks for $0.9\,\Msun$ DA WD models using
     the conductive opacities of \cite{Cassisi2007} and
     \cite{Blouin2020_kap}. \label{fig:wdcooling_kap}}
\end{figure}

Figure~\ref{fig:wdcooling_kap} shows the effect of these new opacities
on \mesa\ WD cooling calculations for a $0.9\,\Msun$ WD with a
H atmosphere, confirming the result of
\citet{Blouin2020_kap} that the cooling time to reach
$\Teff = 4000\,\K$ can be reduced by $\approx 2\,\rm Gyr$.
The difference in cooling times is somewhat smaller for WDs with
He atmospheres, as also confirmed in \mesa\ models for AM CVn
accretors \citep{Wong2021}.
\cite{Salaris2022} provide a more detailed study of the impact of
these updated opacities on WD cooling timescales.
The conductive opacities including the \cite{Blouin2020_kap}
corrections are now the default in \mesa.

\subsection{Radiative Accelerations from the OP Monochromatic Opacities}
\label{sec:radlev} 

The radiative acceleration $g_{{\rm rad},i}$ of a given species $i$ 
represents the acceleration caused by the radiation field.
\mesathree\ (\S9) describes the inclusion of radiative levitation 
from \citet{Hu2011}, which includes evaluation of the
Rosseland mean opacity $\kappa_{\rm R}$ and $g_{{\rm rad},i}$
using the Opacity Project (OP) monochromatic opacity tables \citep{Seaton2005}.
The computation of both of these quantities requires wavelength integrations of monochromatic opacities according to the local mixture, introducing a significant amount of extra time. 
\mesafive\ (\S6.2) briefly describes the steps taken to reduce 
the time required to evaluate $\kappa_{\rm R}$ and $g_{{\rm rad},i}$. 
We now describe a new implementation from \cite{Mombarg2022}.

The OP tables provide monochromatic cross sections $\sigma_i$, equally spaced in
\begin{equation}
v(u) = \frac{15}{4 \pi^4} \int_0^u \frac{x^4 e^{-x}}{(1 - e^{-x})^3}{\rm d}x,
\end{equation}
with a step size $\Delta v$, where 10$^{-3}$ $\leq$ $u\,\equiv\,h\nu/\kB T$ $\leq$ 20. The OP tables contain data for H, He, C, N, O,
Ne, Na, Mg, Al, Si, S, Ar, Ca, Cr, Mn, Fe, and Ni. 
The Rosseland mean opacity for cell $k$ is then given by
\begin{equation}
    \kappa_{{\rm R},k} = \frac{1}{\mu_k} \left( \sum \limits_{n} \frac{1}{\sum \limits_{i} f_{i,k} \sigma_i(v_n)}  \Delta v \right)^{-1},
\end{equation}
where $\mu_k$ is the mean molecular weight and $f_i$ the fractional element abundance of species $i$. The monochromatic opacities are sampled at 10,000 points in the frequency  parameter $v_n$, denoted by index $n$.

In the new implementation, the OP monochromatic data are converted from ($T, n_e$) space to ($T, \rho$) space with
\begin{equation}
\begin{aligned}
    \log (\rho / {\rm g \ cm}^{-3}) & = \log (n_e / {\rm cm}^{-3})  + \log \mu \\
                                    & - \log \Xi - \log (\NA/{\rm g}) ,
\end{aligned}
\end{equation}
where $\Xi = \sum \nolimits_{i} f_i \Xi_i$ is the average number of electrons per atom (given by the OP tables) 
and $\NA$ is Avogadro's constant. 
The tables contain data for 3.5~$\le \log(T/\rm{K}) \le $~8.0 and $\log(\rho/\grampercc)$ bounds that depend on $\log(T/\rm{K})$.
An error is returned for $T$ and $\rho$ beyond these limits.

We select the point in the OP tables with the smallest value of 
\begin{equation}
         \sqrt{\frac{(\log T_i^{\rm OP} - \log T_k)^2}{0.0025} + \frac{(\log \rho_i^{\rm OP} - \log \rho_k)^2}{0.25}} ,
\end{equation}
where the different denominators reflect the different OP table spacings.

From this OP data point we select nearest neighbor points and construct a 
\revision{spline interpolant with a maximum degree of three, i.e., a bicubic interpolant}.
The interpolants  are stored such that the opacity can be re-interpolated at the next time step
as long as $\log (T/\rm{K})$ has changed by less than twice the grid spacing (0.01\,dex),
$\log (\rho/\grampercc)$ has changed less than twice its grid spacing (0.1\,dex), and 
the fractional abundances all satisfy
\begin{equation} \label{eq:threshold}
    \frac{|f_{i,t} - f_{i, t-1}|}{f_{i, t-1}} < \epsilon,
\end{equation}
where $t$ and $t-1$ indicate the current and previous time steps, respectively. 
Empirically, we find for MS models that $\epsilon$\,=\,$10^{-4}$ optimizes the computational efficiency 
without sacrificing accuracy in  $\kappa_{\rm R}$ and $g_{{\rm rad},i}$ \citep{Hui-Bon-Hoa2021}.

The radiative acceleration of a species is given by
\begin{equation}
    g_{{\rm rad},i} = \frac{\mu \kappa_{\rm R}}{\mu_i c} \mathcal{F} \gamma_i,
\end{equation}
where $\mu_i$ is the molecular weight of the species and $\mathcal{F}$ is the radiative flux.
The factor $\gamma_i$ is given by 
\begin{equation} \label{eq:gamma}
    \gamma_i = \sum \limits_{n} \frac{\sigma_i(v_n)[1 - e^{-u(v_n)}] - a_i(v_n)}{\sum_{j} f_{j} \sigma_{j}(v_n)}\Delta v,
\end{equation}
where the index $j$ runs over all species.
The correction terms $a_i(v_n)$ are provided by the OP data.  
The numerator in Equation~(\ref{eq:gamma}) is precomputed and stored.

\begin{figure}
    \centering
    \includegraphics[width=\columnwidth]{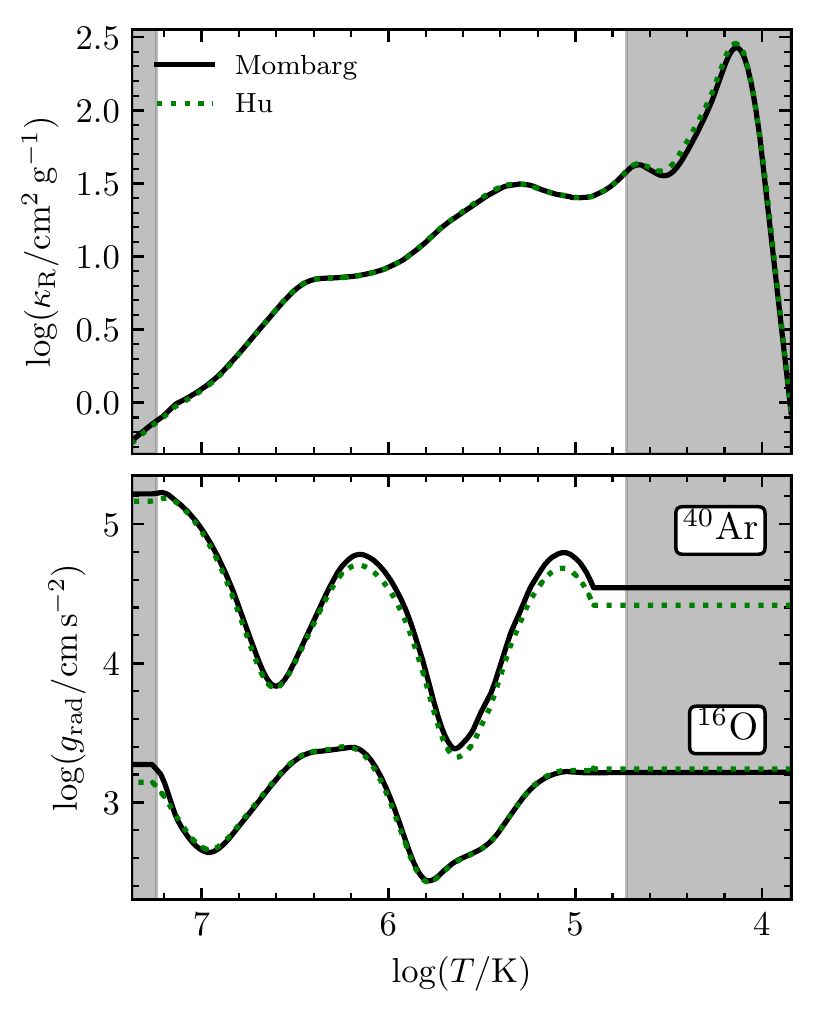}
    \caption{Rosseland mean opacity ($\kappa_{\rm R}$, upper) and the radiative accelerations ($g_{\rm rad}$, lower) 
             of $^{16}$O and $^{40}$Ar in a 2\,$\Msun$ model with an initial $\rm{Z}$\,=\,0.02 at a core H mass fraction of $\rm{X}_{\rm c}$\,=\,0.2. 
             Green dotted curves show the \mesathree\ (Hu) treatment, and black solid curves show the current treatment (Mombarg).
             Shaded areas indicate convective regions.
           }
    \label{fig:compare_Mombarg_vs_Hu}
\end{figure}

In the new implementation, optimized for models with a convective core and radiative envelope, 
the envelope is divided into two equal regions based on the number of cells  
\revision{(regardless of the outer convection cells)}. 
For each region, 
the $\gamma_i$ factors are pre-computed using an average mixture of the region
\begin{equation}
    \left<f_i\right> = \left( \sum \limits_{k = k_1}^{k_2} f_{i,k} \right) \left(\sum \limits_{i} \sum \limits_{k = k_1}^{k_2} f_{i,k}\right)^{-1},
\end{equation}
where $k_1$ and $k_2$ are the first and last cell index of a region. 
Near the boundary of these two regions, 15 cells on each side, the value of $\log g_{{\rm rad},i}$ is blended
\begin{equation}
    \log g_{{\rm rad},i} = \alpha_i \log g_{{\rm rad},i,1} + (1-\alpha_i)\log g_{{\rm rad},i,2},
\end{equation}
where $\log g_{{\rm rad},i,1}$ is computed using the average mixture 
$\left< f_{i,1} \right>$ of the first region, and $\log g_{{\rm rad},i,2}$ from the average mixture of the second region.
\revision{The parameter $\alpha_i$ varies linearly between 0 and 1 as a function of cell number in the region where blending takes place.}

Figure~\ref{fig:compare_Mombarg_vs_Hu} compares $\kappa_{\rm R}$ and $g_{{\rm rad},i}$ from \mesathree\ and the present method. 
The cell-averaged difference and maximum difference ($|\log g_{\rm rad}^{\rm Hu}- \log g_{\rm rad}^{\rm Mombarg}| / \log g_{\rm rad}^{\rm  Hu}$) for $^{16}$O are 0.007 and 0.042, respectively, and for $^{40}$Ar are 0.02 and 0.03, respectively. 
The differences in $g_{{\rm rad}}$ for $^{16}$O are typical for most isotopes.
The new implementation is about 5 times faster for MS stars.

Enabling this capability requires a data file containing the manipulated OP monochromatic data, 
doi:\dataset[https://doi.org/10.5281/zenodo.6761478]{https://doi.org/10.5281/zenodo.6858178}.


\section{Diffusion Coefficients}
\label{sec:diffusion}

\mesathree\ and \mesafour\ describe the implementation
of element diffusion in \mesa\ using the \cite{Burgers1969} equations
with diffusion coefficients based on \cite{Paquette1986}, and
updated with the coefficients of \cite{Stanton2016}.
For strong plasma coupling in the liquid regime (where the multi-component
plasma coupling parameter $\Gamma_{\rm MCP} \equiv q_e^2 \langle Z^{5/3}\rangle/a_e k_{\rm B} T \gtrsim 1$),
these coefficients disagree by a factor of a few with more accurate molecular dynamics (MD) methods
(\mesathree, \citealt{Bauer2020}, \citealt{Caplan2022}).

Until recently, it was
unclear how to generalize these results to the arbitrary plasma mixtures
needed for stellar models. However, \cite{Caplan2022}
show that in the liquid regime ($10 \lesssim \Gamma_{\rm MCP} \lesssim 200$),
diffusion coefficients are approximated to within $\sim 10$\%
accuracy by using a fit to an equivalent one-component plasma (OCP)
coefficient calculated with MD, and then scaling with charge as
\begin{equation}
  \label{eq:Dcap}
  D_j = \left( \frac{Z_j}{\langle Z \rangle} \right)^{-0.6} D_{\rm OCP}~,
\end{equation}
where $D_j$ is the diffusion coefficient of species $j$, $Z_j$ is the
charge of species $j$, $\langle Z \rangle \equiv \sum_j n_j Z_j/n_{\rm ions}$ is the average charge of
ions in the liquid, and $D_{\rm OCP}$ is the diffusion coefficient for
an equivalent OCP with coupling and screening set by the mixture averages.
This scaling is justified by the fact that in the liquid regime, ion
diffusion can be described in terms of spheres experiencing
Stokes-Einstein drift through a viscous liquid
\citep{Bildsten2001,Daligault2006}, with the effective radii of
different ions set by their charge relative to the background plasma
\citep{Bauer2020,Caplan2022}. The $D_{\rm OCP}$ term therefore
captures the physics of the global viscosity experienced by all ions,
while the scaling with charge captures the different effective radii
of ions experiencing Coulomb interactions with the background plasma.

Since the diffusion solver in \mesa\ is cast in terms of the
\cite{Burgers1969} equations, coefficients must be recast
in terms of the binary resistance coefficients $K_{ij}$ between
species $i$ and $j$ rather than the net diffusion coefficients $D_j$.
At strong plasma coupling ($\Gamma_{\rm MCP} \gtrsim 10$), these coefficients must satisfy the
relation%
\footnote{As noted by \cite{Caplan2022}, this expression for $D_j$ in
  terms of $K_{ij}$ is accurate for strong plasma coupling, but
  neglects a higher-order correction for thermal diffusion that can
  reach up to about 20\% for $\Gamma_{\rm MCP} \ll 10$ \citep{Baalrud2014}.}
\begin{equation}
  \label{eq:Dj}
  D_j = \frac{n_j k_{\rm B}T}{\sum_i K_{ij}}~.
\end{equation}
The resistance coefficients $K_{ij}$ must also be symmetric
($K_{ij} = K_{ji}$), motivating
\begin{equation}
  \label{eq:Kij}
  K_{ij} = \frac{n_i n_j (Z_i Z_j)^{0.6} k_{\rm B} T}{n_{\rm ions}
    \langle Z^{0.6} \rangle \langle Z \rangle^{0.6} D_{\rm OCP}}~,
\end{equation}
where $\langle Z^{0.6} \rangle \equiv \sum_j n_j Z_j^{0.6}/n_{\rm ions}$.
The resistance coefficients of
Equation~\eqref{eq:Kij} reduce to the desired net diffusion
coefficients of Equation~\eqref{eq:Dcap} under the summation required
in Equation~\eqref{eq:Dj}.

We calculate the OCP diffusion coefficient $D_{\rm OCP}$ using the
\cite{Caplan2022} fit to high-resolution MD, which is given in terms
of $\Gamma_{\rm MCP}$ and electron screening length
$\lambda_e$. To calculate the value of $\lambda_e$ as input for the
diffusion coefficients, we follow the method described in \mesafour\ based on
\cite{Stanton2016} for non-relativistic electrons at $\rho < 10^6\,
\rm g \, cm^{-3}$. Electrons become relativistic at higher densities,
and we therefore switch to the relativistic screening length
$\lambda_e = \left(2 k_{\rm F} \sqrt{\alpha/\pi}\right)^{-1}$
for densities $\rho > 10^6\, \rm g \, cm^{-3}$, where $\alpha$ is the fine-structure
constant and $k_{\rm F} = (3 \pi^2 n_e)^{1/3}$.

\begin{figure}
\centering
\includegraphics[width=\apjcolwidth]{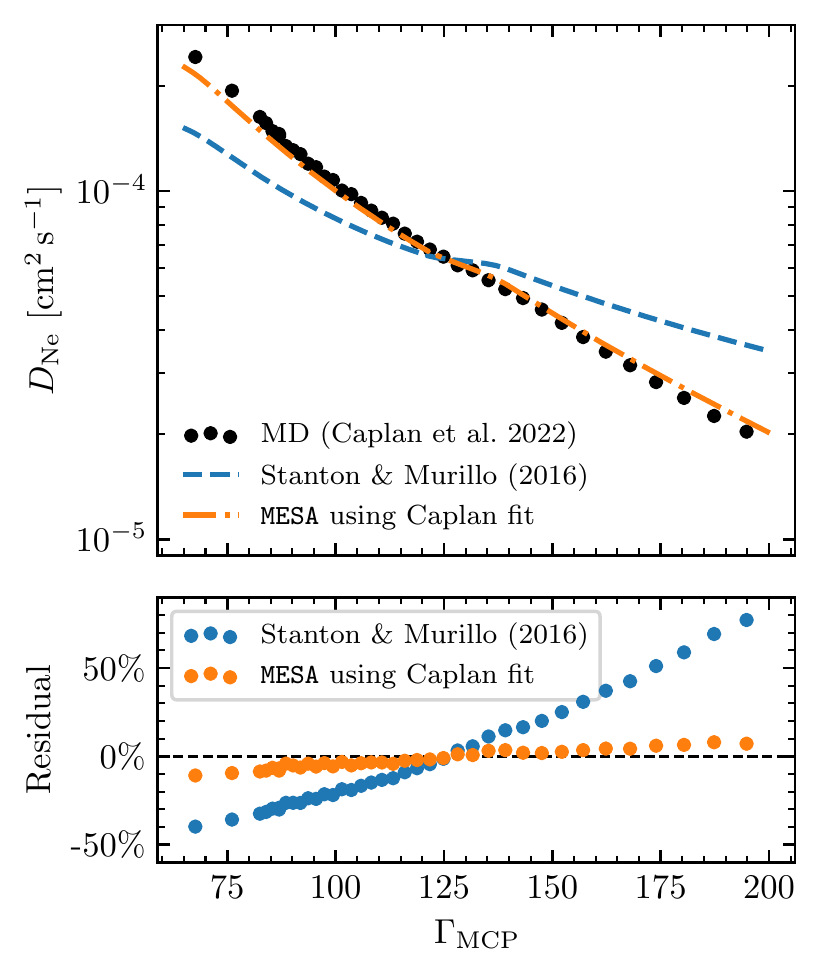}
\caption{Diffusion coefficient of $^{22}$Ne in the liquid interior profile of
  a cooling C/O WD model. The residuals in the lower panel
  are relative to the \cite{Caplan2022} MD for a C/O/Ne plasma
    mixture shown in the top panel.}
\label{fig:Dcoeff}
\end{figure}

Figure~\ref{fig:Dcoeff} shows the $^{22}$Ne diffusion
coefficient in a \mesa\ model of a $0.6\,\Msun$ C/O WD, which
has
historically been a key source of uncertainty in the rate at which
$^{22}$Ne settles toward the center of the WD
\citep{Bildsten2001,Deloye2002,Garcia-Berro2008}.
Such C/O mixtures with trace $^{22}$Ne are in the liquid phase for $\Gamma_{\rm MCP}
\lesssim 200$ \citep{Caplan2020,BlouinPRE2021}, though distillation of $^{22}$Ne
may also occur for sufficient $^{22}$Ne concentration \citep{Blouin2021}.

Figure~\ref{fig:Dcoeff} also shows the MD coefficients from \cite{Caplan2022}
(these closely match the MD coefficients of \citealt{Hughto2010})
along with diffusion coefficients obtained from
Equations~\eqref{eq:Dj} and~\eqref{eq:Kij} for \mesa\ profiles of our
C/O WD model with a similar interior composition.
The coefficients based on \cite{Stanton2016} vary
from the MD results by a factor of two or more, while the residuals
for our implementation of the \cite{Caplan2022} coefficients are 10\%
or less. This represents an order of magnitude improvement on the
uncertainties for diffusion coefficients in the strongly coupled
regime relative to the previous \mesa\ implementation, while also providing
a procedure that is generalizable to compositions other than just
C/O/Ne mixtures (Equation~\ref{eq:Kij}).

We have implemented these \cite{Caplan2022} diffusion coefficients as
the default coefficients in \mesa\ for $\Gamma_{\rm MCP} > 10$, with a smooth
transition from the \cite{Stanton2016} coefficients over the range
$3 < \Gamma_{\rm MCP} <10$.
The diffusion coefficients in the
liquid WD regime are now accurate to $\approx 10\%$
for $10 \lesssim \Gamma_{\rm MCP} \lesssim 200$, a substantial improvement
compared to our previous methods.

When crystallization occurs for $\Gamma_{\rm MCP} \gtrsim 200$, we assume that
freezing into the solid phase causes the diffusion coefficient to go
to zero. We implement this with a smooth turnoff near the
crystallization boundary using the smoothed phase parameter $\phi$
from Skye (\S\ref{sec:Skye}), so that the
diffusion velocities are zero for $\phi \geq 0.5$.


\section{Nuclear physics}
\label{sec:rates}

\MESA\ models calculate the energy generation rates and composition changes due to nuclear burning
over a large range of $T$, $\rho$, and $X_i$. 
The nuclear evolution of the chemical composition dominates the total cost of a model 
(memory + CPU) when the number of isotopes is $\gtrsim$\,30.
Here we report progress on nuclear reaction rates, a new operator split burning option,
and enhanced reaction rate outputs.

\subsection{Reaction rates updates}

\MESA's default reaction rate for $\beryllium[7](e^{-},\nu_e)\lithium[7]$
came from \REACLIB{} \citep{cyburt10}. However, \REACLIB{} is only defined for $T>10^{7}$\,K, and assumes that all atoms are ionized. While this is a reasonable assumption for reactions that occur deep in the stellar interior, significant Li production can occur in stellar envelopes with  $T<10^{7}$\,K where the reaction rate then depends on the ionization balance \citep{schwab2020}.
The new default (as of r22.05.1) for this rate is \citet{simonucci2013} which incorporates ionization contributions.

\REACLIB{} defines a reverse reaction as the endothermic direction.
However, this direction depends on the nuclear masses assumed during the evaluation.
For consistency with the nuclear masses used in \MESA, we define the reverse rate based on the nuclear masses from
\texttt{masslib\_library\_5.data}.
This affects reactions with uncertain nuclear masses; for instance, the reaction $\copper[55]\left(\gamma,p\right)\nickel[54]$
has a $Q=0.293\,\MeV$ (exothermic) from ``rpsm'' (Rauscher 1999, priv. comm) and a $Q=-0.07256\, \MeV$ (endothermic) from ``ths8''\footnote{For full details of the source of the \REACLIB{} data, see \url{https://reaclib.jinaweb.org/labels.php}} ~\citep{cyburt10}.
Incorrectly determining which reaction is endothermic leads to large errors when the reverse rate is computed from detailed balance.

\subsection{Operator split nuclear burning}

By default \MESA\ uses a nuclear reaction network that is fully
coupled to the stellar hydrodynamics (\mesaone), solving for the
changes in the composition simultaneously with the changes in stellar
structure quantities. This approach provides consistency between the nuclear
physics and stellar hydrodynamics, but can place limits on the maximum time step.
When $T \gtrsim 3\times 10^{9}$\,K, the time step is limited by the nuclear burning timescale,
and the composition enters a dynamic equilibrium state where large forward and reverse reaction rates nearly cancel
each other, potentially leading to numerical errors when subtracting large values from one another (\mesathree).
During the final stages of evolution to core-collapse, time steps of $\delta t <10^{-10}$~s
are common \citep{farmer2016}. This can make evolution to core collapse a resource-consuming endeavor.

\MESA\ has the capability to perform an operator-split procedure (\texttt{op\_split\_burn}) to compute the composition change and energy generation.
Cells with high $T$ use operator splitting.
For each such cell, \MESA{} computes the change in composition over the timestep $\delta t$ with a semi-implicit midpoint rule \citep{Bader1983}.
An operator-split cell can then take multiple sub-steps, allowing the composition to evolve  at fixed $T$ and $\rho$ with an adaptive time step.
The nuclear energy generated $\epsnuc$ is calculated by taking the difference between the starting and final compositions,
and energy loss due to neutrinos from nuclear reactions $\epsnu$ is accounted for.

This scheme allows accurate tracking of the net nuclear energy generation rate and composition changes.
However, operator-split burning cannot calculate the partial derivatives of these terms with respect
to $T$ or $\rho$ for the matrix solver.
These partial derivative matrix terms are thus set to zero.
This removes the difficulty of the partial derivatives varying in sign and magnitude over short timescales,
and enables the solver to more robustly find a solution within the requested tolerances.

\begin{figure}
    \centering
    \includegraphics[width=0.48\textwidth]{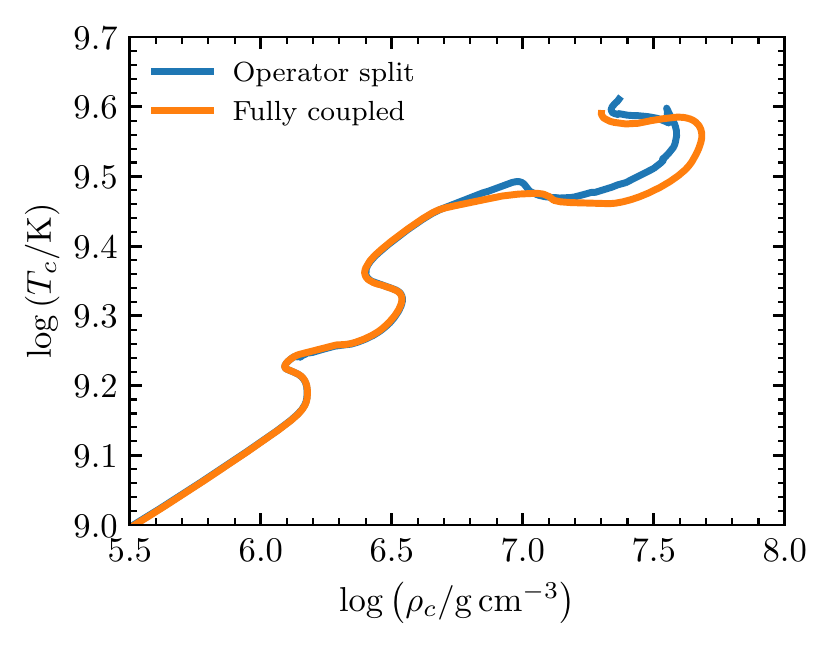}
    \caption{Central $T$-$\rho$ tracks of a 30\,\Msun{} star evolved from ZAMS to formation of the iron core.
    }
    \label{fig:rates_trho_split}
\end{figure}

Figure \ref{fig:rates_trho_split} compares the fully-coupled (unsplit) and operator split solutions in
the central $T$-$\rho$ plane for a 30\,\Msun{}, solar-metallicity model evolved from the ZAMS to the formation of the iron core.
We enable \texttt{op\_split\_burn} for any cell with a $T>10^{9}$~K.
The evolutions are nearly identical up to $T_{c}\approx 3\times10^{9}$~K, after which the \revision{fully-coupled} solution evolves to a slightly larger $\rho$.
These differences are comparable to variations arising from other physics choices, such as the nuclear reaction rate screening prescription (see Appendix A.2 of \mesafive). Evolution up to the formation of the iron core took approximately equal number of timesteps.
Evolving the models further to the onset of core-collapse with ``gold'' tolerances (\mesafive),
the operator split model only required an additional $\approx$\,200 time steps, while the fully-coupled model
failed to reach the onset of core-collapse after an additional $\approx$\,140,000 time steps.

Operator splitting can provide a significant speedup during Si and Fe burning by reducing the number of time steps needed, and may be the only way to make certain problems tractable.
It is however less efficient than a fully-coupled model during earlier stages of evolution due to the additional sub-steps taken by each operator-split cell.
Thus, we suggest considering operator splitting only for models evolving beyond core C-depletion. It is difficult to say which treatment is more accurate; fully-coupled calculations include more physics but can be subject to numerical errors, while \revision{our} operator-split calculations ignore the physics in the partial derivatives, but provide a more numerically stable solution.

\begin{figure}
    \centering
    \includegraphics[width=\linewidth]{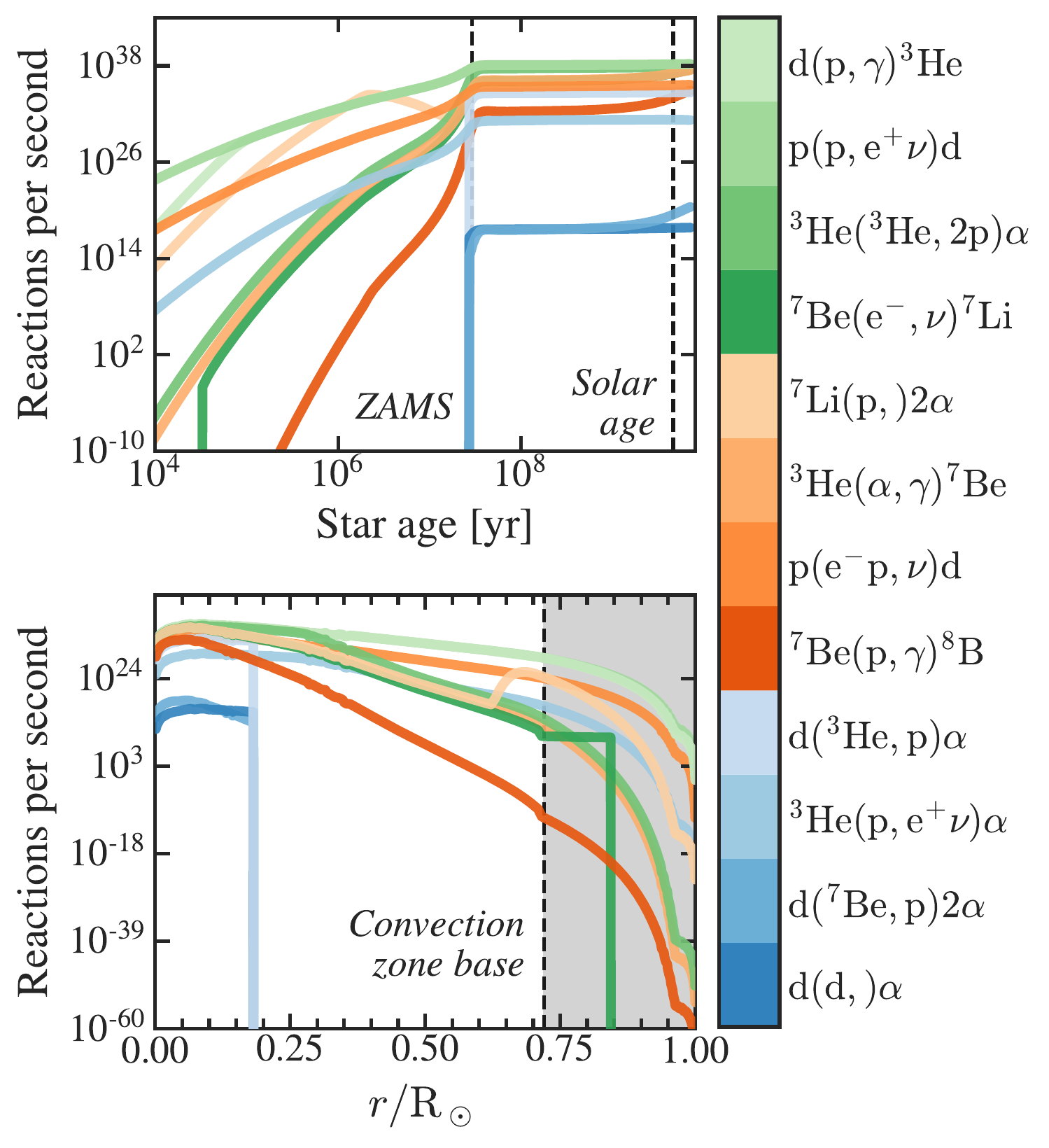}
    \caption{Nuclear reactions rates ongoing in the Sun. {\it Upper:} Evolution of reaction rates from the pre-main sequence to core H exhaustion. {\it Lower:} Profiles at the current solar age as a function of the distance from the solar center. \revision{The reactions in the colorbar are ordered from top to bottom according to their frequency.}}
    \label{fig:solar_rates}
\end{figure}

\begin{figure}
    \centering
    \includegraphics[width=\linewidth]{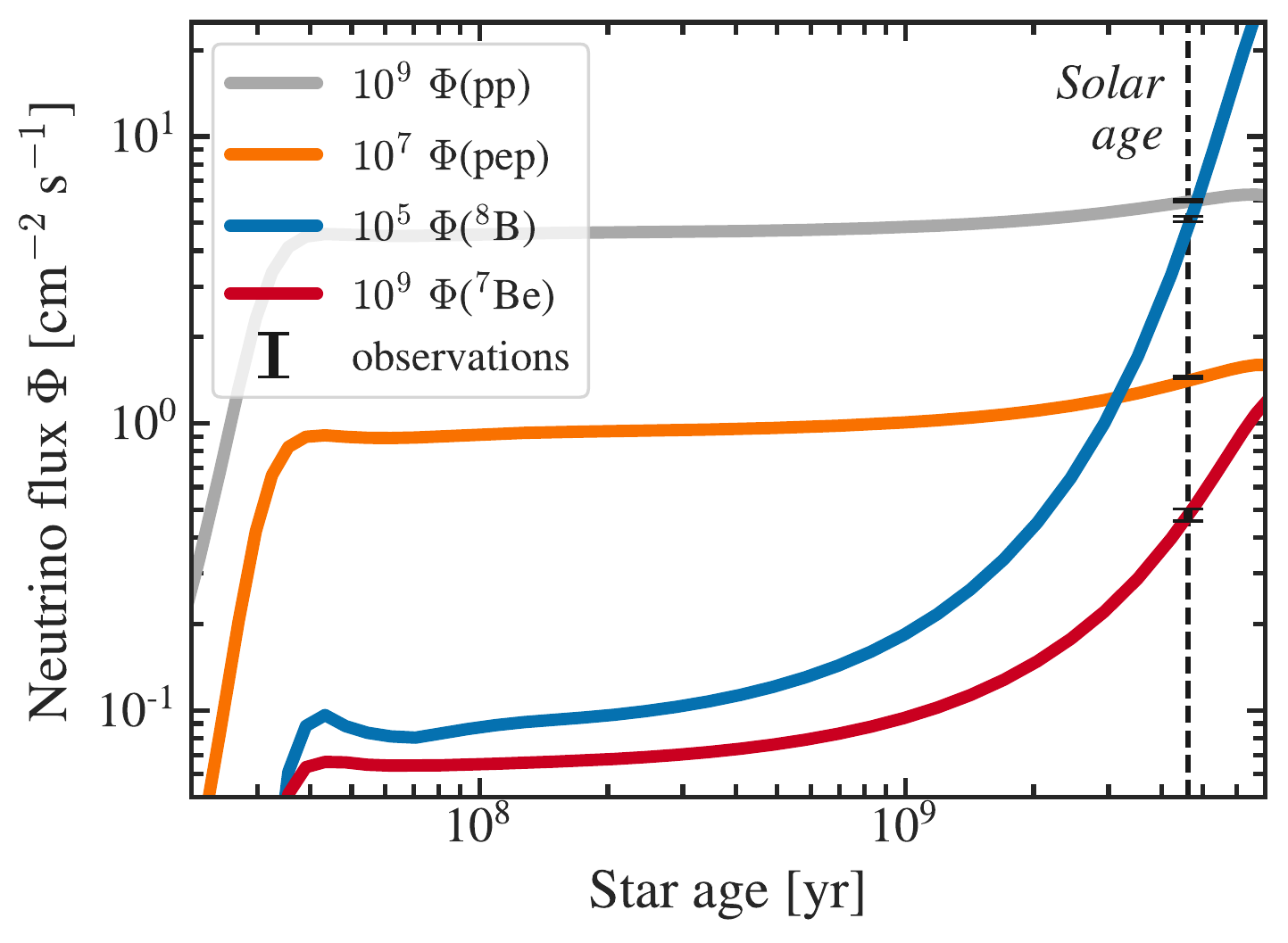}
    \caption{The evolution of solar electron neutrino flux at Earth. The predicted values for the present Sun are in satisfactory agreement with the observed values after correction for flavor effects:
$\Phi(\rm{pp})/10^9 =$~5.934~(\MESA), 5.971~$\pm$~0.035 (Borexino);
$\Phi(\rm{pep})/10^7 =$~1.408~(\MESA), 1.448~$\pm$~0.013 (Borexino);
$\Phi(^7\rm{Be})/10^9 =$~0.476~(\MESA), 0.480~$\pm$~0.023 (Borexino); and
$\Phi(^8\rm{B})/10^5 =$~4.780~(\MESA), 5.160~$\pm$~0.110 (Borexino) cm$^{-2}$~s$^{-1}$.
 }
    \label{fig:solar_neutrino_fluxes}
\end{figure}

\subsection{Reaction rate output}

\MESA\ has the new capability to output individual nuclear reaction rates and related quantities to the history or profile data files.
Using ${\tt add\_raw\_rates}$  will output the rates of all reactions.
Alternatively, ${\tt raw\_rate}$ with the name of a specific reaction will report only that reaction.
Similarly, ${\tt add\_eps\_neu\_rates}$, ${\tt add\_eps\_nuc\_rates}$, and ${\tt add\_screened\_rates}$ provide neutrino energy rates, nuclear energy rates, and screened reaction rates. 

As an example, Figure~\ref{fig:solar_rates} shows the total reaction rates of a $1\,\Msun$ model evolved to the solar age, and the profiles of the reaction rates at the solar age.
Figure~\ref{fig:solar_neutrino_fluxes} further shows the evolution of the solar electron neutrino fluxes and a comparison with the observed solar electron neutrino fluxes
\citep[see also][]{farag_2020_aa} after correction for flavor mixing \citep{2016JHEP...03..132B}. The model agreement with measurements is similar to that obtained using the ASTEC stellar evolution code \citep{bellinger2022}.


\newpage
\section{Constants}
\label{sec:const}

The \MESA\ \texttt{const} module provides mathematical, physical, and
astronomical constants relevant to stellar astrophysics (\mesaone,
Section 4.1).  The values in this module have been updated to reflect
new definitions and conventions.
Some microphysics inputs (e.g., opacity, EOS, reaction rates)
use slightly different constants in constructing their tables or expressions.
Such cases are usually beyond our control, and the updates in this section do not apply to those instances.

As part of the revision of the SI, the values of the physical
constants $e$, $h$, $\kB$, and $\NA$ are now exact \citep{CODATA2017}.
We have adopted these values and ensured that other constants composed
of these and other exact values (e.g., the Stefan-Boltzmann constant)
are defined in a consistent manner.
For other physical constants, (e.g., $G$) we use
CODATA 2018 \citep{CODATA2018}.

\MESA\ follows IAU recommendations for astronomical constants,
currently adopting nominal solar and planetary quantities
from IAU 2015 Resolution B3 \citep{IAU2015B3}.  We follow the
recommended procedure of deriving nominal solar and planetary masses
from the mass parameters $(GM)$ and the adopted value of $G$.  We also
adopt the convention that the nominal radii of planetary bodies are the
equatorial radii.


\section{Infrastructure}
\label{sec:infrastructure}

\subsection{Migration to GitHub}\label{sec:github}

\MESA\ development began in 2007 with a Subversion (SVN) repository
hosted on SourceForge, and from 2017$-$2020 this SVN repository was
hosted by Assembla.  Beginning in December 2020, development shifted
to GitHub, with the root of the new Git repository corresponding to
SVN r15140.  An archival copy of the SVN development history is available
at doi:\dataset[10.5281/zenodo.4745225]{https://doi.org/10.5281/zenodo.4745225}.
The shift to GitHub paves the way for a new era of collaborative \MESA\
development and expanded interaction with the astrophysics community.

As Git repositories can have a non-linear commit history, we will no longer denote public releases with a revision number. In SVN, this counted the number of commits from the beginning of \MESA's development. Instead, public releases will now be identified by the date, in the format YY.MM.I, where YY is the final two-digits of the year, MM is the two digit month number, and I is a version  (usually 1) to distinguish multiple releases in a given month. While we do not recommend publishing an article based on non-released versions of \MESA, if necessary we suggest using the first 7 characters of the Git commit hash as the version number.

\newpage
\subsection{MESA TestHub} \label{sec:testhub}
With the transition from SVN to Git, the versioning and branching scheme used in the development of \MESA{} changed substantially. A major overhaul of the collection and distribution of continuous integration test results on TestHub (\href{https://testhub.mesastar.org}{https://testhub.mesastar.org}) was necessary.
We now highlight these and other changes made since \mesafive{}.

With Git, understanding the relationship of a commit to other commits requires knowledge of the commit graph.  To dynamically obtain and update this information, TestHub uses GitHub webhooks and APIs. Upon pushing to the GitHub remote, GitHub now sends a request to TestHub to update its internal tree of commits. This is done by querying the GitHub API for a current list of branches and the head commits of each. TestHub removes any references to branches that are no longer in GitHub and updates the branches table in the TestHub database, adding commits as needed even if they have not had any tests conducted. This allows easy identification of commits that have not yet been tested. We also now store data about commits, such as the authors, commit messages, and commit times directly in the TestHub database for more convenient access. 

To view the detailed changes made in a given commit, each commit-specific page on TestHub links to the appropriate page on GitHub. Additionally, most views in TestHub now have a branch selector dropdown so developers can look at commits on a specific branch, and we rely on the GitHub API to order these commits within a branch. 

\begin{figure}[!thb]
	\centering
	\includegraphics[width=\columnwidth]{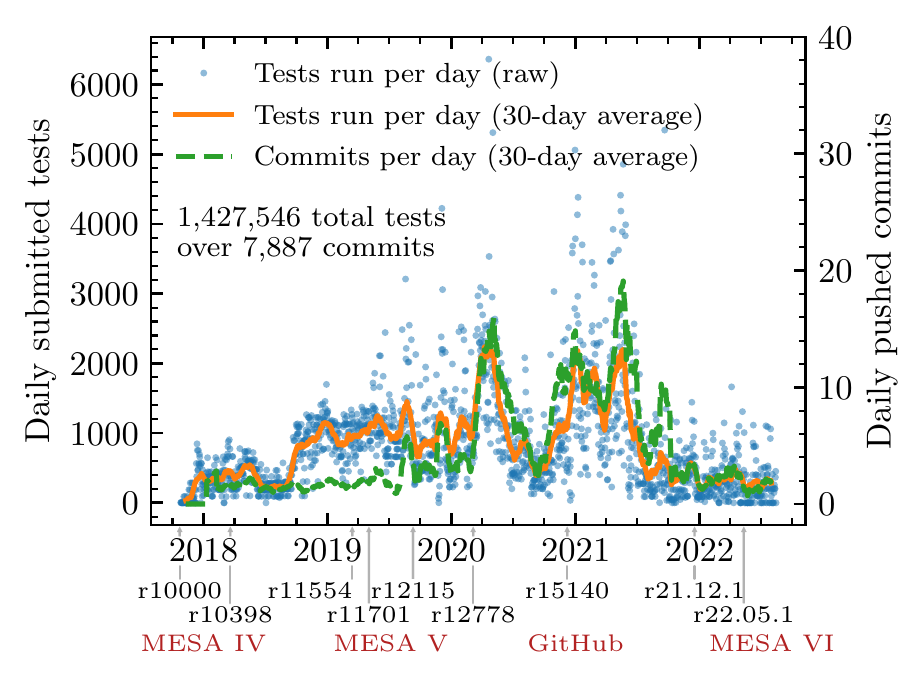}
	\caption{Daily tests submitted since the instantiation of TestHub, spanning the SVN and Git eras. To smooth over the day-to-day variability, we also show 30-day rolling averages of submitted tests and pushed commits. Below the \mbox{x-axis} are dates of public releases and other milestones such as instrument papers and the SVN-to-Git transition.}
	\label{fig:infrastructure_tests_submitted}
\end{figure}

Figure \ref{fig:infrastructure_tests_submitted} shows the total number of individual test instances (i.e., results for a single test case from a single computer) and daily commits submitted since the launch of TestHub in late 2017. Unsurprisingly, the rate of test submissions is strongly correlated with the rate of commits. Substantial deviations occur due to different testing computers coming on- and offline, and because some commits focus on documentation and do not trigger testing runs on automated testing setups. Figure~\ref{fig:infrastructure_tests_submitted} also shows the dates of public releases and key milestones. The switch from SVN to Git was shortly after release r15140, and there is a smooth transition between the eras.

\begin{figure}[!thb]
	\centering
	\includegraphics[width=\columnwidth]{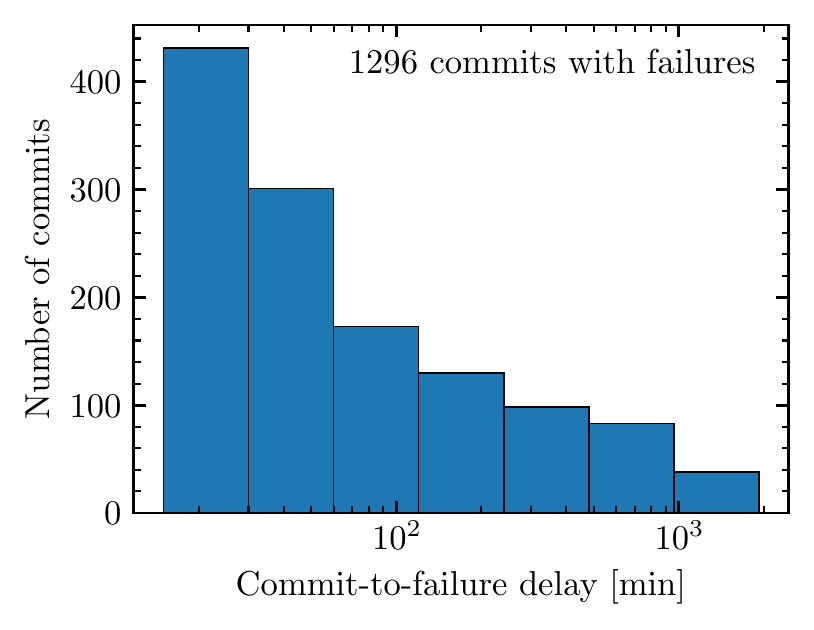}
	\caption{Distribution of delay times from when a commit is pushed to when a failure is first reported.}
	\label{fig:infrastructure_delays}
\end{figure}

The \MESA\ test suite takes $\sim$\,100 core hours to complete when run on a modern workstation. To reduce this runtime while still providing diagnostic information, many tests have optional steps that can be skipped, with the next inlist loading a saved model. The total time for a test suite run is roughly halved when these parts are skipped. Skipping optional inlists has drawbacks in requiring saved models, which can generate spurious failures when skipping inlists. At least one computer runs all inlists on all tests on the main branch. 

We now assess how quickly our distributed and automated continuous integration workflow can identify a failure. Figure~\ref{fig:infrastructure_delays} shows that most failures are detected within an hour of the commit, and more than a third are detected within thirty minutes. This prompt turnaround comes from computing clusters configured to check for new commits every few minutes, and then launch a full test suite run upon detecting a new commit. Some clusters are configured to detect and test commits on any branch of the \mesa\ GitHub repository, allowing full testing coverage during feature development and ensuring that new features pass the test suite before being merged into the main branch.

Awareness of test failure and the commit responsible is useful, but having the detailed output of a failing test is often essential in quickly identifying and rectifying the regression. TestHub now collects information on the runtime, computer architecture, broad failure type (such as a compilation error or a runtime error), and the \code{stdout} and \code{stderr} logs of the compilation and/or test case in the event of a failure. When present, these logs are accessible by links next to the failure indicator. This allows developers to quickly identify what led to the unexpected behavior without having to request more data from the owner of the computer that submitted it. These updates to TestHub improve the pace, efficiency, and quality of \MESA\ source code development.

\subsection{User Contributions}\label{sec:ucontrib}
  
\MESA\ has a Zenodo community%
\footnote{\url{https://zenodo.org/communities/mesa/}}
to encourage users to publicly archive their input and output files.
To make user-contributed routines easier to share while minimizing maintenance, we have additionally created the \code{mesa-contrib} repository%
\footnote{\url{https://github.com/MESAHub/mesa-contrib}}.

Routines in \code{mesa-contrib} can be implemented
via \MESA's hooks with a few Fortran \code{include} statements.
If the build environment defines the location of
\code{mesa-contrib}, then it is included by default as a search path when
\MESA\ is compiled.
Currently, \code{mesa-contrib} contains routines
for the atmospheric $T(\tau)$ relations and corresponding MLT parameters
implemented by \citet{Mosumgaard2018} and
for angular momentum transport by the enhanced Tayler--Spruit dynamo
described by \citet{Fuller2019}.
To ease the burden of maintenance, \code{mesa-contrib} is only intended to work with the latest public \MESA\ release.

\subsection{NuDocker}\label{sec:nudocker}

Reproducibility is a goal of open science and a tenet of scientific research.
Provenance, as the term relates to software instruments \citep{buneman_2001_aa,carata_2014_aa,stodden_2018_aa},
is the ability to record the full history of a result. Scientific
research is generally held to be of good provenance when previous
results, perhaps decades old, can be reproduced.  The aim is to
preserve the final knowledge object and the capability to perform the
scientific actions that are the foundation of the knowledge object.
Thus, we seek to preserve not only the numerical calculation, but
accelerate future research by archiving the computational environment.

Provenance enables reproducing past simulations and performing new calculations with different physics or numerical options. 
A challenge is that compilers change, linked libraries evolve, and operating systems progress.
Many science results have been obtained with older versions of \MESA, which in most cases are not obsolete and therefore remain valuable.
\code{NuDocker}\footnote{\url{https://github.com/NuGrid/NuDocker}}$^,$\footnote{\url{https://doi.org/10.5281/zenodo.3678601}}
provides a solution to the provenance challenge by being able to
run older versions of \MESA\ with age-appropriate compilers, libraries, and operating system using
light-weight, OS-level virtualization (e.g., \code{Docker}\footnote{\url{https://www.docker.com}}).
\code{NuDocker} provides four \code{Docker} images that can be launched with one terminal command,
and has been tested and used in 14 out-of-the-box \MESA\ versions from r4942 to r22.05.1.
A hallmark of \code{Docker} virtualization is the minimal performance penalty
compared to running natively \citep{felter_2014_aa,felter_2015_aa}.

\begin{figure}[!htb]
\centering
\includegraphics[width=\columnwidth]{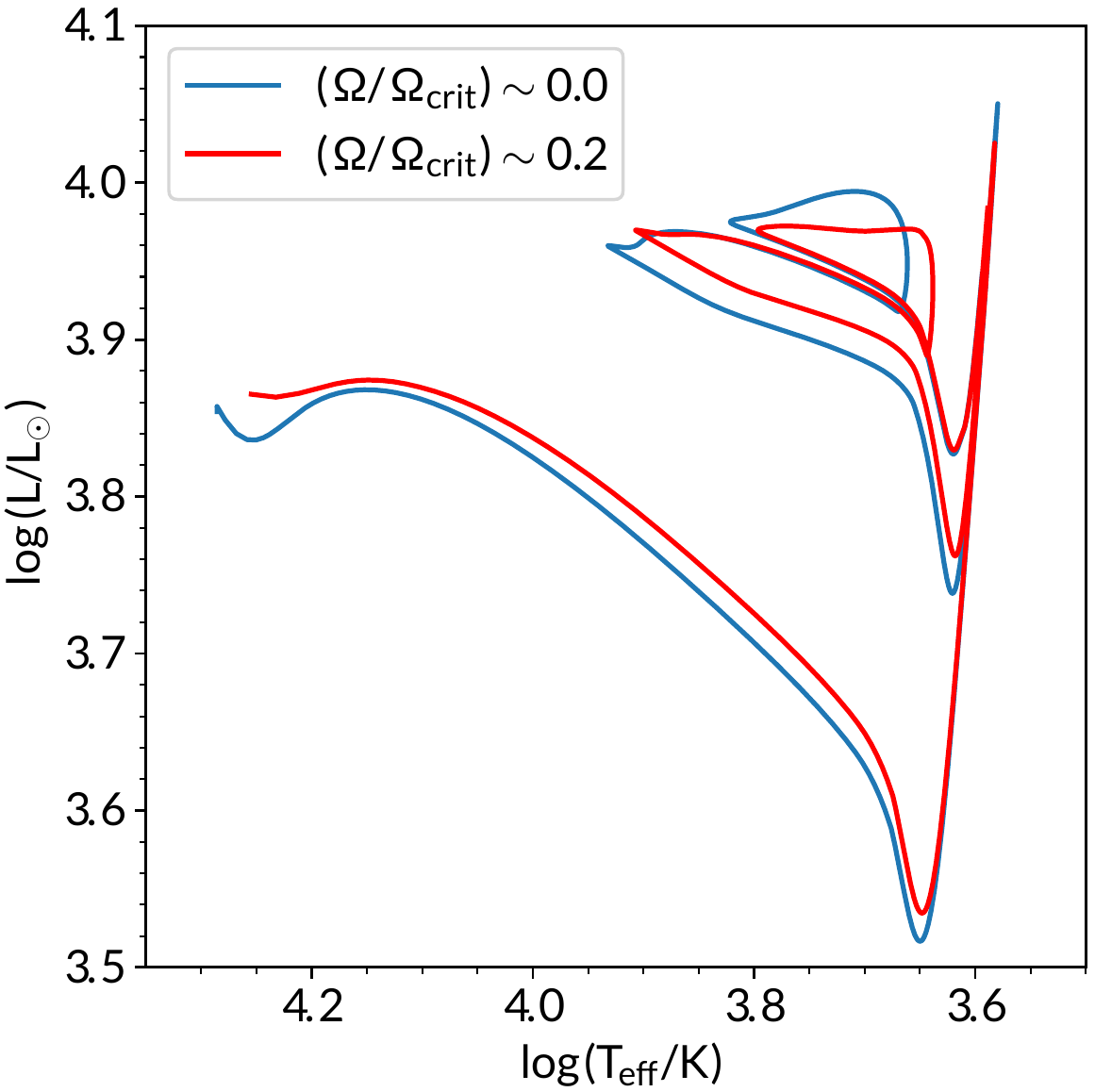}
\caption{Reproduction of Figure~13 from \citet{farmer_2015_aa} using release r6794 in \code{NuDocker}:
HRD of two 8\,\Msun\ models; one non-rotating (blue) and one rotating at $\Omega/\Omega_{\rm crit}$\,$\sim$\,0.2 (red).
The evolutions span from H depletion to He depletion.
}
\label{fig:nudocker}
\end{figure}

As an example of the ability to provide almost decade-old results with \code{NuDocker}, Figure~\ref{fig:nudocker}
reproduces Figure~13 from \citet{farmer_2015_aa} using
the same \MESA\ version 6794 from July 2014 and inlists\footnote{\url{https://zenodo.org/record/2590040}} as in \citet{farmer_2015_aa}.
The virtual containers
allow older versions of \MESA\ to be run with bit-for-bit consistency for all versions after 7503 (see \mesathree),
thereby enabling older versions of \MESA\ to be run on modern hardware, and
preserving the required system environment to enable new research.


\section{Summary}
\label{sec:summary}

We explain significant new capabilities and improvements implemented in \MESA\ since the publication of
\mesaone\ through \mesafive.
Advances in automatic differentiation (\S\ref{sec:auto_diff})
and
time-dependent convection (\S\ref{sec:tdc})
will open opportunities for future investigations in stellar evolution.
Discussion of the current treatment of the energy equation (\S\ref{sec:energy_eq}),
stellar atmospheres (\S\ref{sec:atmosphere}),
and new formalisms for treating starspots and superadiabatic convection (\S\ref{sec:mltconv})
will enhance the study of stellar physics.
Upgrades to the
equation of state (\S\ref{sec:eos}),
opacity (\S\ref{sec:opacity}),
element diffusion coefficients (\S\ref{sec:diffusion}),
nuclear reaction rates (\S\ref{sec:rates}), and
physical constants (\S\ref{sec:const})
will increase the robustness of stellar evolution models.
The transition to GitHub,
upgrades to the \MESA\ TestHub continuous integration framework (\S\ref{sec:infrastructure}),
the opening of a \code{mesa-contrib} repository,
and \code{NuDocker}'s ability to run older versions of \MESA\
will lead to an efficient and distributed model of source code development.
Inlists and related materials for all the figures are available at
doi:\dataset[10.5281/zenodo.6968760]{https://doi.org/10.5281/zenodo.6968760}.



\hbox{ }

We thank Dr.~Bill Paxton for two decades of his amazing talents, extraordinary energy, 
and his generous gift of \MESA\ to the astronomy and astrophysics community.

We also thank 
Amy Mainzer for Figure 4 of \mesathree\ making an appearance in the movie {\it Don't Look Up} (first noted by Jared Goldberg), and
the participants of the \MESA\ Summer Schools for their willingness to experiment with new capabilities
and modalities of delivery. 
\revision{We thank the anonymous referee for a constructive, detailed, and thoughtful report that improved our manuscript.}
Finally, we thank Susie Groves for her heroic assistance in enabling the 
completion of this MESA VI instrument paper under challenging circumstances.

The MESA Project is supported by the National Science Foundation (NSF)
under the Software Infrastructure for Sustained Innovation program grants ACI-1663684, ACI-1663688, and ACI-1663696.
This research was supported in part by the NSF under Grant No.~NSF~PHY-1748958 for the Kavli Institute for Theoretical Physics.
W.H.B acknowledges support from the UK Science and Technology
Facilities Council (STFC) through grant ST/R0023297/1.
G.C.C acknowledges support by the Australian Research Council Centre of Excellence for All Sky Astrophysics in 3 Dimensions (ASTRO 3D), through project number CE170100013 and the Astronomical Society of Australia.
R.F acknowledges support of the University of Amsterdam's Helios cluster which was supported by a European Research Council grant 715063, (PI S.E. de Mink)
F.H acknowledges funding through an NSERC Discovery Grant, through NSERC project award SAPPJ-2021-00032 and 
through the NSF under Grant PHY-1430152 for the JINA Center for the Evolution of the Elements.
The Flatiron Institute is supported by the Simons Foundation.
A.S.J thanks the Gordon and Betty Moore Foundation (Grant GBMF7392) and the National Science Foundation (Grant No. NSF PHY-1748958) for supporting this work.
M.J acknowledges the Lasker Data Science Fellowship awarded by the Space Telescope Science Institute, and thanks Marc Pinnsoneault, Jen van Saders, and Jamie Tayar for many hours of consultation on the Yale Rotating Stellar Evolution Code and its documentation.   
J.S acknowledges support by NASA through Hubble Fellowship grant \# HST-HF2-51382.001-A awarded by the Space Telescope Science Institute, which is operated by the Association of Universities for Research in Astronomy, Inc., for NASA, under contract NAS5-26555, by the A.F. Morrison Fellowship in Lick Observatory, and by the National Science Foundation through grant ACI-1663688.
R.S acknowledges support by the National Science Center, Poland, Sonata BIS project 2018/30/E/ST9/00598.
A.T is a Research Associate at the Belgian Scientific Research Fund (F.R.S.-F.N.R.S.). 
F.X.T acknowledges support by NASA under the Astrophysics Theory Program grant NNH21ZDA001N-ATP, 
and by the NSF under Grant PHY-1430152 for the JINA Center for the Evolution of the Elements.
T.L.S.W thanks support by the Gordon and Betty Moore Foundation through Grant GBMF5076.
J.S.G.M acknowledges support by the KU Leuven Research Counsil (grant C16/18/005: PARADISE).
O.T was supported by a FONDECYT project 321038.
P.M. acknowledges support from the FWO junior postdoctoral fellowship No. 12ZY520N.
This research made extensive use of the SAO/NASA Astrophysics Data System (ADS).


\facilities{
This work used the Extreme Science and Engineering Discovery Environment \citep[XSEDE;][]{XSEDE2014}, 
which is supported by the NSF grant ACI-1548562, specifically comet at the San Diego Supercomputer Center through allocation TG-AST180050.  
We thank Charlie Conroy and the Harvard ITC for providing computational resources for continuous testing of \MESA\ through the FASRC Cannon cluster supported by the FAS Division of Science Research Computing Group at Harvard University.
J.S acknowledges use of the lux supercomputer at UC Santa Cruz, funded by NSF MRI grant AST 1828315,
and thanks Josh Sonstroem and Brant Robertson for supporting this resource.
A.S.J acknowledges use of the rusty supercomputer at the Flatiron Institute, supported by the Simons Foundation, and thanks the Scientific Computing Core for supporting this resource.
W.H.B thanks the University of Birmingham's Advanced Research Computing team for support of the BlueBEAR High-Performance Computing service.
J.S.G.M thanks the VSC (\revision{Vlaams Supercomputer Centrum - Flemish Supercomputer Center}), funded by the Research Foundation - Flanders (FWO) and the Flemish Government - department EWI.
T.L.S.W acknowledges use of computational facilities at UC Santa Barbara funded by NSF grant CNS 1725797, and thanks the Center for Scientific Computing for supporting this resource. 
}

\software{
\AE SOPUS \citep{Marigo09} \url{http://stev.oapd.inaf.it/cgi-bin/aesopus},
\texttt{FreeEOS} \citep{Irwin2004},
\texttt{ipython/jupyter} \citep{perez_2007_aa,kluyver_2016_aa},
\texttt{matplotlib} \citep{hunter_2007_aa},
\texttt{mesaPlot} \citep{mesaplot21}
\code{NuDocker} \url{https://github.com/NuGrid/NuDocker}, \url{https://doi.org/10.5281/zenodo.3678601},
\texttt{NumPy} \citep{der_walt_2011_aa}.
         }


\bibliographystyle{aasjournal}



\end{document}